\documentclass[11pt]{article}

\usepackage[margin=1in]{geometry}

\usepackage{amsmath}
\usepackage{lipsum}
\usepackage{graphicx,psfrag,epsf}
\usepackage{enumerate}
\usepackage{rotating}
\usepackage{multirow,booktabs,threeparttable,color}
\usepackage{url} 
\usepackage{amsfonts, amssymb, amsmath,bm}
\usepackage{url}
\usepackage{hyperref}
\usepackage{fancyhdr}
\usepackage{multicol}
\usepackage{courier} 
\usepackage{float}
\usepackage{subcaption}
\usepackage{titlesec} 
\usepackage{tikz}
\usetikzlibrary{automata,arrows,positioning,calc}
\usepackage{listings}
\usepackage{indentfirst} 
\usepackage{tablefootnote}
\usepackage[english]{babel}
\usepackage{bbm}
\usepackage[longnamesfirst]{natbib}
\usepackage{enumitem}
\usepackage{pdfpages}

\usepackage[margin=10pt,labelfont=bf]{caption}
\usepackage{subcaption}
\usepackage{float}
\usepackage{setspace} 
\usepackage{amsmath,amsthm,amssymb,amsfonts,graphicx,dsfont}
\usepackage[maxfloats=100]{morefloats}
\usepackage[utf8]{inputenc}
\usepackage{booktabs}
\usepackage{epsfig}
\usepackage{url}
\usepackage{bm}
\usepackage{listings}
\usepackage{verbatim}
\usepackage{xcolor}
\usepackage[title]{appendix}
\usepackage{multirow}
\usepackage{longtable}
\usepackage{adjustbox}

\newcommand{\q}{\text{\quad}}
\newcommand{\wh}{\text{where }}

\newcommand{\oo}{\bm{1}}

\newcommand{\PP}{\mathbb{P}}

\newcommand{\sgn}{\textrm{sign}}

\newcommand{\RR}{\mathbb{R}}

\newcommand{\tr}{\mathrm{tr}}

\newcommand{\CP}{\stackrel{p}{\rightarrow}}

\newcommand{\iid}{\stackrel{iid}{\sim}}

\newcommand{\EE}{\mathbb{E}}

\usepackage{rotating} 

\DeclareMathOperator*{\argmax}{arg\,max}
\DeclareMathOperator*{\argmin}{arg\,min}

\newtheorem{definition}{DEFINITION}

\newtheorem{assumption}{ASSUMPTION}
\newtheorem{theorem}{THEOREM}
\newtheorem{app-theorem}{APPENDIX THEOREM}

\newtheorem{corollary}{COROLLARY}
\newtheorem{procedure}{PROCEDURE}

\newcommand\tcaptab[1]{\captionsetup{position=top, font=normalsize, labelfont=bf, textfont=normalfont, justification=centering, margin=0mm, aboveskip=1mm, belowskip=0mm, labelsep=colon, singlelinecheck=false}\caption{#1}}
\newcommand\bnotetab[1]{\captionsetup{position=bottom, font=footnotesize,  textfont=normalfont, margin=1mm, skip=2mm, justification=justified, singlelinecheck=false}\caption*{#1}}

\newcommand\bnotefig[1]{\captionsetup{position=bottom, font=footnotesize,  textfont=normalfont, margin=1mm, skip=2mm, justification=justified, singlelinecheck=false}\caption*{#1}}

\begin{document}
	
	\title{Inference for Large Panel Data with Many Covariates\thanks{\scriptsize We thank Siddhartha Chib
			and Ruoxuan Xiong, as well as conference and seminar participants at Stanford, the California Econometric conference and the NBER-NSF SBIES conference for helpful comments. Jiacheng Zou gratefully acknowledges the generous support by the MS\&E Departmental Fellowship, and Charles \& Katherine Lin Fellowship.}}
	
	\date{December 31, 2022}
	\author{Markus Pelger\thanks{\scriptsize Stanford University, Department of Management Science \& Engineering, Email: mpelger@stanford.edu.}
		\and
		Jiacheng Zou\thanks{\scriptsize Stanford University, Department of Management Science \& Engineering, Email: jiachengzou@stanford.edu}
	}

	\onehalfspacing
	
	\begin{titlepage}
		\maketitle
		\thispagestyle{empty}
		
		\begin{abstract}

			This paper proposes a novel testing procedure for selecting a sparse set of covariates that explains a large dimensional panel. Our selection method provides correct false detection control while having higher power than existing approaches. We develop the inferential theory for large panels with many covariates by combining post-selection inference with a novel multiple testing adjustment. Our data-driven hypotheses are conditional on the sparse covariate selection. We control for family-wise error rates for covariate discovery for large cross-sections. As an easy-to-use and practically relevant procedure, we propose Panel-PoSI, which combines the data-driven adjustment for panel multiple testing with valid post-selection p-values of a generalized LASSO, that allows us to incorporate priors. In an empirical study, we select a small number of asset pricing factors that explain a large cross-section of investment strategies. Our method dominates the benchmarks out-of-sample due to its better size and power.

			\vspace{1cm}
			

		\end{abstract}
	\end{titlepage}

	\onehalfspacing

	\section{Introduction}

	
	Our goal is the selection of a parsimonious sparse model from a large set of candidate covariates that explains a large dimensional panel. This problem is common in many social science applications, where a large number of potential covariates are available to explain the time-series of a large cross-section of units or individuals. An example is empirical asset pricing, where the literature has produced a “factor zoo” of potential risk factors to explain the large cross-section of stock returns. This problem requires a large panel, as a successful asset pricing model should explain the many available investment strategies, resulting in a large panel of test assets. At the same time, there is no consensus about what are the appropriate risk factors, which leads to a statistical selection problem from a large set of candidate covariates. So far, the literature has only provided solutions to one of the two subproblems, while keeping the dimensionality of the other problem small. Our paper closes this gap.

	
	The inferential theory on a large panel with many covariates is a challenging problem. As a first step, we have to select a sparse set of covariates from a large pool of candidates with a regularized estimator. The challenge is to provide valid $p$-values from this estimation that account for the post-selection inference. Furthermore, researchers might want to impose economic priors on which variables should be more likely to be selected. The second challenge is that the panel cross-section results in a large number of $p$-values. Hence, some of them are inadvertently very small, which if left unaddressed leads to ``$p$-hacking''.
	The multiple testing adjustment conditional on the selected subset of covariates from the first step is a novel problem, and requires to redesign what hypotheses should be tested jointly. A naive counting of all tests is overly conservative, and the test design and simultaneity counts should to be conditional on the covariate selection. 
	
	
	This paper proposes a new method for covariate selection in large dimensional panels, tackling all of the above challenges. We develop the inferential theory for large dimensional panel data with many covariates by combining post-selection inference with a new multiple testing method specifically designed for panel data. Our novel data-driven hypotheses are conditional on sparse covariate selections and valid for any regularized estimator. Based on our panel localization procedure, we control for family-wise error rates for the covariate discovery and can test unordered and nested families of hypotheses for large cross-sections. As an easy-to-use and practically relevant procedure, we propose Panel-PoSI, which combines the data-driven adjustment for panel multiple testing with valid post-selection $p$-values of a generalized LASSO, that allows to incorporate priors. In simulations and an empirical study we show that our selection method provides correct false detection control but has substantially higher power than existing approaches.
	
	Our paper proposes the novel conceptual idea of data-driven hypotheses families for panels. This allows us to put forward a unifying framework of valid post-selection inference and multiple testing. Leveraging our data-driven hypotheses family, we adjust for multiple testing with a localized simultaneity count, which increases the power, while maintaining false discovery rate control. An essential step for a formal statistical test is to formulate the hypothesis. This turns out to be non-trivial for a large panel with a first stage selection step for the covariates. It is a fundamental insight of our paper, that the hypothesis of our test has to be conditional on the selected set of active covariates of the first stage. Once we have defined the appropriate hypothesis, we can deal with the multiple testing adjustment, which by construction is also conditional on the selection step.  
	
	Our method is a disciplined approach based on formal statistical theory to construct and interpret a parsimonious model. It goes beyond the selection of a sparse set of covariates as it also provides the inferential theory. This is important as it allows us to rank the covariates based on their statistical significance and can also be applied for relatively short time horizons, where cross-validation for tuning a regularization parameter might not be reliable. We answer the question which covariates are needed to explain the full panel jointly, and can also accommodate ``weak'' covariates or factors that only affect a small subset of the cross-sectional units. 
	

	Our data-driven hypothesis perspective exploits the geometric structure implied by the first stage selection step. Given valid post-selection $p$-values of a regularized sparse estimator from time-series regressions, we collect them across the large cross-section into a ``matrix'' of $p$-values.  Only active coefficients, that are selected in the first stage, contribute $p$-value entries, whereas covariates that were non-active lead to ``holes'' in this matrix. We leverage the non-trivial shape of this matrix to form our adaptive hypotheses. This allows us to make valid multiple testing adjusted inference statements, for which we design a panel modified Bonferroni-type procedure that can control for the family-wiser error rate (FWER) in the discovery of the covariates. As one loosens the FWER requirements, the inferential thresholds admits more and more explanatory variables. Hence, the number of admitted covariates and the FWER control level form an ``false-discovery control frontier''. We provide a method that allows us to traverse the inferential results and determine the least number of covariates that have to be included given a user-specified FWER level. In other words, we can make a statement on the number of covariates or factors needed to explain a panel based on a statistical significance requirement. 

	
	We propose the novel procedure Panel-PoSI, which combines the data-driven adjustment for panel multiple testing with valid post-selection $p$-values of a generalized LASSO. While our multiple testing procedure is valid for any sparsity constrained model, Panel-PoSI is an easy-to-use and practically relevant special case. We propose Weighted-LASSO for the first stage selection regression and provide valid $p$-values through post-selection inference (PoSI), which yields a truncated-Gaussian distribution for an adjusted LASSO estimator. This geometric perspective is less common in the LASSO literature, but has the advantage that it avoids the use of infeasible quantities, in particular the second moment of the large set of potential covariates. The Weighted-LASSO generalizes LASSO by allowing to put weights onto prior belief sets. For example, a researcher might have economic knowledge that she wants to include in her statistical selection method, and impose an infinite prior weight to include specific covariates in the sparse selection model. Our Weighted-LASSO makes several contributions. First, the expression for the truncated conditional distribution with weights become much more complex than for the special case of the conventional LASSO. Second, we provide a simple, easy-to-use and asymptotically valid conditional distribution in the case of an estimated noise variance. 
	
	
	We demonstrate in simulations and empirically that our inferential theory allows us to select better models. We compare different estimation approaches to select covariates and show that our approach better trades off false discovery and correct selections and hence results in a better out-of-sample performance. Our empirical analysis studies the fundamental problem in asset pricing of selecting a parsimonious factor model from a large set of candidate factors that can jointly explain the asset prices of a large cross-section of investment strategies. We consider a standard data set of 114 candidate asset pricing factors to explain 243 double sorted anomaly portfolios. We show that Panel PoSI selects 3 factors which form the best model to explain out-of-sample the expected returns and the variations of the test assets. The selected factors are economically meaningful and include the size and value factors of the Fama-French model. Hence, our statistical selection procedure confirms two of the most widely used asset pricing factors. Our findings contribute to the discussion about the number of asset pricing factors. We confirm that independent of the rotation of the covariates we select 3 factors for a 5\% FWER control.  
	
	
	
	The rest of the paper is organized as follows. Section \ref{sec:lit} relates our work to the literature. Section \ref{sec:model} introduces the model and the Weighted-LASSO. Section \ref{sec_hypothesis} discusses the appropriate hypotheses to be considered for inference on the entire panel. Section \ref{sec_multiple_testing} proposes a joint unordered test for the panel using multiple testing adjustment so that we can maintain FWER control, and shows how to traverse this procedure to acquire the least covariate count associated with each FWER target. In section \ref{sec:ordered}, we consider the case of nested hypotheses, where the covariates follow a fixed ordering, which is of independent interest, and we propose a step-down procedure for this setting that maintains false discovery control. Section \ref{sec:simulation} provides the results of our simulation and Section \ref{sec:empirics} discusses our empirical analysis on a large asset pricing panel data set. Section \ref{sec6} concludes. The Appendix collects a detailed discussion about post-selection LASSO. The proofs and more technical details are available in the Online Appendix.

	\label{sec:intro}
	
	\subsection{Related Literature}
	\label{sec:lit}
	
	The problem of multiple testing is an active area of research with a long history. The statistical inference community has studied the problem of controlling the classical FWER since \cite{Bonf}, and controlling for false-discover rate (FDR) going back to \cite{BH95} and \cite{BY01}. \cite{Bonf} allows for arbitrary correlations in the test statistics because its validity comes from a simple union bound argument, and is in fact the optimal test when statistics are ``close to independent'' under true sparse non-nulls. FDR control on the other hand requires a discussion about the estimated covariance in the test statistics. Recent developments include a stream of papers led by \cite{15-AOS1337} and \cite{rssb.12265}, which constructs a generative model to produce fake covariates and control for FDR. \cite{fithian2022conditional} is a more recent work that iteratively adjusts the threshold for each hypothesis in the family to seek finite sample exact FDR control and dominates \cite{BH95} and \cite{BY01} in terms of power. Another notion on temporal false discovery control has been revived more recently by \cite{doi:10.1287/opre.2021.2135}, who consider the industry practice of constantly checking $p$-values and provide an early stopping in line with \cite{SiegmundSeq} that adjusts for a bias from sequentially picking favorable evidence, whereas we consider a static panel that is not an on-going experiment.

	There are cases where the covariates warrant a natural order such that the hypothesis family possesses a special testing logic. A hierarchical structure in covariates arises when the inclusion of the next covariate only make sense if the previous covariates is included. An example is the use of principal component (PC) factors, where PCs are included sequentially from the dominating one to the least dominating one. We distinguish this from putting weights and assigning importance on features because this variant of family of hypotheses warrants a new definition of FWER. We propose a step-down procedure that can be considered as a panel extension of \cite{rssb.12122}, relying on an approximation of the R\'enyi representation of $p$-values. The step-down control for nested FWER is based on \cite{2336545}, which along with \cite{Bonf} can be seen as comparing sorted $p$-values against linear growth. Our framework contributes to estimating the number of principal component factors in a panel. There are have been many studies that provide consistent estimators for the number of PCs based on the divergence in eigenvalues of the covariance matrix, which include \cite{1468-0262.00273}, \cite{40985808}, \cite{ECTA8968} and \cite{PELGER201923}. Another direction uses sequential testing procedures that presume correct nested family of hypotheses, which include \cite{jbes.2009.07239} and \cite{16-AOS1536}. In contrast, we characterize the least number of covariates (which can also be based on principal components), which should be expected when a FWER rate is provided. The nested version of our procedure is close in nature to a panel version of ``when-to-stop'' problem of a multiple testing procedure.

	The problem of post-LASSO statistical testing for small dimensional cross-sections is studied in a stream of papers including \cite{009053606000000281}, \cite{rssb.12026}, \cite{14-AOS1221} and \cite{10.1214/17-AOS1630}, which consider inference statements by debiasing the LASSO estimator. An alternative stream of post-selection or post-machine learning inference literature includes \cite{annurev-economics-012315-015826}, \cite{kuchibhotla2018valid} and \cite{zrnic2020postselection}, who provide non-parametric post-selection or post-regularization valid confidence intervals and $p$-values. These papers do not make conditional statements and presume that the researcher sets the hypotheses before seeing the data, which we will refer to as data agnostic hypothesis family. We follow a different train of thought that treats LASSO, among a family of conic maximum likelihood estimator, as a polyhedral constraint on the support of the response variable. This geometric perspective that provides inferential theory post-LASSO is pioneered by the work of \cite{lee2016exact} and followed up by \cite{fithian2017optimal} and \cite{tian2018selective}, assuming Gaussian linear models. \cite{markovic2018unifying} extend the results to LASSO with cross-validation, \cite{tian2017selective} discuss a square-root LASSO variant that takes an unknown covariance into consideration. \cite{taylortibshirani2016inference} and \cite{tian2017asymptotics} study asymptotic results that allow to relax the assumptions of Gaussian errors or generalized linear models. This body of literature is often referred to as PoSI, and traverses the Karush-Kuhn-Tucker (KKT) condition of a LASSO optimization problem to show that the LASSO fit can be expressed as a polyhedral constraint on the support of the response variable. We extend this work by allowing to put weights onto prior belief sets, and by bringing it to the panel setting with multiple testing adjustment.

	\section{Sparse linear models}\label{sec:model}
	

	We consider a large dimensional panel data set $\bm{Y} \in \RR^{T\times N}$ which we want to explain with a large number of potential covariates $\bm{X} \in \RR^{T\times J}$. The panel data and explanatory variables are both observed over $T$ time periods.\footnote{Our setting and multiple testing results can be readily extended to the case of unbalanced panel, although we focus on the balanced panel case for now to highlight the core multiple testing insight of our method. We will further discuss on this once we introduce our main procedure in Section \ref{sec_multiple_testing}} The size of the cross-section $N$ and the dimension of the covariate candidate set $J$ are both large in our problem. 
	
	We assume a linear relationship between $\bm{Y}$ and $\bm{X}$:
	\begin{align*}
		Y_{t}^{(n)} = \sum_{j=1}^J X_{t,j} \beta_{j}^{(n)} + \epsilon_{t}^{(n)} \qquad \text{for $n=1,...,N$},
	\end{align*} 
	
	which reads in matrix notation as 
	\begin{equation}\label{1}
		\bm{Y}=\bm{X}\bm{\beta}+\bm{\epsilon}.
	\end{equation}
	We refer to the coefficients $\bm{\beta}$ as loading matrix, where the $n$th column $\beta^{(n)}\in\RR^{J}$ corresponds to the $n$th unit and $\beta^{(n)}_{j}$ denotes the loading of the $n$th unit on the $j$-th covariate. The remainder term $\bm{\epsilon}$ is unexplained noise. Throughout this paper, we use the superscript $\cdot^{(n)}$ to denote cross-sectional variables corresponding to the $n$th unit, subscript $\cdot_j$ for variables corresponding to the $j$th covariate, and subscript $\cdot_t$ for time-series corresponding to the $t$th time period.

	We assume that a sparse linear model can explain jointly the full panel. Formally, a sparse linear model with $s$ active covariates is 
	\begin{equation}\label{2}
		\bm{Y}=\bm{X}_S\bm{\beta}_S+\bm{\epsilon}
	\end{equation}
	where  $s=|S|$ is the cardinality of the set of active covariates $S=\{j:\exists  \beta^{(n)}_{j}\neq 0, n \in \{1,...,N\}$, that is, the set of covariates with non-zero loadings. $\bm{X}_S$ is the subset of covariates that belong to $S$. Our goal is to estimate this low dimensional model, that can explain the full panel, from a large number of candidate covariates, and provide a valid inferential theory.
	
	Note that our sparse model formulation allows for two important properties. First, different units can be explained by different covariates with different loadings. This means that $\beta^{(n)}\neq \beta^{(m)}$ for $n \neq m$ is allowed. For example, a subset of the cross-sectional units might be modeled by different covariates than the remaining part of the panel. Second, we can accommodate ``weak'' covariates. A covariate is included in $S$ if it is required by at least one cross-sectional unit as explanatory variable. In other words, a sparse model can include covariates in $\bm{X}_S$ that explain only a very small subset of the panel $\bm{Y}$.

	The first step is to estimate the sparse models over the time-series for each unit separately due to the heterogeneity in the loadings. In a second step, we provide the valid inferential theory for the loadings on the full panel. The time-series estimation requires an appropriate regularization to select a small subset of covariates that contains all the relevant covariates for each unit. We allow for prior belief weights $\omega_j\in (0,+\infty]$ on the $J$ candidate covariates, so that different $\bm{X}$ can have different relative penalizations, and a global $\lambda \in\RR_+$ scalar penalty parameter. For the $n$th unit, we denote its $\beta^{(n)}$ regularized estimate as $\hat{\beta}^{(n)}$ and the active set $M^{(n)}=\{j:\hat{\beta}^{(n)}_j\neq 0\}$ as the set of $j$'s with non-zero loadings $\hat{\beta}^{(n)}_j$. A general regularized linear estimator solves the following optimization problem 
	\begin{equation}\label{4}
		\hat{\beta}^{(n)}(\lambda,\omega)=\argmin_{\beta} \frac{1}{2T}\|Y^{(n)}-\bm{X}\beta\|_2^2+\lambda\cdot f(\beta,\omega)
	\end{equation}
	for a penalty function $f$ and appropriate weights, where $Y^{(n)}$ is the vector of response variables of $n$th unit. In this paper, we consider the weighted-LASSO estimator with the regularization function 
	\begin{equation}
		f(\beta,\omega)=\sum_{j=1}^J f_j(\beta_j,\omega_j)
		\quad\wh
		f_j(\beta_j,\omega_j)=\begin{cases}
			\frac{|\beta_j|}{\omega_j}& \omega_j<\infty\\
			0& o.w.
		\end{cases}
	\end{equation}
	and weights $\omega_j>0$ for all $j\in \{1,...,J\}$ and $\sum_{j=1}^J\omega_j^{-1}=J$. We consider the penalty $\lambda$ as exogenously provided such that the set $\|\hat{\beta}^{(n)}\|_0=|M^{(n)}|$ is low dimensional.\footnote{In Appendix \ref{lab:appendixA2} we discuss the case where variances of $\bm{\epsilon}$ are unknown and need to be estimated. We provide a specific discussion on $\lambda$'s rate conditions in terms of $J$ and $T$, as is typically required to ensure consistency in the LASSO literature.} Importantly, we do not need to assume that the selected set contains all ``true'' active covariates. Our goal is to provide a valid inferential theory conditional on the selected set. Our estimator generalizes the conventional LASSO with the $l_1$ regularization function of \cite{2346178} by allowing for different relative weighting in the penalty. Importantly, we also allow for an infinite weight, which can be interpreted as a prior on a set of covariates. This allows researchers to take advantage of prior information and for example ensure that a specific set of covariates will always be included. The weighted-LASSO will be particularly relevant in our empirical study, where we can answer the question which risk factors should be added to a given set of economically motivated risk factors. Our weighted-LASSO formulation can also be interpreted as a Bayesian estimator with the canonical Laplacian prior.

	Conventional regression theory will not provide correct inferential statements on the weighted-LASSO estimates. We face two challenges. First, regularized estimation results in a bias, which needs to be corrected. Second and more challenging, post-selection inference changes the distribution of the estimators. When we observe an active $\hat{\beta}^{(n)}_j$ from (\ref{4}), it would be incorrect to simply calculate its $p$-value from a conventional Student $t$-distribution. This invalidity stems from the fact that conditional on observing a LASSO output, ${\beta}^{(n)}_j$ must be large enough in magnitude for its $\hat{\beta}^{(n)}_j$ to be active. In other words, the probability distribution of the estimators is truncated. 
	

	The correct inference has to be conditional on the covariates being selected by the LASSO estimator. Hence, valid $p$-values have to be the tail probability conditional on being in the selection set. The key to quantify such styles of inference is to recognize that a sparsity constrained estimator is typically the result of solving Karush-Kuhn-Tucker (KKT) conditions, which can in turn be geometrically characterized as polyhedral constraints on the support of response variables. This is first established in \cite{lee2016exact}, who provide the stylized results that Post-Selection Inference (PoSI) of debiased non-weighted LASSO estimators can be calculated as polyhedral truncation on $\bm{Y}$. This line of research is also referred to as Selective Inference, for example in \cite{taylor2015statistical}.  We extend this line of literature to allow for the Weighted-LASSO. We derive these results with assumptions common in the PoSI LASSO literature, detailed in Appendix \ref{lab:appendix}, and referred to as conventional regularity conditions for the ease of exhibition.
	
	Theorem \ref{thm1_main} shows how we calculate $p$-values from the post-selection distribution of the debiased estimate $\bar{\beta}_j^{(n)}$. The regularized estimate $\hat{\beta}_j^{(n)}$ has a well-known bias. We debiase the LASSO estimate by a shifting argument. While we use a geometric argument to remove the bias, the bias adjustment takes the usual form in the LASSO literature as for example in \cite{10.3150/11-BEJ410}. The debiased LASSO estimator simply equals a standard OLS estimation on the subset $M^{(n)}$ selected by the Weighted-LASSO.


	\begin{theorem}{\bf Truncated Gaussian Distribution of Feasible Weighted-LASSO}\label{thm1_main}\\
		Under the conventional regularity conditions stated in Assumptions \ref{asu1} and \ref{asu_known} in the Appendix, the debiased estimate $\bar{\beta}_j^{(n)}$ for the $j$-th Weighted-LASSO active covariate of the $n$th unit is conditionally distributed as
		\begin{equation}\label{eq:thm1_formula}
			\bar{\beta}_{j}^{(n)}|\textrm{Weighted-LASSO}			\sim \mathcal{TN}_{\{\eta^\top Y^{(n)}:AY^{(n)}\leq b(Y^{(n)},\omega)\}},
		\end{equation}
		where $\mathcal{TN}_{\mathcal{A}}$ is truncated Gaussian with truncation $\mathcal{A}$, and the weights $\omega$ only appear in $b(Y^{(n)},\omega)$. 
  Under the null hypothesis $H_D$, that the active covariates of unit $n$ have zero coefficients, and conditional on the selection events and the weights, the post-selection $p$-values of the active coefficients follow a uniform distribution, that is, 
		$$p^{(n)}_j \stackrel{H_D|\mathcal{M},\omega}{\sim} \textrm{Unif }[0,1].$$
  Under conventional asymptotic conditions stated in \ref{asu_consist} and \ref{asu_gram} in the Appendix, and for $T\rightarrow \infty$, the same truncated Gaussian distribution holds for feasible $p$-values with estimated noise variance. 
  	\end{theorem}

	
	Theorem \ref{thm1_main} has two key elements. 
	First, the distribution of the linear coefficients is not a usual Gaussian distribution, but it is truncated due to studying post-selection coefficients. This geometric perspective is less common in the LASSO literature, but provides several advantages. One advantage of the geometric approach is that it avoids the use of infeasible quantities, in particular the second moment of the large set of potential covariates. Second, conditional on the selection and under the null hypothesis that the coefficients of the active covariates for unit $n$ are zero, the post-selection $p$-values of the active coefficients follow a uniform distribution. This is important as it implies that we obtain valid post-selection $p$-values, which provide the correct Type-I error control for individual regressions. These individually valid $p$-values are the key for deriving the multiple testing adjustment in large panels.
	
	Appendix \ref{lab:appendixA1} provides the detailed information on constructing $\bar{\beta}$ and the definitions of $\eta, A, b(\omega)$ along with lemmas that lead up to this result. It also shows that, under the assumption of a known noise variance, the distribution result is not asymptotic in $T$, but also valid in finite samples. We can obtain these results because we make the stronger assumption that the noise is normally distributed. When using a consistent sample estimate of the noise variance, we need to require additionally that $T \rightarrow \infty$. Appendix \ref{lab:appendixA1} clarifies the implications of different assumptions in the three Theorems \ref{thm1}, \ref{thm:appendix} and \ref{col2}. Theorem \ref{col2} with estimated noise variance represents the explicit form of Theorem \ref{thm1_main}, which we use for our empirical analysis. 
	It is possible to relax the normality assumption of the noise and instead use a pivot convergence similarly to \cite{tian2017asymptotics} to obtain asymptotically a truncated Gaussian distribution. However, this would not change the nature of our statement.

	Our Weighted-LASSO results make several contributions. First, the expression for the truncated conditional distribution with weights become more complex than for the special case of the conventional LASSO. Second, we provide a simple, easy-to-use and asymptotically valid conditional distribution in the case of an estimated noise variance. Last but not least, we show the formal connection with alternative debiased LASSO estimators by showing that debiasing can be interpreted as one step in a Newton-Ralphson method of solving a constrained optimization.

	Theorem \ref{thm1_main} allows us to obtain valid post-selection $p$-values for Weighted-LASSO coefficients. We obtain these $p$ values from the simulated cumulative distribution function of the truncated Gaussian distribution. Crucially, all results for multiple testing adjustment in panels that we study in the following sections neither require us to use a weighted Lasso estimator nor to use the $p$-values implied by Theorem \ref{thm1_main}. We only require to have a set of valid post-selection $p$-values for sparsity constrained models. These can be obtained with any suitable regularized estimator and post-selection inference. The key element is the selection of a low dimensional subset with $p$-values conditional on this selection. We propose the weighted LASSO conditional inference results as an example of the type of sparsity constraint models we are interested in, and demonstrate a machinery with which we can obtain valid $p$-values for sparsity constrained models. In our empirical studies, we use Weighted-LASSO as our sparsity constrained model since we want to specify strong prior beliefs on a few covariates and it is common practice to use LASSO in the context of our empirical studies. Nonetheless, the testing methods in the next sections accommodate any sparse estimator, and can be detached from inference for Weighted-LASSO.

	
	\section{Data-Driven Hypotheses}\label{sec_hypothesis}

	Our goal is to provide formal statistical tests that allow us to establish a joint model across a large cross-section with potentially weak covariates. This requires us to provide a form of statistical significance test with multiple testing adjustment that properly accounts for covariates that only explain a small subset of the cross-sectional units. This is important as in many problems in economic and finance, there is substantial cross-sectional variation in the explanatory power of covariates, and a model that simply minimizes an average error metric might neglect weaker covariates.
	
	An essential step for a formal statistical test is to formulate the hypothesis. This turns out to be non-trivial for a large panel with a first stage selection step for the covariates. It is a fundamental insight of our paper, that the hypothesis of our test has to be conditional on the selected set of active covariates of the first stage. Once we have defined the appropriate hypothesis, we can deal with the multiple testing adjustment, which by construction is also conditional on the selection step. 
	
	Our hypothesis formulation and test construction only requires valid post-selection $p$-values from a first stage selection estimator as formalized in Section \ref{sec_multiple_testing}.
The results of the next two sections do not depend on a specific model for obtaining these $p$-values and the active set. The results are valid for any model including non-linear ones. The input to the analysis is a $N \times J$ matrix, which specifies which covariates are active for each unit and the corresponding post-selection $p$-values. The Weighted-LASSO is only one possible model, but it can be replaced by any regularized model. We have introduced the sparse linear model as it is the horse race model for many problems in economics and finance, and therefore of practical relevance.

\begin{figure}[t!]
	\tcaptab{Illustrative example of data-driven selection}\label{fig:demo_running}
	\begin{center}\begin{subfigure}[t]{.35\textwidth}
			\includegraphics[width=0.75\linewidth]{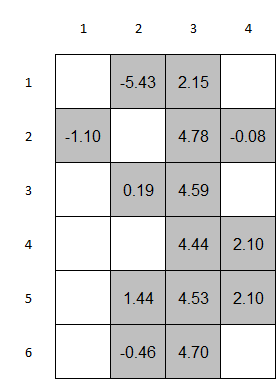}
			\caption{Matrix $\bar{\bm{\beta}}$}
		\end{subfigure}\hspace{2.5cm}
		\begin{subfigure}[t]{.35\textwidth}
			\includegraphics[width=0.75\linewidth]{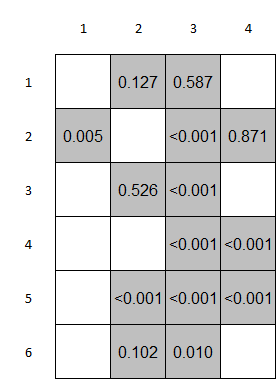}
			\caption{Matrix $\bm{P}$ of $p$-values}
	\end{subfigure}\end{center}
	\bnotetab{This figure illustrates in a simple example the data-driven selection of a linear sparse model. In a first stage, we have estimated a regularized sparse linear model for each of the $N=6$ units with $J=4$ covariates. Each row represents the selected covariates with their estimated coefficients and $p$-values. The columns represent the $J=4$ different covariates. The grey shaded boxes represent the active set, while white boxes indicate the inactive covariates. The numbers are purely for demonstrative purposes.}
\end{figure}

We illustrate the concept of a data-driven hypothesis with a simple example, which we will use throughout this section. For simplicity we assume that we have $J=4$ covariates and want to explain $N=6$ cross-sectional units. In the first stage, we have estimated a sparse model and have obtained the post-selection valid $p$-values for each of the $N$ units. We collect the fitted sparse estimator $\bar{\beta}^{(n)}$ for the $n$th unit in the matrix $\bar{\bm\beta}$.  Note, that this matrix has ``holes'' due to the sparsity for each $\bar{\beta}^{(n)}$.  Figure \ref{fig:demo_running}(a) illustrates $\bar{\bm\beta}$ for this example.

Similarly, we collect the corresponding $p$-values in the matrix $\bm{P}$. For the $n$th unit, we only have $p$-values for those covariates that are active in the $n$th linear sparse model. Thus, Figure \ref{fig:demo_running}(b) also has white boxes showing the same pattern of unavailable $p$-values due to the conditioning on the output of the linear sparse model. These holes can appear at different positions for each unit, which makes this problem non-trivial. This non-trivial shape of either subplot (a) or (b) is completely data-driven and a consequence of linear sparse model selection. We show that the hypothesis should be formed around these non trivial shapes as well, which is why we name it the data-driven hypothesis family.

We want to test which covariates are jointly insignificant in the full panel. A data-agnostic approach would simply test if all covariates are jointly insignificant, independent of the data-driven selection step in the first stage. A data-agnostic hypothesis is unconditional as it does not depend on any model output. However, as we will show, this perspective is problematic for the high-dimensional panel setting with many covariates as it ignores the dimension reduction from the selection step. Therefore, an unconditional multiple testing adjustment accounts for ``too many'' tests, which severely reduces the power.

We propose to form the hypothesis conditional on the first stage selection step. The data-driven hypothesis only tests the significance of the covariates that were included in the selection, and hence can drastically reduce the number of hypothesis. However, given the non-trivial shape of the active set, the multiple testing adjustment for the data-driven hypothesis is more challenging. 

Before formally defining the families of hypothesis, we illustrate them in our running example. The data-agnostic hypothesis $H_A$ for explaining the full panel takes the following form: 
\begin{equation}\small
	\begin{split}
		H_A=\{
		H_{A_{0,1}}&,H_{A_{0,2}},H_{A_{0,3}},H_{A_{0,4}}
		\}\\=\{
		\beta^{(1)}_1=&\beta^{(2)}_1=\beta^{(3)}_1=\beta^{(4)}_1=\beta^{(5)}_1=\beta^{(6)}_1=0,\\
		\beta^{(1)}_2=&\beta^{(2)}_2=\beta^{(3)}_2=\beta^{(4)}_2=\beta^{(5)}_2=\beta^{(6)}_2=0,\\
		\beta^{(1)}_3=&\beta^{(2)}_3=\beta^{(3)}_3=\beta^{(4)}_3=\beta^{(5)}_3=\beta^{(6)}_3=0,\\
		\beta^{(1)}_4=&\beta^{(2)}_4=\beta^{(3)}_4=\beta^{(4)}_4=\beta^{(5)}_4=\beta^{(6)}_4=0\}
	\end{split}
\end{equation}
The data-driven hypothesis $H_D$ only includes the active set and hence equals
\begin{equation}\small
	\begin{split}
		H_D
		=\{
		\beta^{(2)}_1=&0,\\
		\beta^{(1)}_2=&\beta^{(3)}_2=\beta^{(5)}_2=\beta^{(6)}_2=0,\\
		\beta^{(1)}_3=&\beta^{(2)}_3=\beta^{(3)}_3=\beta^{(4)}_3=\beta^{(5)}_3=\beta^{(6)}_3=0,\\
		\beta^{(2)}_4=&\beta^{(4)}_4=\beta^{(5)}_4=0\}
	\end{split}
\end{equation}
Clearly, $H_A$ has a larger cardinality of $|H_A|=24>|H_D|=14$. This holds in general, unless the first stage selects all covariates for each unit, in which case the two hypotheses coincide.

Formally, the data-agnostic family of hypothesis is defined as follows:
\begin{definition}{\bf Data-agnostic family}\\
	The data-agnostic family of hypotheses is
	\begin{equation}
		\begin{split}
			H_A&=	\{
			H_{A_{0,j}}|j\in[J]
			\}\\
			\q \wh H_{A_{0,i}}&=\bigcap_{n\in[N]}H_{A_{0,i}}^{(n)}\text{ and }H_{A_{0,j}}^{(n)}:\beta^{(n)}_j=0.
		\end{split}
	\end{equation}
\end{definition}
It is evident that $H_A$ does not need any model output or exploratory analysis, so it is indeed data-agnostic.

As soon as we use a sparsity constrained model that has censoring capabilities, we no longer observe $(\bm{Y},\bm{X})$ from its data generating process. Consequently, unless our hypotheses depend on how we built the model, or equivalently on how the data was censored, the data-agnostic hypotheses forgo power without any benefit in false discovery control. Therefore, we formulate the hypothesis on the $j$th covariate $H_{0,j}^{(n)}$ only if $j\in M^{(n)}$, that is, it is in the active set of the $n$th unit. Conditional on observing the model output, there is no inference statement to be made about $H_{0,j}^{(n)}$ if $j\notin M^{(n)}$, because its estimator is censored by the model. 

We denote as $\mathcal{K}_j$ the set of units for which the $j$th covariate is active. We define the cross-sectional hypothesis for the $j$th covariate as:
\begin{equation}
	H_{0,j}=\bigcap_{n\in\mathcal{K}_i}H_{0,j}^{(n)}\bigg\rvert \mathcal{M},\q \forall j:\mathcal{K}_j\neq\emptyset .
\end{equation}
By combining all covariates $\{j:\mathcal{K}_j\neq \emptyset\}$ that show up at least once in one of the active sets of our sparse linear estimators, we arrive at a data-driven hypothesis associated with our panel. This is defined as follows:
\begin{definition}{\bf Data-driven family}\\\label{def5}
	The data-driven family of hypotheses conditional on $\mathcal{M}$ is 
	\begin{equation}\label{23}
		H_D=	\{
		H_{0,j}|j:\mathcal{K}_j\neq\emptyset 
		\}.
	\end{equation}
\end{definition}
This demonstrates the non-trivial nature of writing down a hypothesis in high-dimensional panel: we can only collect $\mathcal{K}_j$ - the set of units for which the $j$th covariate is active - after seeing the sparse selection estimation result.

\section{Multiple Testing Adjustment for Data-Driven Hypothesis}\label{sec_multiple_testing}

\subsection{Simultaneity Counts through Panel Localization}

We show how to adjust for multiple testing of data-driven hypotheses. Given the the first stage selection of active covariates in $\mathcal{P}$, we form the data-driven hypothesis $H_D$. The only assumption that we require for the multiple testing adjustment is that we have valid post-selection $p$-values $p_j^{(n)}$ for covariate $j\in M^{(n)}$ and unit $n \in \mathcal{K}_j$. This is formalized in the following assumption:

\begin{assumption}{\bf Valid post-selection $p$-values}\label{assu:valid_p}\\
	We assume that we have valid individual post-selection $p$-values $p_j^{(n)}$ for each unit $n$ and covariate $j$ in the active set in $\bm{P}$. Valid post-selection p-values are defined such that their Type-I error control satisfies
	$$\PP_{H_D|\mathcal{M},\omega}(p^{(n)}_j \leq x)\leq x,\quad \forall x\geq 0.$$
	conditional on the selection and prior weights, and under the null hypothesis that the active covariates are zero. 
\end{assumption}

Valid post-selection $p$-values abstract away from model-specific conditions on how the $p$-values are obtained. Post-selection $p$ values are trivially valid, if they follow a uniform distribution under the null hypothesis conditional on the selection event. This special case can be interpreted as exact valid post-selection $p$-values, since $p^{(n)}_j \stackrel{H_D|\mathcal{M},\omega}{\sim} \textrm{Unif }[0,1]$ implies $\PP_{H_D|\mathcal{M},\omega}(p^{(n)}_j \leq x)= x$. The result, that valid post-selection p-values conditional on the selection event and weights are uniform under the null hypothesis, is first introduced in \cite{tibshirani2016exact} for sparse regressions. In Theorem \ref{thm1_main}, we extend this result to generic sparse regressions with weights. Hence, the post-selection $p$-values of our Weighted-LASSO satisfy Assumption \ref{assu:valid_p}.

Our definition of valid post-selection $p$-values is natural as it simply states that we have correct conditional Type-I error control for each cross-sectional unit and each active covariate. This definition is in line with \cite{tibshirani2016exact} and \cite{heard2018choosing}, but more general. It also allows for more conservative $p$-values, that would have smaller size and worse power. The valid post-selection $p$-values provide the correct Type-I error control for each unit individually, and do not take the multiple testing issue into account. Hence, the $p$-values of Assumption \ref{assu:valid_p} are essentially the results of post-selection inference that is applied separately to each cross-sectional unit $n$. Given those values we show how to correct them to adjust for multiple testing.

Our derivations for the multiple testing adjustment only take advantage of the conditional valid distribution of the individual post-selection $p$-values, and hence we impose this property as the fundamental underlying assumption. Note that that our results do not require a linear model, but hold for any set of valid post-selection $p$-values.\footnote{This notion of valid $p$-values deviates from what is commonly used in regression analysis as its entire statement is based on a conditional distribution, highlighting its ``post-selection'' nature. In other words, the Assumption \ref{assu:valid_p} holds under the null $H_D$ and conditional on the selection event $\mathcal{M}$ and weights $\omega$, as opposed to the classical regression analysis that does not depend on the selection.} The assumptions on the data generating process and asymptotic regime are implicitly included in Assumption \ref{assu:valid_p}. Theorem \ref{thm1_main} is a specific example that imposes Gaussian errors and $T \rightarrow \infty$. Hence, as long as a researcher has a selection estimation approach that provides valid post-selection $p$-values, our multiple testing results are applicable.

Our goal is to reject members of $H_D$ while controlling the Type I error, and the common way to measure such an error is the family-wise error rate. This is the same underlying logic that is used to define confidence intervals and determine significance of covariates in a conventional setup. The crucial difference is that we need to account for multiple testing given the large number of cross-sectional units. The family-wise error rate (FWER) is defined as follows:

\begin{definition}{\bf Family-wise error rate}\label{def6}\\
	Let $V$ denote the number of rejections of $H_{0,j}^{(n)}|\mathcal{M}^{(n)}$ when the null hypothesis is true. The family-wise error rate (FWER) is $\PP_{H_D|\mathcal{M},\omega}(V\geq 1)$.
\end{definition}

Similar to the conventional definition, we simply count the number of Type I false rejections $V$, and define FWER as the probability of making at least one false rejection. Importantly, the FWER accounts for the fact that we might repeatedly test a specific covariate for multiple cross-sectional units rather than just for one unit. Our contribution to FWER control in the panel setting is thus to take into consideration both the multiplicities in units and covariates when we deal with the ``matrix'' of $p$-values $\bm{P}$. To achieve this goal, we propose a new simultaneity account for the $j$th covariate, calculated as
\begin{equation}\label{eq28:simcount}
	N_j=\sum_{n\in \mathcal{K}_j}|M_n|
\end{equation}

\begin{figure}[t!]
	\tcaptab{Simultaneity counts $N_i$ in the illustrative example}\label{fig:demo1}
	\begin{center}
		\begin{subfigure}[t]{.24\textwidth}
			\includegraphics[width=1\linewidth]{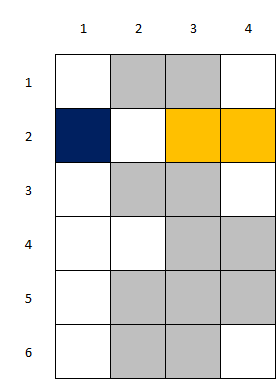}
			\caption{$N_1=3$}
		\end{subfigure}
		\begin{subfigure}[t]{.24\textwidth}
			\includegraphics[width=1\linewidth]{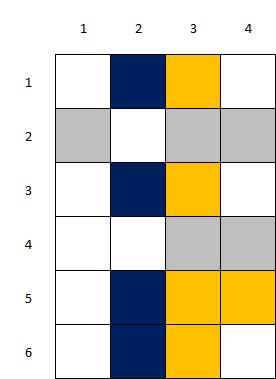}
			\caption{$N_2=9$}
		\end{subfigure}
		\begin{subfigure}[t]{.24\textwidth}
			\includegraphics[width=1\linewidth]{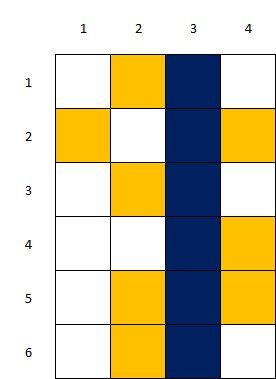}
			\caption{$N_3=14$}
		\end{subfigure}
		\begin{subfigure}[t]{.24\textwidth}
			\includegraphics[width=1\linewidth]{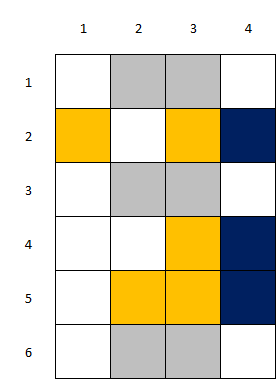}
			\caption{$N_4=8$}
		\end{subfigure}
	\end{center}
	\bnotetab{This figure shows the simultaneity counts $N_i$ in the illustrative example. The subplots represent the simultaneity counts for the $J=4$ covariates. The blue boxes indicate the active set $\mathcal{K}_j$ of the $j$ covariates, while yellow boxes indicate the ``co-active'' covariates of the $j$th covariate. The simultaneity counts are the sum of yellow and blue boxes.}   
\end{figure}

Figure \ref{fig:demo1} illustrates the simultaneity counting for our running example with $N=6$ units and $J=4$ covariates. The blue boxes represent the active set for a specific covariate. The yellow boxes indicate the ``co-active'' covariates, which have to be accounted for in a multiple testing adjustment. In the case of the first covariate $j=1$, only the second unit $n=2$ has selected this covariate. This second unit has also selected covariate $j=3$ and $j=4$, which are jointly tested with the first covariates. Hence, they are ``co-active'', and the simultaneity count equals $N_1=3$. Intuitively, $N_j$ represents all relevant comparisons for the $j$th covariate because it counts how many covariates are active with the $j$th covariate in the regressions. Hence, $N_j$ quantifies the number of ``multiple tests'' for each covariate.

In subplot \ref{fig:demo1}(a), we see that $\mathcal{K}_1=\{2\}$ for the 1st covariate, indicated by the blue box, because it is only active in the second unit's regression. The multiple testing adjustment needs to consider all yellow boxes, and $N_1=3$ is thus the total count of 1 blue and 2 yellow boxes. Similarly, for the second covariate, $\mathcal{K}_2=\{1,3,5,6\}$, so we shade boxes yellow for the 2nd, 3rd and 5th units and obtain $N_2=9$. We can already see that our design of simultaneity counts takes all relevant pairwise comparisons into considerations, but avoids counting the white boxes - which would cause overcounting and result in over-conservatism. 

Our multiplicity counting is a generalization of the classical Bonferroni adjustment for multiple testing. A conventional Bonferroni method for the data-agnostic hypothesis $H_A$ has a simultaneity count of $|H_A|=N\cdot J=24$ for testing each covariate. A direct application of a vanilla Bonferroni method to the panel of all selected units and the data-driven hypothesis $H_D$, would use a simultaneity count of $|H_D|=14$ for testing each covariate. Our proposed multiplicity counting is a refinement that leverages the structure of the problem, and takes the heterogeneity of the active sets for each covariate into account. Our count has only $N_1=3$, $N_2=9$ and $N_4=8$ for the covariates $j=1,2$ and $4$. Only for covariate $j=3$ is the simultaneity count the same as a vanilla Bonferroni count applied to $H_D$, i.e. $N_3=14$. 

In addition to the simultaneity count of each covariate, we need an additional ``global'' metric for our testing procedure. We define a panel cohesion coefficient $\rho$ as a scalar that measures how sparse or de-centralized the proposed hypotheses family is:
\begin{equation}
	\rho = \left(\sum_{1\leq j\leq J:\mathcal{K}_j\neq\emptyset}\frac{|\mathcal{K}_j|}{N_j}   \right)^{-1}
\end{equation}
The panel cohesion coefficient $\rho$ is conditional on the data-driven selection of the overall panel. It is straightforward to compute once we observe the sparse selection of the panel. This coefficient takes values between $J^{-1}$ and 1,\footnote{We prove this bound in the Online Appendix, without leveraging sparsity of first-stage models but rather as an algebraic result with intuitive interpretation.} where larger values of $\rho$ imply that the active set is more dependent in the cross-section. This can be interpreted as that the panel $Y$ has a stronger dependency due to the covariates $X$. Intuitively, in the extreme case when $\rho=J^{-1}$, the panel can be separated into $J$ smaller problems, each containing a subset of response units explained by only one covariate. Thus the panel would be very incohesive, and could be studied with $J$ separate tests. In the other extreme, if $\rho $ approaches 1, the first-stage models include the same active covariates for all units. We consider this as a very cohesive panel. If $\rho$ is between theses bounds, the panel is cohesive in a non-trivial way such that some units can be explained by some covariates and there is no clear separation of the panel into independent subproblems. 

\begin{figure}[t!]
	\tcaptab{Illustration of the cohesion coefficient}\label{fig:cohe}
	\begin{center}
		\begin{subfigure}[t]{.24\textwidth}
			\includegraphics[width=1\linewidth]{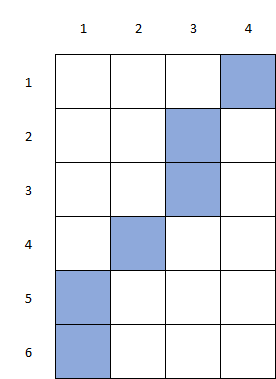}
			\caption{$\rho=J^{-1}=0.25$}
		\end{subfigure}
		\begin{subfigure}[t]{.24\textwidth}
			\includegraphics[width=1\linewidth]{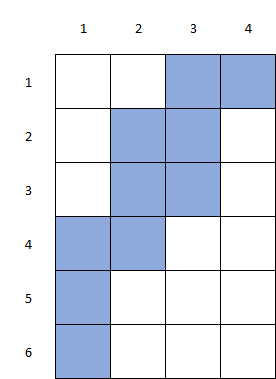}
			\caption{$\rho=0.44$}
		\end{subfigure}
		\begin{subfigure}[t]{.24\textwidth}
			\includegraphics[width=1\linewidth]{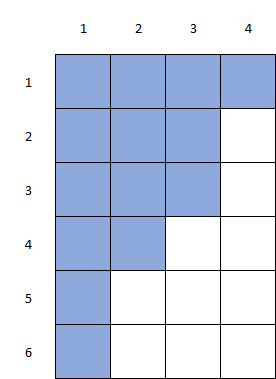}
			\caption{$\rho=0.76$}
		\end{subfigure}
			\begin{subfigure}[t]{.24\textwidth}
				\includegraphics[width=1\linewidth]{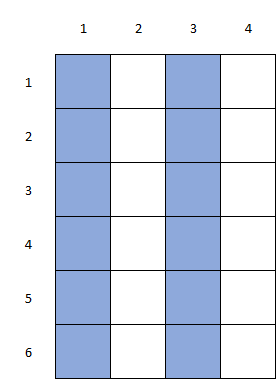}
				\caption{$\rho=1$}
			\end{subfigure}
		\end{center}
		\bnotetab{This figure illustrate the cohesion coefficient $\rho$ in four examples. It shows the smallest, largest and in-between cases of $\rho$. The columns represent the $J=4$ different covariates.The blue boxes indicate the active sets for each panel.}   
	\end{figure}

Figure \ref{fig:cohe} illustrates the panel cohesion coefficient with examples. The subplots show four active sets that are different from our running example. The left subplot \ref{fig:cohe}(a) shows the extreme case of $\rho=J^{-1}$, where the panel is the least cohesive. The right subplot \ref{fig:cohe}(d) illustrates the other extreme for $\rho=1$, where the panel is the most cohesive. The middle subplots \ref{fig:cohe}(b) and (c) correspond to the complex cases of a medium cohesion coefficient.

	Our novel simultaneity count and cohesiveness measure are the basis for modifying a Bonferroni test for FWER-controlled inference. Theorem \ref{thm_MT} formally states the FWER control. The proof is in Appendix \ref{app:proof}. 
	
	\begin{theorem}{\bf FWER control}\label{thm_MT}\\
		Under Assumption \ref{assu:valid_p}, 
		the following rejection rule has FWER$\leq \gamma$ on $H_D$:  
		\begin{equation}
			\min_{n\in \mathcal{K}_j} \left \{p^{(n)}_j \right\}\leq \rho\frac{\gamma}{ N_j}\Rightarrow\text{ Reject $H_{0,j}$},
		\end{equation}
		where $p^{(n)}_j$ are valid post-selection $p$-values for the active covariates $j$ of unit $n$, and $\rho $ is the panel cohesion coefficient.
	\end{theorem}
	
	Theorem \ref{thm_MT} is based on an algebraic union bound argument, that leverages the structure of the panel hypotheses $H_D$. The assumptions on the data generating process and asymptotic distribution are implicitly included in the valid post-selection $p$-values.

	This completes the joint testing procedure. First, we calculate $p$-values after running a sparse linear time-series regression. Second, we use the output of the sparse linear estimation to write down a hypothesis and, third, we provide a FWER control inference procedure by combining the $p$-values across the cross-section and test the hypothesis.
	
	The difference between a naive Bonferroni and our FWER control is particularly pronounced for weak covariates that affect only a subset of the cross-sectional units. Given a FWER control level of $\gamma$, the rejection threshold for a naive Bonferroni test is $\frac{\gamma}{J N}$ for every covariate. The rejection threshold for our FWER control is always higher, and differs in particular when $N_j$ is small and $\rho$ is large. This is the case for weak covariates in a cohesive panel. 
	
	As it is common in statistical inference, we focus on Type I error control. Type II error rates require the specification of alternatives. While we do not provide formal theoretical results for the power of our inference approach, we show comprehensively in the simulation and empirical part, that our approach has substantially higher power than conventional approaches.

	We point out that the validity of our procedure holds for unbalanced panels as well. This is because even when there are different number of observations for the $n$th and $m$th units, i.e. $T_n\neq T_{m}$ for $n \neq m$, they can still be estimated separately in the first stage of the regularized regression. The hypothesis testing and selection of a parsimonious model only requires the matrix $\bm{P}$ of valid $p$-values, which can be based on samples of different sizes.

	\subsection{Least Number of Covariates: Traversing the Threshold}

	The typical logic of statistical inference is to determine which covariates we should admit from $X_M$, given a significance level $\gamma$. We use $K$ to denote the number of selected covariates. When $\gamma$ is specified as a lower quantity, we expect $K$ to decrease as well, that is,  the rejection becomes harsher.
	
	As the number of admitted covariates of our procedure is monotonically increasing in $\gamma$, we want to ask the following converse question: How do we need to set $\gamma$ such that we reject $K$ covariates? Concretely, we want to find:
	
	\begin{equation}\label{24}
		\gamma^*(K)=\sup \left\{\gamma | K=\sum_{j=1}^{J} \oo \left(\min_{n\in \mathcal{K}_j} \left \{p_j^{(n)} \right \}\leq \rho \frac{\gamma}{N_j} \right) \right\} .
	\end{equation}

	Let $p_j=\min_{n\in \mathcal{K}_j}\{p_j^{(n)}\}$ be the $1$st order statistic for $j=1,...,J$. Then (\ref{24}) is simply the $K$-th order statistics of $N_j p_j / \rho$:
	\begin{equation}
		\gamma^*(K)=\min\{N_ip_i/\rho |
		\exists j_1,j_2,...,j_K\in \{1,...,J \}: N_ip_i\geq N_{j_k}p_{j_k}
		\}.
	\end{equation}
	
	Since this minimization scan is monotone, we can determine how many covariates at least should be admitted, given a control level, which is similar to the ``SimpleStop'' procedure described in \cite{16-AOS1536}. The following corollary formalizes this inversion method that finds the least number of covariates to admit:
	
	\begin{corollary}{\bf Least number of covariates}\label{col3}\\
		Under Assumption \ref{assu:valid_p}, given the FWER level $\gamma$, there exists a unique number $K^*(\gamma)$ such that
		\begin{equation}
			K^*(\gamma)=
			\begin{cases}
				\argmax_{0\leq K\leq J}\gamma^*(K)\leq\gamma & \exists K:\gamma^*(K)\leq\gamma\\
				d & o.w.\\
			\end{cases}
		\end{equation}
	\end{corollary}
	
	The statement simply states that the simplest linear model should have at least $K^*(\gamma)$ covariates for a given $\gamma$. Note that it is possible that, for example, $\gamma^*(5)$ and $\gamma^*(6)$ are both equal to $0.05$, while $\gamma^*(7)>0.05$. In this case the minimum number of covariates is $K^*(0.05)=6$ because it does not hurt FWER-wise to include 6 covariates in the model. Hence, we are making a slightly different statement than that there would be exactly $K^*(\gamma)$ covariates in the true linear model. The number of covariates is obviously conditional on the set of candidate covariates $\bm{X}$, and we can only make statements for this given set. 
	
	In our empirical study we consider candidate asset pricing factors $\bm{X}$ to explain the investment strategies $\bm{Y}$. More generally, the linear model that we consider is often referred to as a factor model. Therefore, we will also refer to the selected covariates as factors, and use these two expressions as synonyms moving forward. This directly links our procedure to the literature on estimating the number of factors to explain a panel. A common approach in this literature is to use statistics based on the eigenvalues of either $\bm{Y}$ or $\bm{X}$ to make statements about the underlying factor structure. Our approach is different, as it provides significance levels for the selected factors and FWER control for the number of factors.
	

	Table \ref{tab:toy} illustrates the estimation of the number of factors and their ranking with our running example introduced in Figure \ref{fig:demo_running}. We calculate the simultaneity counts $N_j$'s as given in (\ref{eq28:simcount}) and demonstrated in Figure \ref{fig:demo1}, and $p_j$ as the smallest $p$-values associated with the $j$th covariate. Then, the rejection rule in Theorem \ref{thm_MT} is based on whether a pre-specified level $\gamma$ satisfies $p_j<\frac{\rho \gamma}{N_j}$, which is equivalent to $\rho ^{-1}\cdot N_j\cdot p_j <\gamma$. 
	
	\begin{table}[t!]
		\tcaptab{Sorted $p$-values for the running example}\label{tab:toy}
		\centering
		\begin{tabular}{cc|cc|cc}
			\toprule		Factor ($j$) & $p_j$ &\multicolumn{2}{c|}{Simultaneity count for $H_D$} & \multicolumn{2}{c}{Conventional Bonferroni for $H_A$}\\ 
			&    & $\rho^{-1} \cdot N_j$    &$\rho^{-1} \cdot N_j \cdot p_j$ & $J\cdot N$&$J\cdot N\cdot p_j$\\ \midrule
			3 &$<0.001$ & 22.1&$<0.001$&24& 0.002\\
			4   &$<0.001$& 11.1& 0.001&24 & 0.003\\
			1   & 0.005 & 4.7& 0.024&24&0.120 \\
			2   & 0.002& 14.3 & 0.028&24&0.051\\ \bottomrule
		\end{tabular}
		\bnotetab{This table constructs ``significance'' levels for the running example introduce in Figure \ref{fig:demo_running}. We compare the simultaneity count for the data-driven hypotheses $H_D$ and a onventional Bonferroni count for data-agnostic hypotheses $H_A$. The products $N_j \cdot p_j$, respectively $J\cdot N\cdot p_j$, can be interpreted as the significance levels for the corresponding approach. Given a FWER control $\gamma$ all factors with $\rho^{-1}\cdot N_j \cdot p_j$ (respectively $J\cdot N\cdot p_j$) below this threshold are selected.}
	\end{table}
	
	Thus, the natural ranking of the covariates is to sort all covariates in descending order of the $\rho ^{-1}\cdot N_j\cdot p_j $ values as shown in Table \ref{tab:toy}. It is then trivial to determine $K^*(\gamma)$ for any choice of $\gamma$. For example, for $\gamma=1\%$, we would select factors 3 and 4, but not 1 or 2. On the other hand, for $\gamma>2\%$, we would include all four factors. Hence, the ranking of $\rho ^{-1}\cdot N_j\cdot p_j $ directly maps into $K^*(\gamma)$. Moreover, the ordered list of $\rho ^{-1}\cdot N_j\cdot p_j $ provides an importance ranking of the factors. Furthermore, the number $N_j$ reveals if significant factors are ``weak''. In our case, factor 1 has $N_1=3$, which indicates that it affects only a small number of hypothesis. Its $p$-value $p_1$ is sufficiently small to still imply significance in terms of FWER control.
	
	For comparison, Table \ref{tab:toy} also includes the corresponding analysis for the data-agnostic hypothesis and a conventional Bonferroni correction. The Bonferroni analysis uses the same $p$-values but a different multiple testing adjustment. In our case, the $p$ values would be multiplied by $J \cdot N=24$ as this corresponds to the total number of hypothesis tests. This will obviously make the inference substantially more conservative. Indeed, even for a FWER control of $\gamma=4\%$, we would only select factors 3 and 4. We would need to raise the FWER control to $\gamma=12\%$ to include factor 1. Hence, weak factors, like factor 1, are more likely to be discarded by the data-agnostic hypothesis with conventional multiple testing adjustment. 
	
	We emphasize that a data-agnostic hypotheses with conventional Bonferroni correction does provide correct FWER control, but it is overly conservative, and does not sufficiently leverage information already observable in LASSO estimation results. By construction, the data-agnostic Bonferroni approach will test a larger number of hypothesis, which means that the corresponding ``significance levels'' will always be lower or equal to our data-driven simultaneity count. Second, the data-agnostic Bonferroni approach does not differentiate the ``strength'' of the factors, while our approach provides a selection-based heterogeneous adjustment of the $p$-values. This is essential for detecting weak factors.

	Having introduced all building blocks of our novel method to detect covariates, we put the entire procedure together as ``Panel-PoSI'':
	
	\begin{procedure}{\bf Panel-PoSI}\\
		The Panel-PoSI procedure consists of the following steps:
		\begin{enumerate}
			\item For each unit $n=1,...,N$ unit, we fit a linear sparse model $\hat{\beta}^{(n)}$ given $(\bm{X},\bm{Y},\lambda,\omega)$. We suggest cross-validation to select the LASSO penalty $\lambda$. We construct the sparse estimators $\bar{\beta}^{(n)}$ and the corresponding $p$-values for the active covariates for each unit, and collect them in the ``matrix'' of $p$-values $\bm{P}$. 
			\item We collect the panel-level sparse model selection event $\mathcal{M}$ and construct the data-driven hypothesis $H_D$.
			\item  Given the FWER control level $\gamma$ and based on the the simultaneity counts $N_j$'s of active covariates, we make inference decision for the sparse model. We can rank covariates in terms of their significance and select a parsimonious model that explains the full panel.
		\end{enumerate}
	\end{procedure}

	As we have now all results in place, we can summarize the advantages of our procedure. First, we want to clarify that our goals and results are different from just some form of optimal shrinkage selection. Selecting a shrinkage parameter with some form of cross-validation in a regularized estimator like LASSO does not provide the same insights and model that we do. A shrinkage estimator can either be applied to each unit separately, as we do it in our first step, or to the full panel in a LASSO panel regression. The separate covariate selection for each cross-sectional unit does not answer the question which covariates are needed to explain the full panel jointly. A shrinkage selection on the full panel for some form of panel LASSO can neglect weaker factors, as those receive a low weight in the cross-validation objective function. Second, tuning parameter selection with cross-validation requires a sufficiently large amount of data. Our approach is attractive as we can do the complete analysis on the same data. That means, an initial LASSO is used to first reduce the number of covariates, but this set is then further trimmed down using inferential theory. Hence, we can construct a parsimonious model even for data with a relatively short time horizon, but large cross-sectional dimension. Third, the statements that we can make are much richer than a simple variable selection. We can formally assess the relative importance of factors in terms of their significance. The model selection is directly linked to a form of significance level, which allows us to assess the relevance of including more factors. Last but not least, we can also make statements about the strength of factors. In summary, Panel-PoSI is a disciplined approach based on formal statistical theory to construct and interpret a parsimonious model.


	\section{Ordered Multiple Testing on Nested Hypothesis Family}\label{sec:ordered}

	So far, our hypothesis family $H_D$ has no hierarchy and consequently, we have not imposed a sequential structures on the admission order of covariates of $\bm{X}$. However, there are cases where the covariates or factors warrant a natural order such that the family possesses a special testing logic. A hierarchical structure in covariates arises when the inclusion of the next covariate only make sense if the previous covariates is included. One example would be if the next covariates refines a property of the previous covariate. Another case is the use of principal component (PC) factors. The conventional logic is to include PCs sequentially from the dominating one to the least dominating one. This is similar to the motivation for \cite{16-AOS1536}, but different from them, we treat the PCs as exogenous without taking the estimation of PCs explicitly into account. In this section, we will use exogenous PCs as hierarchical covariates, as this is the main example in our empirical study. However, all the results hold for any set of exogenous hierarchical covariates.

	Without loss of generality, we presume $\bm{X}$ has the $k$th column as the $k$th nested factor. A $k$-order nested model is of the following form
	\begin{equation}
		k-\text{order nested model}: \bm{Y}=\bm{X}_{[k]}\bm{\beta}_{[k]}
	\end{equation}
	where $[k]=\{1,...,k\}$ is the set that includes indices up to $k$. For example, a hierarchical 3-order model corresponds to the case where variables $\bm{X}_{\{1,2,3\}}$ are included, but not for the rest of the covariates in $\bm{X}$. When formulating our hypothesis family, we must represent the sequential testing structure, as reflected in our definition of nested families of hypotheses:
	\begin{definition}{\bf Data-driven nested family}\label{def9}\\
		The data-driven nested family of hypotheses conditional on $\mathcal{M}$ is
		\begin{equation}
			H_{N}=\{H_{N,k}:k=0,1,...,J\},\q
			H_{N,k}=	\bigcap_{j\in\mathcal{K}_k}
			H_{N,k}^{(n)}\bigg\rvert \mathcal{M},\q
			H_{N,k}^{(n)}:\{k':\beta_{k'}^{(n)}\neq 0,k'\leq k\}.
		\end{equation}
	\end{definition}
	
	$H_{N,0}$ completes the case when no rejection on any factor is made. Whenever $H_{N,k}$ is true, then $H_{N,k'}$ is also true for $k<k'\leq J$. Moreover, in the cases where $\mathcal{K}_k=\emptyset$ but $\mathcal{K}_{k'}\neq\emptyset$ with $k<k'$, the notation ensures that the hypothesis $H_{N,k}$ is included in $H_N$ simply because $\mathcal{K}_{k'}$ is present. In other words, if a less dominating hypothesis $H_{N,k'}$ is suggested by data (that is, its active set is non-empty $\mathcal{K}_{k'}\neq\emptyset$), $H_N$ would automatically include all $H_{N,k}$ for $k\leq k'$.
	
	The FWER control property needs to be adapted to the nested nature of this family. \cite{16-AOS1536} argue that the proper measurement is to control for ordered factor count over-estimation with level $\gamma$, as follows:
	
	\begin{definition}{\bf FWER for nested family}\label{def10}\\
		For a test that rejects $H_{N,k}$ for $k=1,2,...,\hat{k}$ of $H_{N}$, the FWER control at the level $\gamma$ satisfies $\PP(\hat{k}\geq s)\leq \gamma$, where $s$ is the true factor count.
	\end{definition}
	
	Given the hierarchical logic embedded in the model, we need  the following assumption, which is more restrictive than Assumption \ref{assu:valid_p}:
	
	\begin{assumption}{\bf Tail $p$-values}\label{asu8}\\
		Under $	H_{N,k}^{(n)}$, it holds that $p^{(n)}_{k'}\iid\text{Unif }[0,1]$ for all $k'>k$.
	\end{assumption}
	
	Assumption \ref{asu8} only needs to hold for the tail hierarchical covariates, but requires them to be independent, whereas Assumption \ref{assu:valid_p} for the unordered tests does not require independence. In the case of PCs, it only applies to the lower order tail PC factors that should not be included for a given null hypothesis. For example, if the true model is $H_{N,5}$, we only need $p^{(n)}_{k'}\iid \text{Unif}[0,1]$ for $k'>5$, which is a usual type of assumption in this literature such as in \cite{rssb.12122}. Moreover, because the nested nature guarantees that the higher-order PCs are more likely to be null, a step-down procedure is expected to increase the power relative to a step-up procedure. 
	
	As our focus is to control for false discoveries, we also need to propose new simultaneity counts in the nested family setting. Concretely, we consider first taking a union to obtain the active unit set ${\mathcal{K}}_k^{\text{order}}$ and then calculate conservative simultaneity counts $N^{\text{order}}_k$, with $|M_j|$ as the number of units that the $j$th variable is active in:
	\begin{equation}
		{\mathcal{K}}^{\text{order}}_k=
		\bigcup_{k'\in\{k,k+1,...,J\}} \mathcal{K}_{k'},\q
		{N}_k^{\text{order}}=\sum_{j\in {\mathcal{K}}_k^{\text{order}}}|M_j| .
	\end{equation}
	It is possible for some $|M_k|$ to be 0 (for instance, the $k$th PC could be inactive for all units), but its ${N}_k^{\text{order}}$ would be 0 if and only if higher-order PCs all have $|M_{k'}|=0$ for $k'>k$. Note, that we are not assuming that $|M_j|$ is decreasing in $j$. Hence, our data-driven approach imposes only a nested structure in the hypotheses, but not a nested structure for the number of active covariates.
	
\begin{figure}[t!]
		\centering
		\tcaptab{Example of hierarchical simultaneity counts ${N}_k^{\text{order}}$ for $H_{N}$} \label{fig:demo2}
		\begin{center}
			\begin{subfigure}[t]{.24\textwidth}
				\includegraphics[width=1\linewidth]{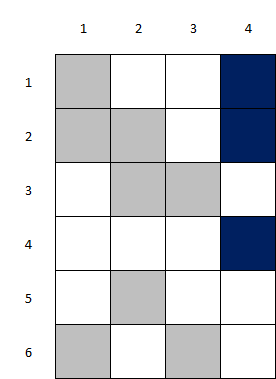}
				\caption{$N_4^{\text{order}}=3$}
			\end{subfigure}
			\begin{subfigure}[t]{.24\textwidth}
				\includegraphics[width=1\linewidth]{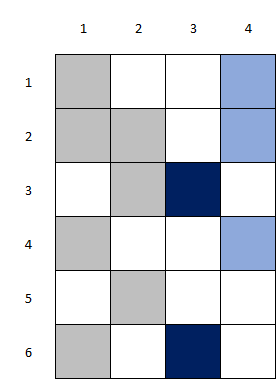}
				\caption{$N_3^{\text{order}}=5$}
			\end{subfigure}
			\begin{subfigure}[t]{.24\textwidth}
				\includegraphics[width=1\linewidth]{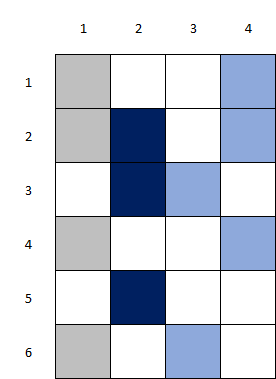}
				\caption{$N_2^{\text{order}}=8$}
			\end{subfigure}
			\begin{subfigure}[t]{.24\textwidth}
				\includegraphics[width=1\linewidth]{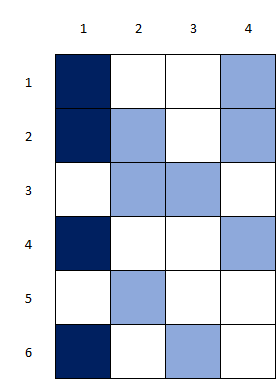}
				\caption{$N_1^{\text{order}}=12$}
			\end{subfigure}
		\end{center}
		\bnotetab{This figure shows the simultaneity counts $N_i^{\text{order}}$ in an illustrative example. The subplots represent the simultaneity counts for the $J=4$ covariates and $N=6$ units. The dark blue columns present the active factors, while the light blue columns capture factors of higher-order. The sub-plots from left-to-right represent our calculation order from the highest-order factor to the 1st factor.}
	\end{figure}

	
	Figure \ref{fig:demo2} illustrates the process of our step-down simultaneity count. From the left, we start with factor $k=4$ and move step-wise down to factor $k=1$ on the right. The dark blue columns present the active factors, while the light blue columns capture factors of higher-order. 
	In the left-most sub-figure, we only need to account for the 4th PC, implying ${N}_4^{\text{order}}=3$, whereas in the mid-left sub-figure, the 3rd PC has ${N}_3^{\text{order}}=2+3=5$. Eventually, in the right-most sub-figure, we have swept through the entire panel and the 1st PC has a simultaneity count of ${N}_1^{\text{order}}=12$.

	Now we can introduce a step-down procedure adapted to the nested structure of $H_{N}$:
	
	\begin{procedure}{\bf Step-down rejection of nested ordered family $H_{N}$} \label{procedure:nested}\\
		The step-down rejection procedure consists of the following steps:
		\begin{enumerate}
			\item For each $k \in \{1,...,J\}$ calculate the ordered simultaneity count ${N}_k^{\text{order}}$.
			\item For each $k \in \{1,...,J\}$ calculate the approximated R\'enyi representation ${Z}_k^{\text{order}}$ and its transformed reversed order statistics ${q}_k^{\text{order}}$:
			\begin{equation}
				{Z}_k^{\text{order}}=\sum_{i=k}^J\sum_{n\in\mathcal{K}_i}
				\frac{ \ln(p^{(n)}_k) }{{N}_1^{\text{order}}-{N}^{\text{order}}_{i+1}\oo\{i\neq J\}}
				,\q {q}^{\text{order}}_k=\exp(-{Z}^{\text{order}}_k)
			\end{equation}
			\item Reject hypothesis $1,2,...,\hat{k}$, where $\hat{k}=\max\{k:{q}^{\text{order}}_k\leq\frac{\gamma {N}^{\text{order}}_k}{JN}\}$.
		\end{enumerate}
	\end{procedure}
	This procedure will have FWER control at level $\gamma$ as stated in the following theorem:
	
	\begin{theorem}{\bf FWER control for ordered hypothesis}\label{thm4}\\
		Under Assumption \ref{asu8}, Procedure \ref{procedure:nested} has FWER control of $\gamma$ for the ordered hypothesis $H_{N}$.
	\end{theorem}
	
	
	The proof is deferred to the Online Appendix. This design extends Procedure 2 from \cite{rssb.12122} and ``Rank Estimation'' from \cite{16-AOS1536}, both of which focus on a single sequence of $p$-values rather than the panel setting.
	
	In Step 2, we use Assumption \ref{asu8} to transform $p$-values into $\ln(p^{(n)}_k)$, which are i.i.d. standard exponential random variables. Since the family $H_{N}$ has $J$ members, we need to modify our simultaneity count and in a sense condense the panel into a sequence of statistics associated with the ordered covariates. We built a staircase sequence of conservative simultaneity counts ${N}_k^{\text{order}}$ in Step 1 to accumulate the number of $p$-values we use up to the $k$th ordered covariate, starting from the end. By the R\'enyi representation of \cite{Rnyi1953OnTT}, the ${Z}_k^{\text{order}}$ of Step 2 approximate exponential order statistics and the ${q}_k^{\text{order}}$ approximate uniform order statistics. The nature of these approximations is to create a more conservative rejection, the technical details of which are examined in the proof in our Online Appendix. Finally, we run the order statistics through a step-down procedure proposed by \cite{2336545} so that we find the $\hat{k}$ largest number of ordered covariates rejected by the data with FWER control. Also note that even if the global null, i.e. $H_{N,0}$, is true, and every linear sparse model active set is empty, that is ${N}^{\text{order}}_1=0$, the procedure in Step 3 is still valid because we do not reject $H_{N,1}$.

	\section{Simulation}\label{sec:simulation}

	%
	%
	%
	%
	%
	%
	%
	%
	%
	%
	%
	%
	%
	%

	We demonstrate in simulations that our inferential theory allows us to select better models. We compare different estimation approaches to select covariates and show that our approach better trades off false discovery and correct selections and hence results in a better out-of-sample performance.
	
	Table \ref{tab:summary} summarizes the benchmark models. Our framework contributes among three dimensions: the selection step for the sparse model, the construction of the hypothesis and the multiple testing adjustment. We consider variations for these three dimensions which yields in total six estimation methods. By varying the different elements of the estimators, we can understand the benefit of each component.

	\begin{table}[H]
		\tcaptab{Summary of estimation methods} \label{tab:summary}
		{\small
			\resizebox{\textwidth}{!}{
				\renewcommand{\arraystretch}{1.2}
				\centering
				\begin{tabular}{@{}llllll@{}}
					\toprule
					Name & Abbreviation & Selection & Hypothesis & Multiple Testing & Rejection rule \\ \midrule
					Naive OLS  &  N-OLS  &OLS without LASSO     & Agnostic $H_A$     & No adjustment           &  $ p^{\text{OLS}}< \gamma $      \\
					Bonferroni OLS  &  B-OLS &OLS without LASSO     & Agnostic $H_A$     & No adjustment           &  $ p^{\text{OLS}}< \frac{\gamma}{J N} $      \\
					Naive LASSO  &  N-LASSO &LASSO without PoSI     & Agnostic $H_A$     & No adjustment           &  $ p^{\text{LASSO}}< \gamma $      \\
					Bonferroni Naive LASSO  &  B-LASSO &LASSO without PoSI    & Agnostic $H_A$     & Bonferroni         &  $ p^{\text{LASSO}}< \frac{\gamma}{J N} $      \\
					Bonferroni PoSI  &  B-PoSI & LASSO with PoSI    & Agnostic $H_A$     & Bonferroni         &  $ p^{\text{PoSI}}< \frac{\gamma}{J N} $      \\
					Panel PoSI  &  P-PoSI & LASSO with PoSI    & Data-driven $H_D$     & Simultaneity count         &  $ p^{\text{PoSI}}< \frac{\rho \gamma}{N_i} $      \\
					\bottomrule
				\end{tabular}
		}}
		\bnotetab{This table compares the different methods to estimate a set of covariates from a large dimensional panel. For each method, we list the name and abbreviation. The selection refers to the regression approach for each univariate time-series. The hypothesis is either agnostic or data-driven given the selected subset of covariates. The multiple testing adjustment includes no adjustment, a conventional Bonferroni adjustment and our novel simultaneity count for a data-driven hypothesis. The rejection rules combine the valid p-values and multiple testing adjustment. $p^{\text{OLS}}$ is the p-value for a conventional t-statistics of an OLS estimator. $p^{\text{LASSO}}$ is the p-value without removing the lasso bias or adjusting for post-selection inference, that is, it is simply the OLS p-values using the selected subset of regressors. $p^{\text{PoSI}}$ is the debiased post-selection adjusted p-value based on Theorem \ref{thm1_main}.}
	\end{table}

	Our baseline model is Panel PoSI, which uses post-selection inference LASSO, and a simultaneity count for a data driven hypothesis. The first component that we modify is the selection of the sparse model. A simple OLS regression without shrinkage does not produce a sparse model. This gives us the methods Naive OLS and Bonferroni OLS. A conventional LASSO results in a sparse selection, but the p-values are not adjusted for the post-selection inference and the bias adjustment. The corresponding models are the Naive LASSO and the Bonferroni Naive LASSO. The second component is the hypothesis, which is agnostic for methods besides Panel PoSI. For the comparison models, we either consider no multiple testing adjustment or the conventional Bonferroni adjustment. Under the multiple testing adjustment we obtain the Bonferroni OLS, the Bonferroni Naive LASSO and the Bonferroni PoSI. The outcome of all the estimations are adjusted p-values for the covariates, which we use to select our model for a given target threshold. For a given value of $\gamma$ we include a covariate if its adjusted p-value is below the critical values summarized in the last column of Table \ref{tab:summary}.
	
	We simulate a simple and transparent model. Our panel follows the linear model
	\begin{align*}
		Y_{t}^{(n)}=\sum_{j=1}^J X_{t,j}\beta_{j}^{(n)}+\epsilon_{t}^{(n)}\qquad \text{for $t=1,...,T$, $n=1,...,N$ and $j=1,..,J$.}
	\end{align*}
	The covariates and errors are sampled independently as normally distributed random variables:
	\begin{align*}
		X_{t,j}\iid N(0,1), \qquad \epsilon_{t}\iid N(0,\Sigma) .
	\end{align*}
	The noise is either generated as independent noise with covariance matrix $\Sigma= \sigma^2 I$ or as cross-sectionally dependent noise with non-zero off-diagonal elements $\Sigma_{ij}=\kappa$ and diagonal elements $\Sigma_{ii}=\sigma^2$. Note that our theorems for PoSI assume homogeneous noise, while dependent noise violates our assumptions. Hence, the dependent noise allows us to test how robust our method is to misspecification. We set $\sigma^2=2$ and $\kappa=1$, but the results are robust to many choices. The covariance $\Sigma$ as treated as unknown and hence has to be estimated.

	\begin{figure}[t!]
		\centering
		\tcaptab{Design of loadings $\bm{\beta}$} \label{fig:simulation-design}
		\begin{center}
			\includegraphics[width=0.4\linewidth]{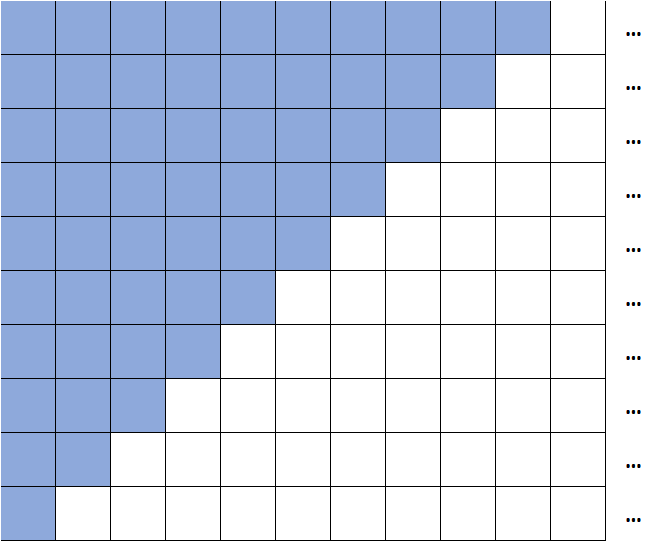}
		\end{center}
		\bnotetab{This figure demonstrates the setting of our simulations with $10$ factors, where loadings are shaded based on whether they are active. In this staircase setting, the first factor affects all units, the 2nd factor affects $90\%$, and so on, and lastly the 10th factor affects $10\%$ of all units.}
	\end{figure}

	We construct the active set based on the staircase structure depicted in Figure \ref{fig:simulation-design}. Of the $J$ covariates in $X$, we have $K=10$ active independent factors. Figure \ref{fig:simulation-design} demonstrates the setting for the $10$ factors, where loadings are shaded based on whether they are active. The first factor affects all units, the 2nd factor affects $90\%$, and so on, and lastly the 10th factor affects $10\%$ of all units. This setting is relevant, and also challenging from a multiple testing perspective. It results in a large cohesion coefficient $\rho$, which makes the correct FWER control even more important. The loadings are sampled from a uniform distribution, if they are in the active set:
	\begin{align*}
		\beta_{j}^{(n)}\iid \textrm{Unif} \left[-\frac{1}{2},\frac{1}{2} \right] \quad \text{for $j$ in the active set}, \qquad \qquad \beta_{j}^{(n)}=0 \quad \text{for $j$ outside the active set.}
	\end{align*}

	\begin{table}[t!]
		\tcaptab{Simulation Comparison between Selection Methods}\label{tab:sim_results}
		\centering
			\renewcommand{\arraystretch}{1.0}
			{\small
				\begin{tabular}{@{}llllr@{}}
					\toprule \midrule
					\multicolumn{5}{c}{Independent noise}   \\ \midrule \midrule
					Method & \# Selections & \# False Selections &  \# Correct Selections &  OOS $R^2$ \\ \midrule 
					\multicolumn{5}{c}{FWER $\gamma=5\%$}   \\ \midrule
					Panel PoSI             & 10.8 & 2.8  & 7.9  & 10.0\%    \\
					Bonferroni PoSI        & 4.7  & 0.0  & 4.7  & 8.0\%     \\
					Bonferroni Naive LASSO & 0.0  & 0.0  & 0.0  & 0.0\%     \\
					Naive LASSO            & 0.2  & 0.0  & 0.2  & 0.4\%    \\
					Bonferroni OLS         & 1.0  & 0.0  & 1.0  & 1.7\%    \\
					Naive OLS              & 99.2 & 89.2 & 10.0 & -144.2\%  \\      \midrule
					\multicolumn{5}{c}{FWER $\gamma=1\%$}   \\ \midrule
					Panel PoSI             & 8.6  & 1.1  & 7.5 & 10.6\%  \\
					Bonferroni PoSI        & 2.7  & 0.0  & 2.7 & 5.2\%   \\
					Bonferroni Naive LASSO & 0.0  & 0.0  & 0.0 & 0.0\%   \\
					Naive LASSO            & 0.1  & 0.0  & 0.1 & 0.3\%    \\
					Bonferroni OLS         & 0.2  & 0.0  & 0.2 & 0.5\%    \\
					Naive OLS              & 46.4 & 36.5 & 9.9 & -19.3\% \\     \midrule \midrule
					\multicolumn{5}{c}{Cross-sectionally dependent noise}   \\ \midrule \midrule
					Method & \# Selections & \# False Selections &  \# Correct Selections &  OOS $R^2$ \\ \midrule 
					\multicolumn{5}{c}{FWER $\gamma=5\%$}   \\ \midrule
					Panel PoSI             & 10.1 & 2.2  & 7.9  & 8.0\%   \\
					Bonferroni PoSI        & 4.4  & 0.0  & 4.4  & 7.2\%    \\
					Bonferroni Naive LASSO & 0.0  & 0.0  & 0.0  & 0.0\%   \\
					Naive LASSO            & 0.4  & 0.0  & 0.4  & 0.5\%    \\
					Bonferroni OLS         & 0.9  & 0.0  & 0.9  & 1.3\%    \\
					Naive OLS              & 83.7 & 73.7 & 10.0 & -83.8\%  \\      \midrule
					\multicolumn{5}{c}{FWER $\gamma=1\%$}   \\ \midrule
					Panel PoSI             & 7.9  & 0.6  & 7.3 & 10.3\% \\
					Bonferroni PoSI        & 2.4  & 0.0  & 2.4 & 3.9\%   \\
					Bonferroni Naive LASSO & 0.0  & 0.0  & 0.0 & 0.0\%   \\
					Naive LASSO            & 0.0  & 0.0  & 0.0 & 0.0\%   \\
					Bonferroni OLS         & 0.3  & 0.0  & 0.3 & 0.4\%   \\
					Naive OLS              & 31.0 & 21.2 & 9.8 & -6.8\% \\      
					\bottomrule
				\end{tabular}
			}
		\bnotetab{This table compares the selection results for different methods in a simulation. For each method we report the number of selected covariates, the number of falsely selected covariates and the number of correctly selected covariates. We also report the out-of-sample $R^2$ of the models that estimated with the selected covariates on the out-of-sample data. All results are averages of 100 simulations. The rejection FWER is set to $\gamma=5\%$ or $\gamma=1\%$. We simulate a panel of dimension $N=120$, $J=100$, $T=300$. The first half of time-series observations is used for the in-sample estimation and selection, while the second half serves for the out-of-sample analysis. The panel is generated by 10 independent factors. The active set of the factors follows the staircase structure of Figure \ref{fig:simulation-design}. The first factor affects all units, the second 90\%, and lastly the 10th factor affects 10\%. The unknown error variance is estimated based as a homogenous sample variance. The noise is either generated as independent noise with covariance matrix $\Sigma= \sigma^2 I$ or as cross-sectionally dependent noise with $\Sigma_{ij}=\kappa$ and $\Sigma_{ii}=\sigma^2$ for $\sigma^2=2$ and $\kappa=1$.}
	\end{table}
	
	We simulate a panel of dimension $N=120$, $J=100$ and $T=300$ with $K=10$ active factors. The first half of the time-series observations is used for the in-sample estimation and selection, while the second half serves for the out-of-sample analysis. All results are averages of 100 simulations. We use the covariates selected on the in-sample data for regressions out-of-sample. Our focus is on the inferential theory, and not on the bias correction for shrinkage. Hence, we first use the inferential theory on the in-sample data to select our set of covariates. Second, we use the selected subset of covariates in an OLS regression on the in-sample data to obtain the loadings. Last but not least, we apply the estimated loadings of the selected subset to the out-of-sample data to obtain the model fit. Note that this procedure helps a Naive LASSO, which in contrast to PoSI LASSO does not have a bias correction. The out-of-sample explained variation is measured by $R^2$, which is the sum of explained variation normalized by the total variation. The rejection FWER is set to $\gamma=5\%$ or $\gamma=1\%$. The LASSO shrinkage penalty $\lambda$ is selected by 5-fold cross-validation on the in-sample data.

	Table \ref{tab:sim_results} compares the selection results for the different methods. For each method we report the number of selected covariates, the number of falsely selected covariates and the number of correctly selected covariates. We also report the out-of-sample $R^2$. The upper panel shows the results for independent noise, while the lower panel collects the results for cross-sectionally dependent noise.
	
	PanelPoSI clearly dominates all models. It provides the best trade-off between correct and false selection, which results in the best out-of-sample performance. In the case of $\gamma=5\%$ and independent noise, Panel PoSI selects 10.8 factors in a model generated by 10 factors. 7.9 of these factors are correct. A simple Bonferroni correction is overly conservative. The Bonferroni PoSI selects only 4.7 correct factors. While this overly conservative selection protects against false discovery, it omits over half of the relevant factors which lowers the out-of-sample performance. Using post-selection inference is important, as a naive lasso provides wrong p-values which makes the overly conservative selection even worse. The other extreme is to have neither shrinkage nor multiple testing adjustment. As expected the naive OLS has an extreme number of false selections with a correspondingly terrible out-of-sample performance. 
	
	As expected, tightening the FWER control to 1\% lowers the number of false rejections, but also the number of correct selections. It reveals again that Panel PoSI provides the best inferential theory among the benchmark models. Panel PoSI selects 7.5 correct covariates, while it controls the false rejections at 1.1. The overly conservative Bonferroni methods select even fewer correct covariates, which further deteriorates the out-of-sample performance. The gap in OOS $R^2$ between Panel PoSI and Bonferroni PoSI widens to 5.4\%. All the other approaches cannot be used for a meaningful selection.
	
	Panel PosI performs well, even when some of the underlying assumptions are not satisfied. The lower panel of Table \ref{tab:sim_results} shows the results for dependent noise. As the dependence in the noise is relatively strong, it can be interpreted as omitting a relevant factor in the set of candidate covariates $X$. Even thought the PoSI theory is developed for homogeneous noise, Panel PoSI continues to perform very well. In contrast, the comparison methods perform even worse, and the Bonferroni approaches select even fewer correct covariates.

\section{Empirical Analysis}\label{sec:empirics}

%
%
%
%

\subsection{Data and Problem}

Our empirical analysis studies a fundamental problem in asset pricing. We select a parsimonious factor model from a large set of candidate factors that can jointly explain the asset prices of a large cross-section of investment strategies. Our data is standard and obtained from the data libraries of Kenneth French and \cite{HouEtAl}.

We consider monthly excess returns from January 1967 to December 2021, which results in a time dimension of $T=660$. Our test assets are the $N = 243$ double-sorted portfolios of Kenneth French's data library summarized in Table \ref{DSSource} in the Appendix. The candidate factors are $J=114$ univariate long-short factors based on the data of \cite{HouEtAl}. We include all univariate portfolio sorts from their data library that are available for our time period, and construct top minus bottom decile factor portfolios. In addition, we include the five Fama-French factors of \cite{FAMA20151} from Kenneth French's data library.

Our analysis projects out the excess return of the market factor. We are interested in the question which factors explain the component that is orthogonal to market movements. Hence, we regress out the market factor from the test assets and use the residuals as test assets. We also do not include a market factor in the set of long-short candidate factors. The original test assets have a market component as they are long only portfolios. Our results are essentially the same when we include the market component in the test assets, with the only difference that we would need to include the market factor as an additional factor in our parsimonious models. The market factor would always be selected by all models as significant, but this by itself is neither a novel nor interesting result.  

We present in-sample and out-of-sample results. The in-sample analysis uses the first 330 observations (January, 1967 to June, 1994), while the out-of-sample results are based on the second 330 observations (July, 1994 to December, 2021). As in the simulation, we first use the inferential theory on the in-sample data to select our set of covariates. Second, we use the selected subset of covariates in an OLS regression on the in-sample data to obtain the loadings. Last but not least, we use the estimated loadings on the selected subset of factors for the out-of-sample model. The LASSO penalty $\lambda$ is selected via 5-fold cross-validation on the in-sample data.\footnote{Our cross-validation follows the one-standard-deviation rule for selecting parsimonious models, that is, the largest choice of $\lambda$ within 1 standard error of minimizing the squared errors. This is the default setting of popular implementations like \texttt{glmnet} and argued for in \S3.4 of \cite{hastie2009elements}. We select $\lambda$ from the grid $\exp(a)\cdot \log J/\sqrt{T}$ with $a=-8,...,8$. This grid choice satisfies the Assumptions in \cite{chatterjee2014assumptionless} and hence Assumption A.4.} Hence, LASSO represents a first-stage dimension reduction tool, and we need the inferential theory to select our final sparse model.

We allow our selection to impose a prior on two of the most widely used asset pricing models. More specifically, we estimate models without a prior, and two specific priors that impose an infinite weight on the Fama-French 3 factors (FF3) and the Fama-French 5 factors (FF5). This prior as part of PoSI LASSO enforces that the FF3 and FF5 factors are included in the active set. Note that because we work with data orthogonal to the market return, we do not include the market factor in the prior, but only the size and value factors for FF3 and in addition the investment and profitability factor for FF5. We denote these weights by $\omega_{\text{FF3}}$ and $\omega_{\text{FF5}}$. This is an example where the researcher has economic knowledge that she wants to include in her statistical selection method.

We evaluate the models with standard metrics. The root-mean-squared error (RMSE) is based on the squared residuals relative to the estimated factor models. Hence, in-sample the models are estimated to minimize the RMSE. The pricing error is the economic quantity of interest. It is the time-series mean of the residual component of the factor model, and corresponds to the mean return that is not explained by the risk premia and exposure to the factors. In summary, we obtain the residuals as $\hat \epsilon =Y_{t,n}- X_S \hat \beta_S$ for the selected factors, where the loadings are estimated on the in-sample data. The metrics are the RMSE and mean absolute pricing error (MAPE):
\begin{align*}
	\text{RMSE}= \sqrt{ \frac{1}{N \, T}\sum_{i=1}^N \sum_{t=1}^T \hat \epsilon ^2}, \qquad \text{MAPE} = \frac{1}{N } \sum_{i=1}^N \left | \frac{1}{T} \sum_{t=1}^T \hat \epsilon  \right |.
\end{align*}

In addition to Panel PoSI without and with the FF3 and FF5 priors, we consider the benchmark methods of Table \ref{tab:summary}. We compare Panel PoSI (P-PoSI), Panel PoSI with infinite priors on FF3 and FF5 (P-PoSI $\omega_{\text{FF3}}$ respectively $\omega_{\text{FF5}}$), Bonferroni Naive LASSO (B-LASSO), Naive LASSO (N-LASSO), Bonferroni OLS (B-OLS) and Naive OLS (N-OLS). Our main analysis sets the FWER control to the usual $\gamma=5\%$.

%
%
%
%
%

\subsection{Asset Pricing Results}

Panel PoSI selects parsimonious factor models with the best out-of-sample performance among the benchmarks. For the FWER rate of $\gamma=5\%$ the number of factors differs substantially among the different methods. Panel PoSI selects 3 factors. Imposing infinite priors on FF3 or FF5 results in 4 and 5 factors for P-PoSI $\omega_{\text{FF3}}$ respectively $\omega_{\text{FF5}}$. In contrast, the alternative approaches select too many factors. Bonferroni Naive LASSO includes 10, Naive Lasso 70, Bonferroni OLS 107 and Naive OLS 114. These over-parametrized models lead to overfitting of the in-sample data.

\begin{figure}[t!]
	\tcaptab{RMSE across cross-sections}
	\label{fig:RMSE}
	\begin{subfigure}[t]{.5\textwidth}\centering
		\includegraphics[width=.9\linewidth]{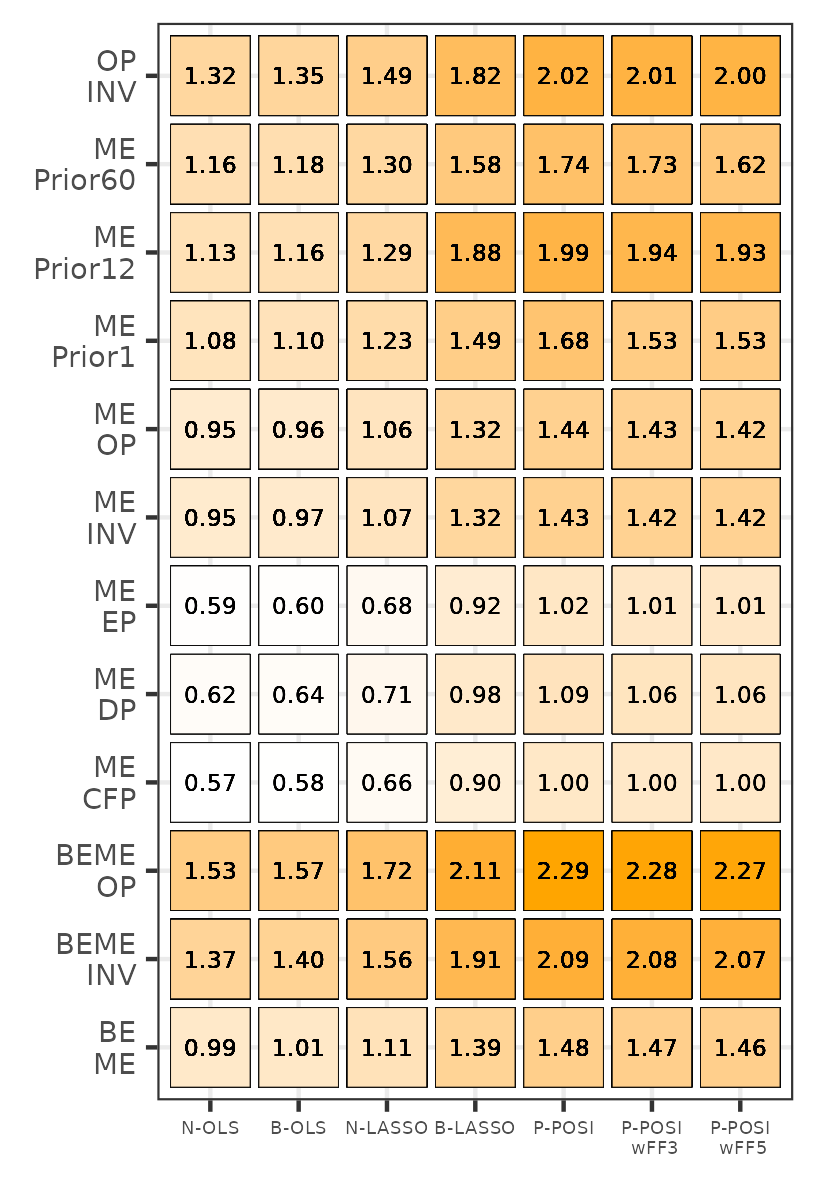}
		\caption{In-sample}
	\end{subfigure}
	\begin{subfigure}[t]{.5\textwidth}\centering
		\includegraphics[width=.9\linewidth]{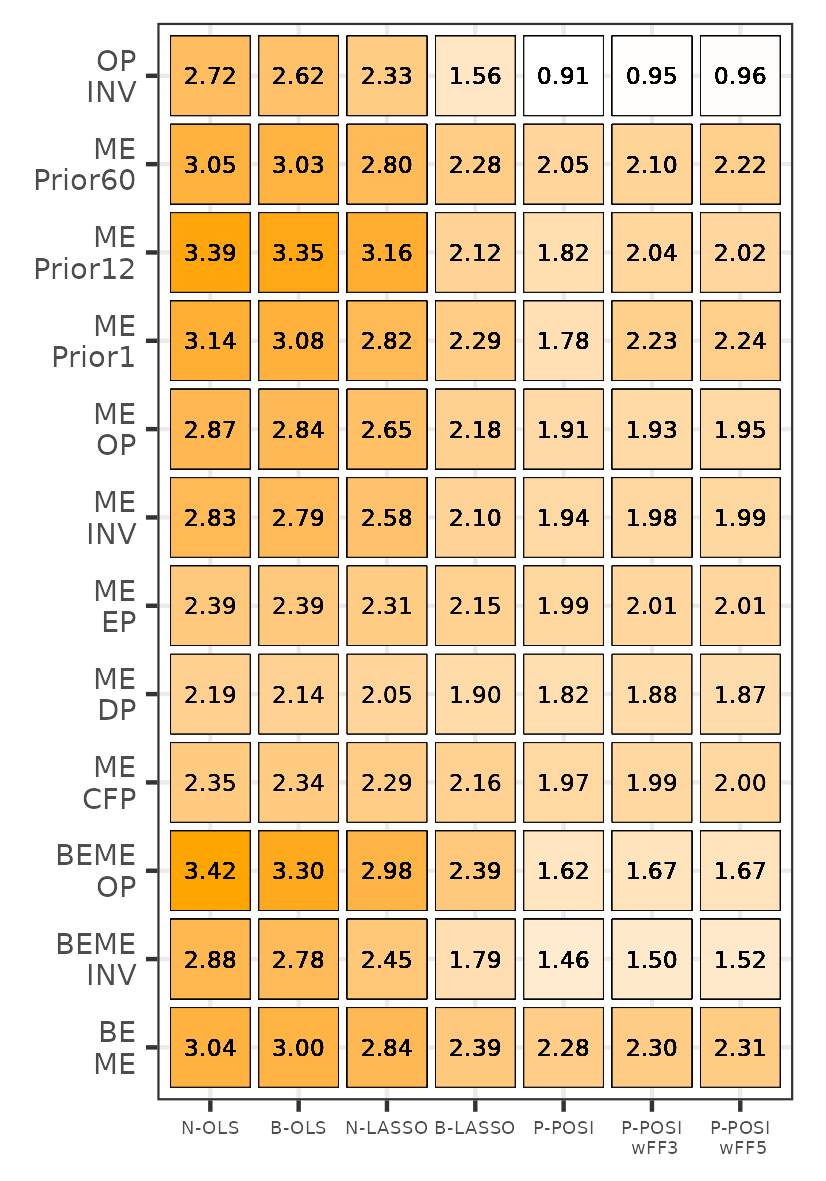}
		\caption{Out-of-sample}
	\end{subfigure}
	\bnotetab{
		This figure shows the in-sample and out-of-sample root-mean-squared errors (RMSE) for each cross-section of test assets for different factor models. The test assets are the $N=243$ double-sorted portfolios, and we show the RMSE for each set of double-sorts. The rejection FWER is set to $\gamma=5\%$ The candidate factors are the 114 univariate factor portfolios. The time dimension is $T=660$. We use the first half for the in-sample estimation and selection, while the second half serves for the out-of-sample analysis. We compare Panel PoSI (P-PoSI), Panel PoSI with infinite priors on FF3 and FF5 (P-PoSI $\omega_{\text{FF3}}$ respectively $\omega_{\text{FF5}}$), Bonferroni LASSO (B-LASSO), Naive LASSO (N-LASSO), Bonferroni OLS (B-OLS) and Naive OLS (N-OLS).          
	}		
\end{figure}

\begin{figure}[t!]
	\tcaptab{MAPE across cross-sections}
	\label{fig:MAPE}
	\begin{subfigure}[t]{.5\textwidth}\centering
		\includegraphics[width=.9\linewidth]{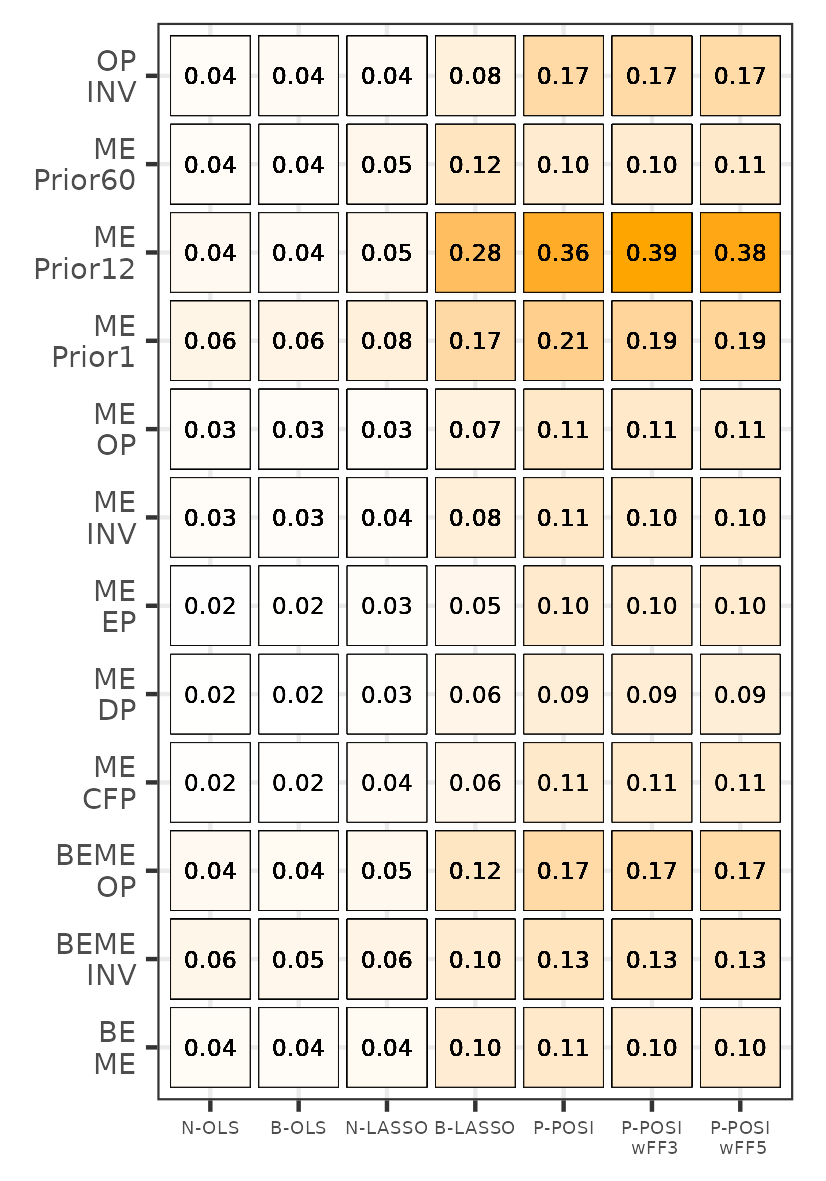}
		\caption{In-sample}
	\end{subfigure}
	\begin{subfigure}[t]{.5\textwidth}\centering
		\includegraphics[width=.9\linewidth]{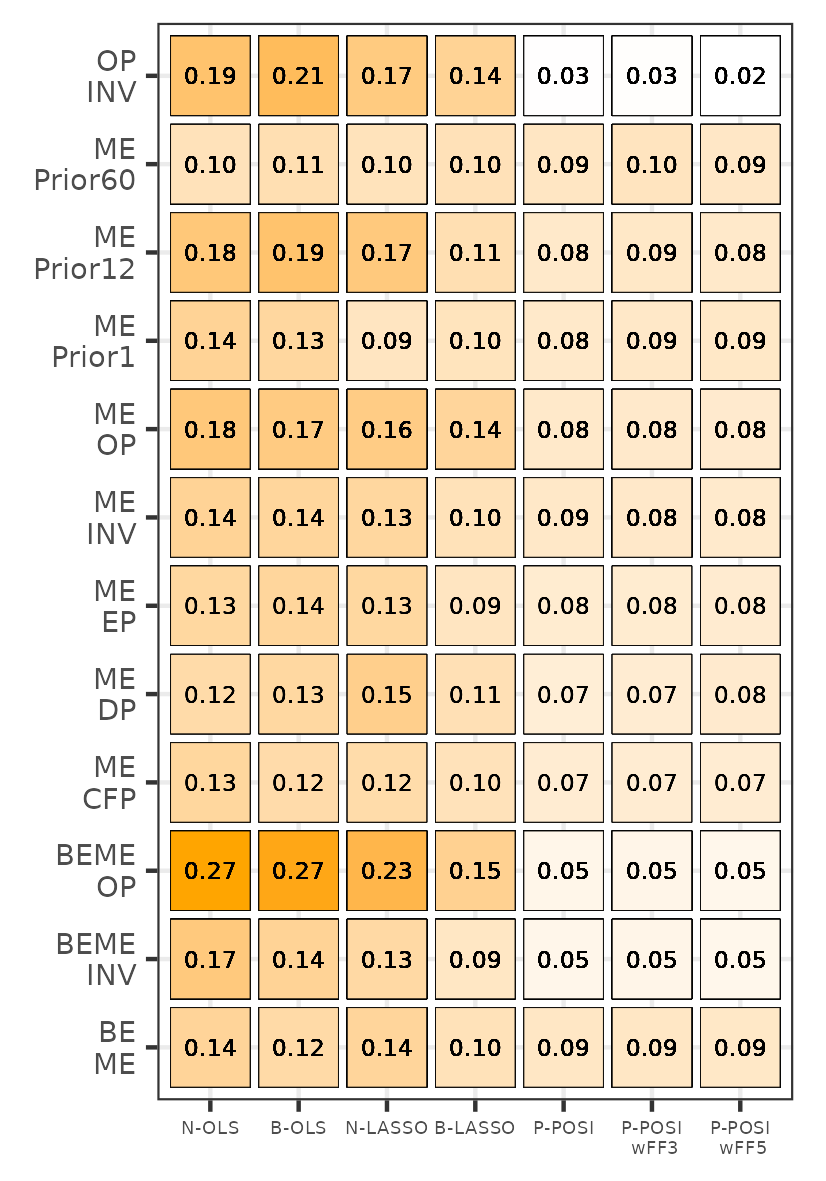}
		\caption{Out-of-sample}
	\end{subfigure}
	\bnotetab{
		This figure shows the mean absolute pricing errors (MAPE) for each cross-section of test assets for different factor models. The test assets are the $N=243$ double-sorted portfolios, and we show the average $|\alpha|$ for each set of double sorts. The rejection FWER is set to $\gamma=5\%$ The candidate factors are the 114 univariate factor portfolios. The time dimension is $T=660$. We use the first half for the in-sample estimation and selection, while the second half serves for the out-of-sample analysis. We compare Panel PoSI (P-PoSI), Panel PoSI with infinite priors on FF3 and FF5 (P-PoSI $\omega_{\text{FF3}}$ respectively $\omega_{\text{FF5}}$), Bonferroni LASSO (B-LASSO), Naive LASSO (N-LASSO), Bonferroni OLS (B-OLS) and Naive OLS (N-OLS).          
	}		
\end{figure}

Figure \ref{fig:RMSE} shows in-sample and out-of-sample RMSE for each set of double-sorts. The composition of the double sorts is summarized in Table \ref{DSSource} in the Appendix. The in-sample performance in the left subfigure has the expected result that more factors mechanically decrease the RMSE. The important findings are in the right subfigure with the out-of-sample RMSE. The uniformly best performing model is Panel PoSI without any priors. In fact, imposing a prior on the Fama-French factors increases the out-of-sample RMSE. The conventional LASSO and OLS estimates have substantially higher RMSE, which can be more than twice as large.

The Panel PoSI models also explain the average returns the best. In Figure \ref{fig:MAPE}, we compare the mean absolute pricing errors among the benchmarks for each set of double sorts. Importantly, the pricing errors are not used as in objective function of the estimation, and hence the fact that the models with the smallest RMSE explain expected returns is an economic finding supporting arbitrage pricing theory. Our Panel PoSI has the smallest out-of-sample pricing errors, which can be up to six times smaller compared to the OLS estimates. Including the Fama-French factors as a prior does not improve the models, except for the profitability and investment double sort, which uses the same information as two of the Fama-French factors.

\begin{table}[t!]
	\tcaptab{Selected factors with Panel PoSI}
	\label{DSHL}
	\centering 
	\begin{tabular}{l|ccc|c}
		\toprule
		Factor  & $N_j$    & $p_j$    & $\rho^{-1} N_jp_j$ & Order \\ \toprule
		\multicolumn{5}{c}{No prior } \\\midrule
		Size (SMB)                      & 1824 & \textless{}0.00001 & \textless{}0.0001    & 1 \\
		Dollar Trading Volume (dtv\_12) & 2099 & \textless{}0.00001 & \textless{}0.0001    & 2 \\
		Value (HML)                     & 1191 & \textless{}0.00001 & 0.0280               & 3 \\
		Short-Term Reversal (srev)      & 1050 & 0.00001            & 0.0974               & 4 \\
		Forecast Revisions (rev\_1)     & 242  & 0.00018            & 0.2782               & 5 \\
		Investment (CMA)                & 998  & 0.00112            & \textgreater{}0.9999 & 6 \\
		Profitability (RMW)             & 797  & 0.00123            & \textgreater{}0.9999 & 7\\\midrule
		\multicolumn{5}{c}{FF3 prior ($\omega_{\text{FF3}}$) } \\\midrule
		Size (SMB)                      & 2802 & \textless{}0.00001 & \textless{}0.0001    & 1 \\
		Value (HML)                     & 2802 & \textless{}0.00001 & \textless{}0.0001    & 2 \\
		Dollar Trading Volume (dtv\_12) & 779  & \textless{}0.00001 & 0.0017               & 3 \\
		Short-Term Reversal (srev)      & 1106 & \textless{}0.00001 & 0.0049               & 4 \\
		Profitability (RMW)             & 819  & 0.00006            & 0.2527               & 5 \\
		Investment (CMA)                & 874  & 0.00087            & \textgreater{}0.9999 & 6\\ \midrule 
		\multicolumn{5}{c}{FF5 prior ($\omega_{\text{FF5}}$) }  \\\midrule
		Size (SMB)                      & 2911 & \textless{}0.00001 & \textless{}0.0001 & 1 \\
		Value (HML)                     & 2911 & \textless{}0.00001 & \textless{}0.0001 & 2 \\
		Forecast Revisions (rev\_1)     & 230  & \textless{}0.00001 & 0.0005            & 3 \\
		Short-Term Reversal (srev)      & 1140 & \textless{}0.00001 & 0.0052            & 4 \\
		Dollar Trading Volume (dtv\_12) & 661  & \textless{}0.00001 & 0.0072            & 5 \\
		Profitability (RMW)             & 2911 & 0.00001            & 0.1937            & 6 \\
		Investment (CMA)                & 2911 & 0.00001            & 0.1996            & 7 \\
		Gross profits-to-assets (gpa)   & 1151 & 0.00013            & 0.8382            & 8\\ 
		\bottomrule
	\end{tabular}
	\bnotetab{This table reports ranking of factors based on their FWER bound for no prior, and infinite weight priors on the Fama-French 3 and 5 factors. The test assets are the $N=243$ double-sorted portfolios and the candidate factors are $J=114$ univariate long-short factors. The rows are ordered based on sorted ascending $\rho^{-1} N_jp_j$, which corresponds to the FWER bound.}
\end{table}

The Panel PoSI models select economically meaningful factors. Table \ref{DSHL} reports the ranking of factors based on their FWER bound without prior and infinite prior weights on the Fama-French 3 and 5 factors. The rows are ordered based on sorted ascending $\rho^{-1} N_jp_j$, which corresponds to the FWER bound. It allows us to infer the number of factors for different levels of FWER control values. Setting $\gamma=5\%$ leads to 3, 4 and respectively 5 factors, while a $\gamma=1\%$ results in 2, 4 and 5 factors, respectively. 

In addition to their significance, we can infer the relative importance of factors. The baseline PoSI with $\gamma=5\%$ selects a size, dollar trading volume and value factor. The size and value factors are among the most widely used asset pricing factors. Their selection is in line with their economic importance and confirms the Fama-French 3 factor model. The dollar trading volume factor is less conventional, but is correlated with many assets in our cross-sections. The size factor is the most important as measured by the FWER bound, that is, the product of the number of relevant assets and its minimum p-value are the smallest. The short term reversal factor is less important and would require a FWER control of 10\% to be included.

Imposing a prior affects the p-values of PoSI and the simultaneity count. For example, the cohesiveness coefficient increases from $\rho=0.16$ for no priors to $\rho=0.18$ in the case of the two priors. Hence, the FWER bounds of all factors can change when we impose a prior. The FF3 prior increases the significance of the short-term reversal factor, which is widely used in asset pricing. Interestingly, even for a FF5 prior, the profitability and investment factors remain insignificant.

\subsection{Number of Factors}

Our method contributes to the discussion about the number of asset pricing factors. Many popular asset pricing models suggest between three and six factors. Our approach allows a disciplined estimate for the number of factors based on inferential theory. The level of sparsity of a linear model also depends on the rotation of the covariates. Therefore, we also study the principal components (PCs) of the covariates $X$ as candidate factors. In this case, we use the step-down procedure, which we refer to as ``Ordered PoSI'' or O-POSI for short.

\begin{figure}[t!]
	\centering	\tcaptab{Number of selected factors for different FWER}	
	\label{fig:number}\centering
	\begin{subfigure}[t]{.48\textwidth}\centering
		\includegraphics[width=0.8\linewidth]{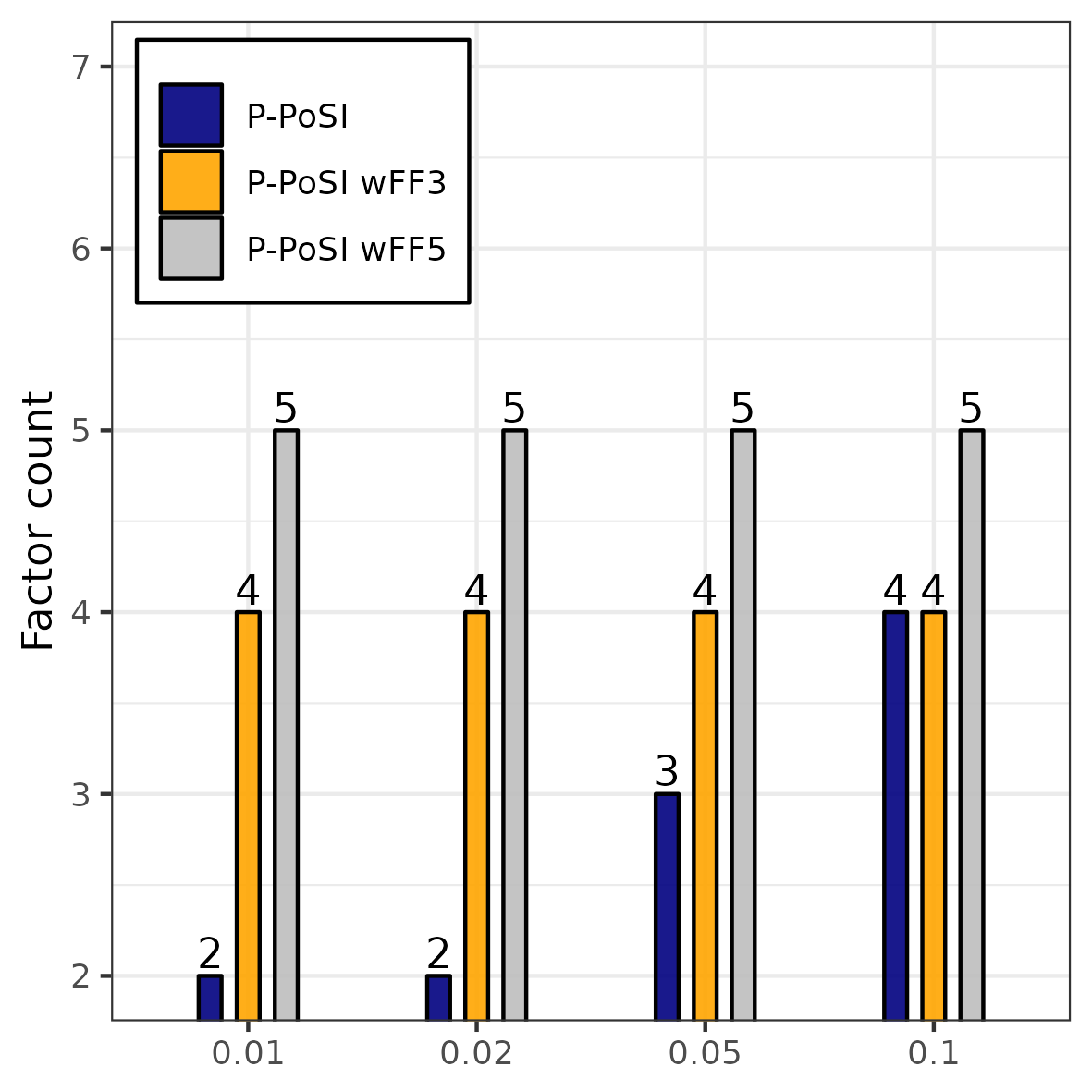}
		\caption{Univariate factors with priors (P-POSI)}
	\end{subfigure}
	\begin{subfigure}[t]{.48\textwidth}\centering
		\includegraphics[width=0.8\linewidth]{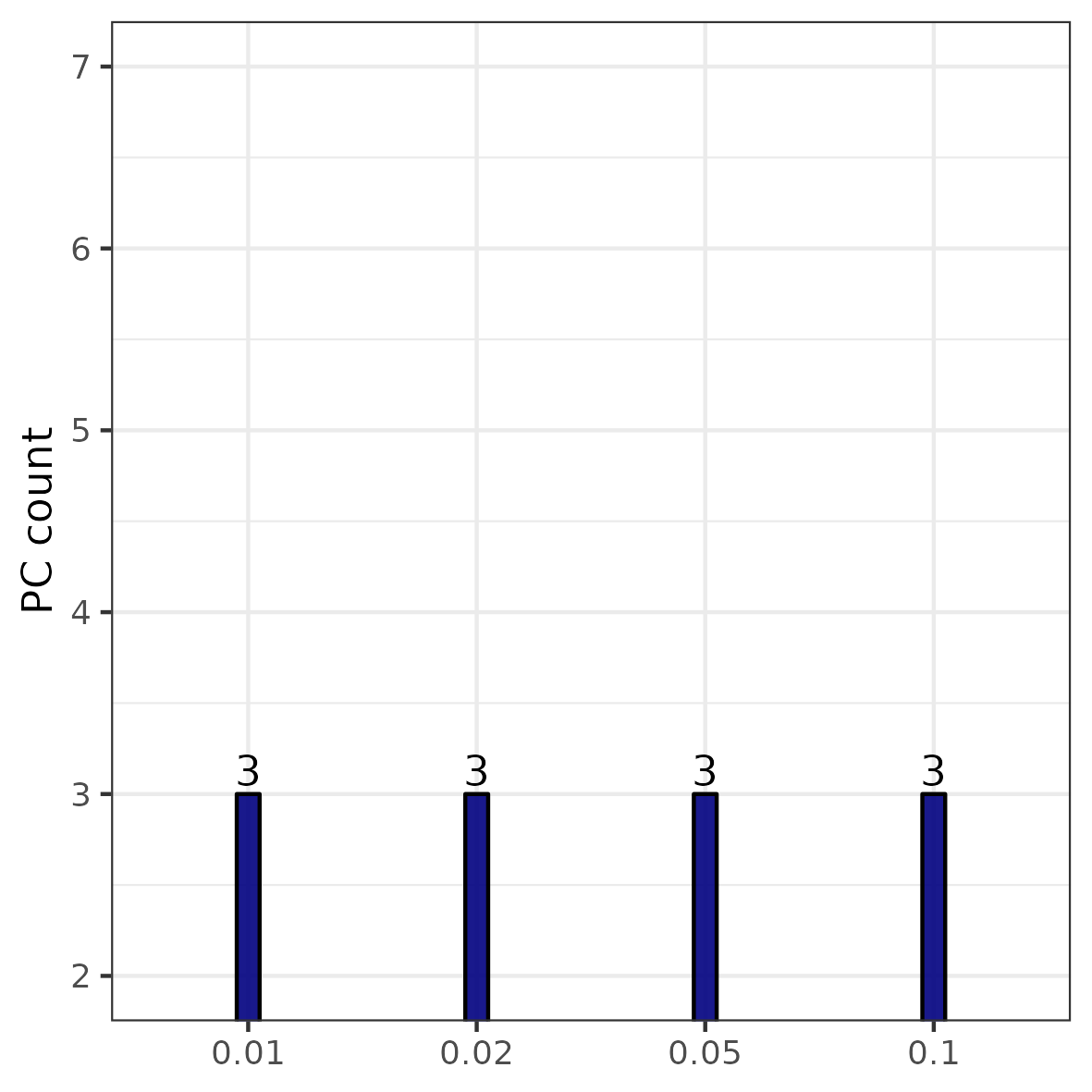}\\
		\caption{PCA rotated factors (O-POSI)}
	\end{subfigure}\\
	\bnotefig{This figure shows the number of selected factors to explain the test assets of double-sorted portfolios for different FWER rates $\gamma$. The factor count is obtained by traversing $K^*(\gamma)$ for $\gamma$ ranging from 0.01 to 0.1. The left subfigure uses univariate high-minus-low factors as candidate factors. We consider the case of no prior, and the cases of an infinite weight on the Fama-French 3 factor model ($\omega_{\text{FF3}}$) and an infinite weight on the Fama-French 5 factor model ($\omega_{\text{FF5}}$). The right subfigure uses the PCA rotation as candidate factors with the step-down procedure Ordered PoSI (O-POSI).}
\end{figure}

\begin{table}[t!]
	\tcaptab{Number of selected factors for different methods}
	\label{tab:number}
	\centering
	{\small
			\renewcommand{\arraystretch}{1.2}
			\begin{tabular}{@{}llll@{}}
				\toprule
				& HL & PCs   & HL + PCs  \\ \midrule
				Panel PoSI             & 3   & 3    & 4 \\
				Bonferroni PoSI        & 2   & 3    & 2 \\
				Bonferroni Naive LASSO & 10  & 29   & 10 \\
				Naive LASSO            & 70  & 50  &76 \\
				Bonferroni OLS         & 107 & 13 & 117 \\
				Naive OLS              & 114 & 50  & 164\\                  
				\bottomrule
			\end{tabular}
		}
		\bnotetab{
			This figure shows the number of selected factors to explain the test assets of double-sorted portfolios for different methods and different sets of candidate factors. The rejection FWER is set to $\gamma=5\%$. The factor count is obtained by traversing $K^*(\gamma)$ for $\gamma$. The number of factors is selected on the in-sample data. For PCs, we use the step-down method for the nested hypothesis. 
		}
	\end{table}

	Figure \ref{fig:number} shows the number of factors for different FWER rates $\gamma$. The factor count is obtained by traversing $K^*(\gamma)$ equal to 0.01, 0.02, 0.05 and 0.1. Panel PoSI without priors selects 2 factors for $\gamma=0.01$ and 3 for $\gamma=0.05$. Once, we impose an infinite weight on the Fama-French 3 factors, we select 4 factors for all FWER levels, while the prior on the Fama-French 5 factors results in a 5 factor model for all FWER levels. The Ordered PoSI with PCA rotated factors selects 3 factors for all FWER levels. In summary, our results confirm that depending on the desired significance, the number of asset pricing factors for a good model seems to be between 2 and 4. Note that our analysis is orthogonal to the market factor, which would also be added to the final model. Thus, the final model would have between 3 and 5 factors.

	Table \ref{tab:number} further confirms our findings about the number of asset pricing factors. We compare the number of factors for $\gamma=5\%$ selected either from the univariate high-minus-low factors (HL), their PCA rotation or the combination of the high-minus-low factors and their PCs. Panel PoSI selects consistently 3 factors from the long-short factors and their PCs. When combined, PoSI selects 4 factors, which is plausible as the optimal sparse model can be different for this larger set of candidate factors. The Bonferroni PoSI is overly conservative and selects only 2 HL factors. The models based on Naive LASSO or OLS select excessively many factors independent of the rotation. Overall, the findings support that parsimonious asset pricing models can be described by three to four factors. Of course, any discussion about the number of asset pricing factors is always subject to the choice of test assets and candidate factors.

\section{Conclusion} \label{sec6}


This paper proposes a new method for covariate selection in large dimensional panels. We develop the conditional inferential theory for large dimensional panel data with many covariates by combining post-selection inference with a new multiple testing method specifically designed for panel data. Our novel data-driven hypotheses are conditional on sparse covariate selections and valid for any regularized estimator. Based on our panel localization procedure, we control for family-wise error rates for the covariate discovery and can test unordered and nested families of hypotheses for large cross-sections. We provide a method that allows us to traverse the inferential results and determine the least number of covariates that have to be included given a user-specified FWER level. 

As an easy-to-use and practically relevant procedure, we propose Panel-PoSI, which combines the data-driven adjustment for panel multiple testing with valid post-selection p-values of a generalized LASSO, that allows to incorporate weights for priors. In an empirical study, we select a small number of asset pricing factors that explain a large cross-section of investment strategies. Our method dominates the benchmarks out-of-sample due to its better control of false rejections and detections.


\appendix

\renewcommand{\thesubsection}{\Alph{section}.\arabic{subsection}}
\setcounter{table}{0}
\setcounter{figure}{0}
\renewcommand{\thetable}{A.\arabic{table}}
\renewcommand{\thefigure}{A.\arabic{figure}}

\newtheorem{thm}{Theorem}[section]
\newtheorem{lem}{Theorem}[section]
\newtheorem{defin}{Theorem}[section]
\newtheorem{ass}{Theorem}[section]

\newtheorem{theorem_app}[thm]{Theorem}
\newtheorem{lemma_app}[lem]{Lemma}
\newtheorem{assumption_app}[ass]{Assumption}
\newtheorem{def_app}[defin]{Definition}

\section{Post-selection Inference with Weighted-LASSO}\label{lab:appendix}

\subsection{Weighted-LASSO: Linear Truncation Results}\label{lab:appendixA1}

This appendix collects the assumptions and formal statements underlying Theorem \ref{thm1_main}. We present the results for the Weighted-LASSO, which includes the conventional LASSO as a special case. In order to ensure uniqueness of the LASSO solution, we impose the following condition, which is standard in the LASSO literature:

\begin{def_app}{\bf General position}\label{def1}\\
	The matrix $\bm{X} \in\RR^{T\times J}$ has columns in general position if the affine span of any $J_0+1$ points $(\sigma_1X_{i_1},...,\sigma_{J_0+1}X_{i_{J_0+1}})$ in $\RR^T$ for arbitrary $\sigma_1,...\sigma_{d_0+1}\in\{\pm1\}$ does not contain any element of $\{\pm X_i:i\notin\{ i_1,...,i_{J_0+1}\} \}$, where $J_0<J$
    and $X_i$ denotes $i$th column of $\bm{X}$.
\end{def_app}

This position notion will help us to avoid ambiguity in the LASSO solution. Note that this condition is a much weaker requirement than the full-rank of $\bm{X}$, and states that if one constructs a $J_0$-dimensional subspace, it must contain at most $J_0+1$ entries of $\{\pm X_1,...,\pm X_J \}$. Even though this appears to be a complicated and mechanical condition, by a union argument it turns out that with probability 1, if the entries of $\bm{X}\in\RR^{T\times J}$ are drawn from a continuous probability distribution on $\RR^{T \times J}$ then $\bm{X}$ is in general position.\footnote{See \cite{cpa.20132} and \S2.2 of \cite{10.1214/13-EJS815} for more discussions on uniqueness and general position.} Then, we will be able to discuss the LASSO solution for general design with relative ease, thanks to Lemma 3 of \cite{10.1214/13-EJS815}. It shows that if $\bm{X}$ lie in general position,  it is sufficient to have a unique LASSO solution regardless of the penalty scalar $\lambda$. This condition will later be used in establishing our Lemma \ref{lem2}.

We can now state the formal assumptions:
\begin{assumption_app}\label{asu1} {\bf Unique low dimensional model}
	\vspace{-0.1cm}
	\begin{enumerate}[label=(\alph*)]
		\item {\bf Low dimensional truth}:\\
		The data satisfies $Y=X_S\beta_S+\epsilon$ where $|S|=O(1)$.
		\item {\bf General position design}:\\
		The covariates $\bm{X}$ have columns in general position as given by Definition \ref{def1}.
	\item {\bf Sufficient observation}:\\
The covariate dimension $J$ is smaller than the number of time-series observations $T$.
		
	\end{enumerate}
\end{assumption_app}

We start our analysis with the simpler model of known error variance, and later extend it to the case of estimated unknown variance.

\begin{assumption_app}\label{asu_known} {\bf Gaussian residual with known variance}\\
	The residuals are distributed as $\epsilon\sim \mathcal{N}(0,\Sigma)$ where $\Sigma$ is known.
\end{assumption_app}


Before formalizing the inferential theory, we need to clarify the quantity for which we want to make inference statements. As stated before, we only test the hypothesis on a covariate if its LASSO estimate turns out active. This is exactly the approach how researchers in practice conduct explorations in high-dimensional datasets. In other words, we focus on $\hat{\beta}_M$ and quantities associated with it, where $M$ denotes the active set of selected covariates. 

We study the inferential theory of the ``debiased estimator", which is a shifted version of the LASSO fit as defined below. We show that this debiased estimator is unbiased, consistent and follows a truncated Gaussian distribution, with profound connections to the debiased LASSO literature such as \cite{10.1214/17-AOS1630}, but has different properties by a subtle different descent direction. More concretely, given $\mathcal{M}$, clearly $\hat{Y}=X_M\hat{\beta}_M$ is the fitted value since $\hat{\beta}_{-M}=0$, where $-M$ is the complement of the set $M$. We let $\hat{\epsilon}_M:=Y-X_M\hat{\beta}_M$ be the residual from the LASSO estimator. By considering only the partial LASSO loss of $\ell(Y,X_M,\lambda,\beta)$ and given we are currently at the LASSO estimator $\hat{\beta}$, the Newton step is $X_M^+\hat{\epsilon}$ following \cite[\S~9.5.2]{boyd2004convex}, where we denote $X_M^+=(X_M^\top X_M)^{-1}X_M^\top $ as the pseudo-inverse of the active submatrix of $X$. The invertibility of $X_M^\top X_M$ either is observed when we are in the fixed design regime or happens almost surely when we are dealing with continuous quantities, as a consequence of Assumption \ref{asu1}(b) as argued in \cite{10.1214/13-EJS815} and \cite{lee2016exact}. Now we can formally define the main object of our inferential theories:

\begin{def_app}{\bf Debiased Estimator}\label{def2}\\
	The debiased Weighted-LASSO estimator $\bar{\beta}_M$ given $\mathcal{M}$ is given by
	\begin{equation}\label{11}
		\bar{\beta}_{M}=\hat{\beta}_{M}+X_M^+\hat{\epsilon}_M
	\end{equation}
\end{def_app}

It is now evident why some of the literature refers to the debiased estimator also as the one-step estimator: given that $\hat{\beta}_M$ solves the Karush-Kuhn-Tucker (KKT) condition and reaches the optimal sub-gradient for the full loss $\ell(Y,X,c,\beta)$, our debiased estimator $\bar{\beta}_M$ is the result of moving {one} more Newton-Ralphson method {step} after $\hat{\beta}_M$, but only taking $X_M$ rather than $X$ as a whole into the likelihood loss function. Hence, the update step is actually only a partial update from the LASSO solution point. Intuitively, $\bar{\beta}_M$ should still be close to solving the KKT conditions, and would exactly solve the KKT conditions if $X_M$ happen to be the true covariates (i.e. $X_M=X_S$). 

If we were to take a Newton's method step with gradient and Hessian calculated with the entirety of data $X$, or equivalently taking a full update from the stationary point, we will recover the $\hat{\beta}_{M}^d$ proposed in \cite{10.1214/17-AOS1630}. The material difference is that the full-update would require the $J\times J$ precision matrix $\Omega=\Gamma^{-1}$, where $\Gamma=X^\top X$ if $X$ assumed fixed or $\Gamma=\EE[X^\top X]$ if $X$ assumed to be generated from a stationary process. Using $\ell(Y,X_M,\lambda,\beta)$ instead of $\ell(Y,X,\lambda,\beta)$, our debiased estimator would not need the full Hessian, which is leveraging LASSO's screening property and uses $(X_M^\top X_M)^{-1}X_M^\top $ (i.e. $X_M^+$) as a much lower-dimensional alternative of $\Omega X^\top $.

Without loss of generality, we assume that the covariate indexed $i\leq |M|$ is part of $M$, and we can always rearrange the columns of $X$ to have the first $|M|$ covariates as active. Let $\eta=(X_M^+)^\top e_j\in\RR^T$ be a vector where $e_j\in\RR^{|M|}$ is a vector with 1 at $j$th coordinate and 0 otherwise.
Hence, the $\eta$ vector is the linear mapping from $Y$ to the $j$th coordinate of an OLS estimator. In particular, the debiased estimator and the response satisfy the following relationship:

\begin{lemma_app}\label{lem1}
	{\bf Debiased Estimator is OLS-post-LASSO}\\
	The debiased estimator is a linear mapping of $Y$. Specifically, given $\eta=(X_M^+)^\top e_j$:
	\begin{equation}\label{16}
		\bar{\beta}_j=\eta^\top Y
	\end{equation}
	Moreover, $\bar{\beta}_{M}$ is the OLS estimate of regressing $X_M$ on $Y$:
	\begin{equation}
		\bar{\beta}_{M}=\argmin_\beta\frac{1}{2T}\|Y-X_M\beta\|_2^2.
	\end{equation}
\end{lemma_app}

The proof of Lemma \ref{lem1} is deferred to the Online Appendix. Although its proof is simple, this lemma reveals that our debiased estimator is the same as the least-square after LASSO estimator proposed in \cite{10.3150/11-BEJ410}.

Our strategy to obtain a rigorous statistical inferential theory with $p$-values is as follows. First we perform an algebraic manipulation to transform $\hat{\beta}_M$ into $\bar{\beta}_M$ in the linear form of (\ref{16}). Then, we follow the strategy in \cite{lee2016exact} to traverse the KKT subgradient optimal equations for general $\bm{X}$ by writing it equivalently into a truncation in the form of $\{ AY\leq b\}$, as we will do in Lemma \ref{lem2}. Finally we will circle back to $\hat{\beta}_M$ by the linear mapping between $\bar{\beta}_M$ and $Y$ and the distributional results induced by the fact that $Y$ is truncated by $\{A Y\leq b\}$.

For our Weighted-LASSO, the KKT sub-gradient equations are 
\begin{equation}
	X^\top (X\hat{\beta}-Y)+\lambda\begin{bmatrix}s\\v\end{bmatrix}\odot\omega^{-1}=0\q \wh 
	\begin{cases}
		{s_j}=\sgn(\hat{\beta}_j)&\text{ if }\hat{\beta}_j\neq 0,\omega_j<\infty\\
		v_j\in[-1,1]&\text{ if }\hat{\beta}_j=0,\omega_j<\infty
	\end{cases}
\end{equation}

In other words, when $\omega$ is specified, the KKT conditions can be identified using the tuple of $\{M,s\}$, where $M$ is the active covariates set and $s$ is the signs of LASSO fit. This is a consequence of how LASSO KKT condition can separate the slacks into $s$ for active variables and $v$ for inactive variables. Let $\mathcal{J}$ be the set of $j$'s corresponding to $\omega_j$'s that are infinite. When $\mathcal{J}\neq \emptyset$, we would simply need $s_j<\infty$ for $i\in\mathcal{J}$ because $\lambda{s}_j/\omega_j=0$ is guaranteed. We rigorously characterize the KKT sub-gradient conditions as a combinations of signs and infinity norm bounds conditions by the following lemma, which parallels Lemma 4.1 of \cite{lee2016exact}:

\begin{lemma_app}\label{lem2}
	{\bf Selection in norm equivalency}\\
	Consider the following random variables
	\begin{equation}
		\begin{split}
			w(M,s,\omega)&=(X_M^\top X_M)^{-1}(X_M^\top Y-\lambda s\odot\omega^{-1}_M)\\
			u(M,s,\omega)&=\omega_{-M}\odot
			\left(
			X_{-M}^\top(X_M^+)^\top s\odot\omega^{-1}_M
			+\frac{1}{\lambda}X_{-M}^\top (I-P_M)Y\right)
		\end{split}
	\end{equation}
	where $P_M=X_MX_M^+$ is the projection matrix. The Weighted-LASSO selection can be written equivalently as
	\begin{equation}
		\{M,s\}=\{\sgn(w(M,s,\omega))=s,\|u(M,s,\omega)\|_\infty<1 \}
	\end{equation}
\end{lemma_app}

Using this characterization, we are then able to provide the distributional results for the debiased estimators. Consider $\xi=\Sigma\eta(\eta^\top \Sigma\eta)^{-1}\in\RR^{T}$ as a covariance-scaled version of our $\eta$, and a mapping of $\bm{Y}$ using residual projection matrix: $ z=(I-\xi\eta^\top )\bm{Y}$. Note that $z$ can be calculated once we observe $(\bm{X},\bm{Y})$, so it can be conditioned on were we to do so. We will soon see that the truncation set will depend on the variable $z$, but this does not cause any issues thanks to the following lemma, the proof of which is deferred to the Online Appendix:

\begin{lemma_app}{\bf Ancillarity in truncation}\label{lem3}\\
	The projected $z$ and the debiased estimator $\bar{\beta}_j$ are independently distributed.
\end{lemma_app}

As a result of Lemma \ref{lem3}, when describing the distribution of $\bar{\beta}_j$, we can use $z$ in its truncation conditions as long as we condition on $z$ as well. To simplify notation, we can collect all quantities we need to condition on into $\tilde{\mathcal{M}}:=((M,s),z,\omega, X)$. Now we can combine Lemmas \ref{lem1}, \ref{lem2}, \ref{lem3} to arrive at the truncated Gaussian statements for the debiased estimator similar to \cite{lee2016exact}, but for weighted-LASSO:

\begin{theorem_app}{\bf Truncated Gaussian}\label{thm1}\\
	Under Assumptions \ref{asu1} and \ref{asu_known} for $j\in M$,  $\bar{\beta}_{j}$ is conditionally distributed as:
	\begin{equation}
		\bar{\beta}_{j}|\tilde{\mathcal{M}}			\sim \mathcal{TN}(\beta_{j},\eta^\top \Sigma\eta;[V^{-}(z),V^{+}(z)])
	\end{equation}
	where $\mathcal{TN}$ is a truncated Gaussian with mean $\beta_{j}$, variance $\eta^\top \Sigma\eta$ and truncation set $[V^{-}(z),V^{+}(z)]$. $\beta_{j}$ denotes the $i$th entry of the true $\beta$. The vector of signs is $s=\sgn(\hat{\beta}_M)\in\RR^{|M|}$ and the truncation set depends on
	\begin{align*}
		&A=\begin{bmatrix}
			\lambda^{-1}X_{-M}^\top (I-P_M)\\
			-\lambda^{-1}X_{-M}^\top (I-P_M)\\
			-\text{diag}(s)X_M^+
		\end{bmatrix}\in\RR^{(2J-|M|)\times T},  
		\;\;\;  b=\begin{bmatrix}
			\omega_{-M}^{-1}-X_{-M}^\top(X_M^+)^\top s\odot\omega^{-1}_M\\
			\omega_{-M}^{-1}+X_{-M}^\top(X_M^+)^\top s\odot\omega^{-1}_M\\
			-\lambda\cdot \text{diag}(s)(X_M^\top X_M)^{-1}s\odot\omega^{-1}_M
		\end{bmatrix}\in\RR^{2J-|M|} \\
		&V^{-}(z)=\max_{j:(A\xi)_j<0}\frac{b_j-(Az)_j}{(A\xi)_j}, \qquad
		V^{+}(z)=\min_{j:(A\xi)_j>0}\frac{b_j-(Az)_j}{(A\xi)_j}.
	\end{align*}	
\end{theorem_app}
Assumption \ref{asu1} of Gaussian errors can be relaxed and replaced by an appropriate asymptotic normal distribution based on a central limit theorem. The arguments for this extension are discussed in \cite{tian2017asymptotics}. The general structure of the statement would remain unchanged, and the coefficients would asymptotically follow the same truncated Gaussian distribution. We adopt Assumption \ref{asu1} to simplify our discussion, and point out where the weights $\omega$ appear in the post-selection inference.      


Notice that Theorem \ref{thm1} is decoupled across $M$, which is to say we are able to deal with $1$-dimensional statistics. We arrive at this form because the construction of $(V^-,V^+)$ over the extreme points of the linear inequality system (or vertices of the polyhedral) has decomposed the dimensionality of the truncation. This decoupling is of significant practical value, in that it would be otherwise a non-trivial task to calculate a statistic of multivariate (in our case $|M|$-dimensional) truncated Gaussian and then marginalize over $|M|-1$ dimensions.

\subsection{Weighted-LASSO Quasi-Linear Truncation with Estimated Variance}\label{lab:appendixA2}

This section generalizes the distribution results to the practically relevant case when the noise variance is unknown and has to be estimated. This becomes a challenging problem for post-selection inference. We replace Assumption \ref{asu_known} by the following assumption:

\begin{assumption_app}{\bf Gaussian residual with simple unknown variance}\label{asu_unknown}\\
	The residuals are distributed as $\epsilon_j\iid \mathcal{N}(0,\sigma^2)$ where $\sigma^2$ is unknown.
\end{assumption_app}

The simple structure of unknown variance of Assumption \ref{asu_known} is common in the post-selection inference literature as for example in \cite{lee2016exact} and \cite{tian2017selective}. A feasible conditional distribution replaces $\sigma^2$ with an estimate. Under Assumption \ref{asu_known}, we can estimate the variance using LASSO residuals and then reiterate the previous truncation arguments. The most common standard variance estimator is
\begin{equation}\label{eq21:sd.est}
	\hat{\sigma}^2(Y)=\|Y-X_M\bar{\beta}_M\|_2^2/(T-|M|).
\end{equation}

In classical regression analysis, the normally distributed estimated coefficient divided by an estimated standard deviation follows a $t$-statistic. Hence, we would expect that a truncated normal debiased estimator divided by a sample standard deviation might yield a truncated $t$-distribution. However, the arguments are substantially more involved. Simply using $\hat{\sigma}(Y)$ of (\ref{eq21:sd.est}) in the expression $\eta^\top\Sigma\eta$ of Theorem \ref{thm1} changes the truncation. Specifically, $Y$ having truncated support means $\hat{\sigma}(Y)^2$ is not $\chi^2$-distributed supported on the entire $\RR_+$, which makes the support of $\bar{\beta}/\hat{\sigma}(Y)$ non-trivial. Therefore, in order to correctly assess the truncation of the studentized quantity,  we have to disentangle how much truncation is implied in $\hat{\sigma}(Y)^{-1}$ and $\bar{\beta}$ simultaneously. Geometrically, as $\hat{\sigma}(Y)$ is a non-linear function of $Y$ and $\bar{\beta}$, the truncation on $Y$ is in fact no longer of the simple linear form $\{AY\leq b\}$ such as in Theorem \ref{thm1}.

Instead of a polyhedral induced by affine constraints, we have a ``quasi-affine constraints'' form of $\{C Y\leq \hat{\sigma}(Y) b\}$ because LASSO KKT conditions preserve the estimated variance throughout the arguments. Thus, both sides of the inequality $CY\leq \hat{\sigma}(Y)b$ have $Y$, and in right-hand-side the $\hat{\sigma}(Y)$ is non-linear in $Y$. A significantly more complex set of arguments are needed compute the exact truncation, which is equivalent to solve for a $|M|$-system of non-linear inequalities rather than linear inequalities that constrain the support of $Y$ for inference on each $\bar{\beta}_j$. Theorem \ref{thm:appendix} shows the appropriate truncation based on those arguments:

\begin{theorem_app}{\bf Truncated $t$-distribution for estimated variance}\label{thm:appendix} \\
	Under Assumptions \ref{asu1} and \ref{asu_unknown}, and the null hypothesis that $\beta_j=0$, the conditional distribution of the studentized statistic follows
	\begin{equation}
			\frac{\bar{\beta}_j}{\|\eta\|\hat{\sigma}(Y)}\sim TT_{d;\Omega},
	\end{equation}
	where $TT$ is a truncated t-distribution with $d$ degrees of freedom and truncation set $\Omega$. The truncation set $\Omega=\bigcap_{j\in M}\{t:t\sqrt{W}\nu_j+\xi_j\sqrt{d+t^2}\leq -\theta_j\sqrt{W}\}$ is an $|M|$-intersection of simple inequality-induced intervals based on the following quantities, where the active signs are denoted as $s=\sgn(\hat{\beta}_M)\in\RR^{|M|}$: 
	\begin{align*}
		\theta_j &=
		(\lambda s_j/\hat{\sigma}(Y) )\cdot  e_j^\top \left((X_M^\top X_M)^{-1}s\odot \omega^{-1}\right) \;\; \text{ for }j\in M,\\
		C&=-diag(s) X_M^+\in\RR^{|M|\times T}, \qquad \nu=C\eta/\|\eta\|_2\in\RR^{|M|}, \qquad \xi=C (P_M-\eta\eta^\top /\|\eta\|_2^2 )Y\in\RR^{|M|}, \\
		d&=\tr(I_T-P_M), \qquad W=\|(I-P_M)Y\|_2^2+\|\eta^\top Y\|_2^2/\|\eta\|_2^2.
	\end{align*}

\end{theorem_app}

The quantities $\theta$ and $C$ describe the quasi-linear constraints, whereas $\nu$ and $\xi$ transform them into the form of $\Omega$. Note that the $\Omega$ set is obtained from solving a low-dimensional set of quadratic inequalities that do not necessarily yield a single interval after intersection. We provide a proof of this result in the Online Appendix.

Using Theorem \ref{thm:appendix} in practice poses several challenges. First, the computations are much more involved, especially as each $\beta_j$ requires calculation of $\Omega$ which includes $|M|$ actual constraints, each of which involves solving a simple but still non-linear inequality. It is non-trivial to ensure that the numerical stability holds at every step of the calculations. Second, since $\Omega$ is not necessarily an interval, it is harder to interpret the truncation and also calculate the cumulative density function through Monte-Carlo simulations when there is a non-trivial truncation structure. Third, in fact, the authors in \cite{tian2017selective} recommend a regularized likelihood minimizing variance estimator that deviates from the simple $\hat{\sigma}(Y)$, which would in turn involves more numerical integration and optimization steps. Last but not least, this result was proposed initially for studying scale-LASSO. Our goal is to provide a set of tools that can be useful for a wide range of applications including the LASSO with $l_2$ squared norm loss rather than un-squared norm loss. These implementation difficulties are also discussed in more detail in the Online Appendix, which provides the accompanying proofs and the exact forms of the truncations.

We provide a practical solution based on an asymptotic normal argument. We impose the following two standard assumptions. First, we assume that we have a consistent estimator of the residual variance:
\begin{assumption_app}{\bf Consistent estimator $\hat{\sigma}^2(Y)$} \label{asu_consist}\\
The residual variance estimator is consistent $\hat{\sigma}^2(Y)\CP\sigma^2$ as $T\to\infty$ given the selection of covariates.
\end{assumption_app}
This general assumption includes many common scenarios such as the results specified in Corollary 6.1 of \cite{BG}, or in Theorem 2 of \cite{chatterjee2014assumptionless}. 

Second, we need to impose standard assumptions on the asymptotic moments of the covariates and asymptotic rates. Assuming full rank for the second moment of the selected set of covariates ensures ensures that we can run a regression with the active set. This standard assumption is also referred to as the Irrepresentable Condition (\cite{zhao2006model}) in the LASSO literature. It simply states that the linear regression with the selected covariates is well-defined in the limit. Moreover, we assume that the number of covariates $J$ grows linearly with the number of time-series observations $T$, which is the largest possible rate compatible with Assumption \ref{asu1}(c). Last but not least, we assume that $\lambda$ shrinks sufficiently fast. Our rate assumption on $\lambda$ follows the usual choice in this literature as for example in Corollary 3 of \cite{10.1214/12-STS400} and Corollary 6.1 of \cite{BG}. This $\lambda$ rate is also adopted in \cite{tian2017asymptotics}, and \cite{10.3150/11-BEJ410} refer to this as ``typically used in the literature''. In addition, $\lambda$ shrinking at this rate leads to diminishing $\lambda\sqrt{\log(J)/T}\to 0$ as $J,T$ grow and consistency of $\hat{\sigma}^2(Y)$ of (\ref{eq21:sd.est}) in the setting of \cite{chatterjee2014assumptionless}

\begin{assumption_app}\label{asu_gram} {\bf Asymptotic behavior}
	\vspace{-0.1cm}
	\begin{enumerate}[label=(\alph*)]
		\item {\bf  Full-rank second moment}:\\
	 Given the selected set of covariates $M$, the sample second moment matrix $\frac{1}{T} X_M^\top X_M $ of the active covariates $X_M$ converges to a full-rank matrix as $T \rightarrow \infty$.
		\item {\bf Growth rates}:\\
The rates satisfy $J=O(T)$ and $\lambda=O \left (\sqrt{\frac{\log J}{T}} \right)$ as $T\to\infty$.
	\end{enumerate}
\end{assumption_app}
Under these additional assumptions, the next theorem shows that the truncated Gaussian distribution results hold for the feasible Weighted-LASSO with estimated noise variance.


\begin{theorem_app}{\bf Asymptotic truncated normal distribution as approximation}\label{col2}\\
	Suppose Assumptions \ref{asu1}, \ref{asu_unknown}, \ref{asu_consist} and \ref{asu_gram} hold. For $T \rightarrow \infty$ and under the null hypothesis that $\beta_j=0$ and conditional on the selection events and the weights, the studentized quantity $\bar{\beta}_j/\|\eta\|\hat{\sigma}(Y)$ converges to a truncated Gaussian distribution. Moreover, the conditional $p$-values calculated from $TN_{\Omega}(\bar{\beta}_j/\|\eta\|\hat{\sigma}(Y))$ with an estimated noise variance satisfy Assumption \ref{assu:valid_p}, where $TN_{\Omega}$ is a standard normal distribution truncated to  $\Omega=[{V}^-(z)/\|\eta\|_2\hat{\sigma}(Y);{V}^+(z)/\|\eta\|_2\hat{\sigma}(Y)]$, and ${V}^{-}(z)$ and ${V}^{+}(z)$ are the same as in Theorem \ref{thm1}.
	\end{theorem_app}

The asymptotic result has several advantages. First, it is intuitive since it parallels the classical OLS inference with a $t$-statistic converging to Gaussianity. Secondly, it is computationally more tractable than results of Appendix Theorem \ref{thm:appendix}. With this result, one can obtain asymptotically valid post-selection $p$-values. The uniform distribution of the post-selection p-values conditional on the selection and under the null hypothesis $H_D$, are a direct consequence of the distribution result.\footnote{\cite{tian2017selective} propose an alternative estimator for the noise variance that also leads to a truncated Gaussian distribution. Their estimator for $\sigma^2$ is obtained from solving a more complicated regularized pseudolikelihood problem, instead of the conventional sample variance estimator in equation (\ref{eq21:sd.est}).} Our asymptotic distribution is derived conditional on the selection that is obtained from the finite sample. This is the same perspective as taken in \cite{tian2017asymptotics}, and the resulting $p$-value statements echo their Lemma 4. It is a challenging open question to derive the asymptotic distribution when the selection event changes as $T$ grows.

\section{Proof of Theorem \ref{thm_MT}}\label{app:proof}

\textbf{Proof:} We show how to count false discoveries under the data-driven null hypothesis conditional on the selection and then evaluate the probability of at least one false discovery under $H_D$. By design of $H_D$, we only need to consider false selections for the covariates that are active for some units. For $\mathcal{K}_j\neq \emptyset$, we denote by $j\in H_M$ that the $j$th covariate is tested in $H_M$. 

First, we separate the data-driven hypothesis into individual components. Within $H_D$, we consider all the hypotheses associated with the $j$th covariate denoted as $H_{D,j}$. The hypothesis $H_{D,j}$ represents an intermediate level of hypotheses between the panel-level $H_D$ and unit-covariate individual null $H_{0,j}^{(n)}|\mathcal{M}^{(n)}$. When written as a set intersection, $H_{D,j}$ equals $\bigcap_{n\in\mathcal{K}_j}H_{0,j}^{(n)}|\bigcap_{n\in \mathcal{K}_j}\mathcal{M}^{(n)}$, that is, $H_{D,j}$ is joint over all the units where the $j$th covariate is active. 

Second, we count the number of false discoveries. Under the null hypothesis $H_D$, we denote the number of false discovery in $H_{D,j}$ as $V_j$. Hence, the total number of false discoveries is the sum over the false discoveries of all covariates given by
\begin{equation}\label{eq:breakupV}
	\begin{split}
		\PP_{H_D} \left(V\geq 1|\mathcal{M} \right)		=	\PP_{H_D} \left(\sum_{j\in H_M} V_j\geq 1|\bigcap_{n\in \mathcal{K}_j}\mathcal{M}^{(n)} \right).
	\end{split}
\end{equation}
Each $V_j$ can be further broken down into the false discovery against the unit-covariate null hypotheses. Each individual potential false discovery is a random event, and the sum of false discoveries greater or equal to 1 corresponds to the union of these random events. The union of these random events has the conditional distribution
\begin{align}
	\PP_{H_D} \left(\sum_{j\in H_M} V_j\geq 1|\bigcap_{n\in \mathcal{K}_j}\mathcal{M}^{(n)} \right)
	= & \PP_{H_D}\left(\bigcup_{j\in H_M} \left\{ \bigcup_{n\in\mathcal{K}_i} \left\{\text{Rejection made based on } p_{j}^{(n)}|\bigcap_{n\in \mathcal{K}_j}\mathcal{M}^{(n)} \right\} \right\} \right) \nonumber\\
	= & \PP_{H_D} \left( \bigcup_{j\in H_M} \left \{ \bigcup_{n\in\mathcal{K}_i} \left\{ p_{j}^{(n)}\leq \rho\frac{\gamma}{ N_j}|\bigcap_{n\in \mathcal{K}_j}\mathcal{M}^{(n)} \right \} \right \} \right ).
\end{align}
The second line simply follows from the design of our rejection procedure. Boole's inequality implies the following union bound:
\begin{equation}\label{eq:boole}
	\begin{split}
		\PP_{H_D} \left (\bigcup_{j\in H_M} \left\{ \bigcup_{n\in\mathcal{K}_i} \left\{ p_{j}^{(n)}\leq \rho\frac{\gamma}{ N_j}|\bigcap_{n\in \mathcal{K}_j}\mathcal{M}^{(n)} \right \} \right \} \right)
		\leq & \sum_{j\in H_M}\sum_{n\in\mathcal{K}_j}\PP \left(p_{j}^{(n)}\leq \rho\frac{\gamma}{ N_j}|\bigcap_{n\in \mathcal{K}_j}\mathcal{M}^{(n)} \right).
	\end{split}
\end{equation}
Third, we take advantage of Assumption \ref{assu:valid_p}. Under Assumption 1, it holds that $\PP_{H_D}(p_{j}^{(n)}\leq \rho\frac{\gamma}{N_j})\leq  \rho\frac{\gamma}{N_j} $, which appears in  the right-hand side of (\ref{eq:boole}). Thus, combining equations (\ref{eq:breakupV})$\sim$(\ref{eq:boole}) yields:
\begin{align}
	\PP_{H_D}(V\geq 1|\mathcal{M})	
	\leq    \sum_{j\in H_M}\sum_{n\in\mathcal{K}_j}\rho\frac{\gamma}{ N_j} =  \gamma\cdot \rho\cdot \sum_{j\in H_M}\frac{1}{N_j}\sum_{n\in\mathcal{K}_j}1
	=  \gamma\cdot \rho\cdot \sum_{j\in H_M}\frac{|\mathcal{K}_j|}{N_j}
	=\gamma.
\end{align}
The second-to-last equation uses and also explains the definition of $\rho$. This completes the proof.\hfill\textbf{[QED]}

\section{Appendix: Empirics}

\begin{table}[H]
\tcaptab{Compositions of DS portfolios}\label{DSSource}
\centering\scriptsize	
\begin{tabular}{c|c||c|c||c|c||c|c}
\toprule
Sorted by     & \# portfolios &Sorted by     & \# portfolios&Sorted by     & \# portfolios&Sorted by     & \# portfolios\\ \midrule
BEME, INV     & 25& ME, CFP       & 6&   ME, INV       & 25& ME, Prior1  & 25 \\ 
BEME, OP      & 25& ME, DP        & 6&   ME, OP        & 25& ME, Prior12 & 25\\ 
ME, BE        & 25& ME, EP        & 6&   OP, INV       & 25& ME, Prior60 & 25\\\bottomrule
\end{tabular}
\bnotetab{This table lists the composition of double sorted portfolios that we use as test assets in our empirical study. All the double sorted portfolios are from Kenneth French's data library.}
\end{table}

\singlespacing
\bibliographystyle{econometrica}
{\small
\bibliography{main}
}
\onehalfspacing

%
%

\includepdf[pages={1-20}]{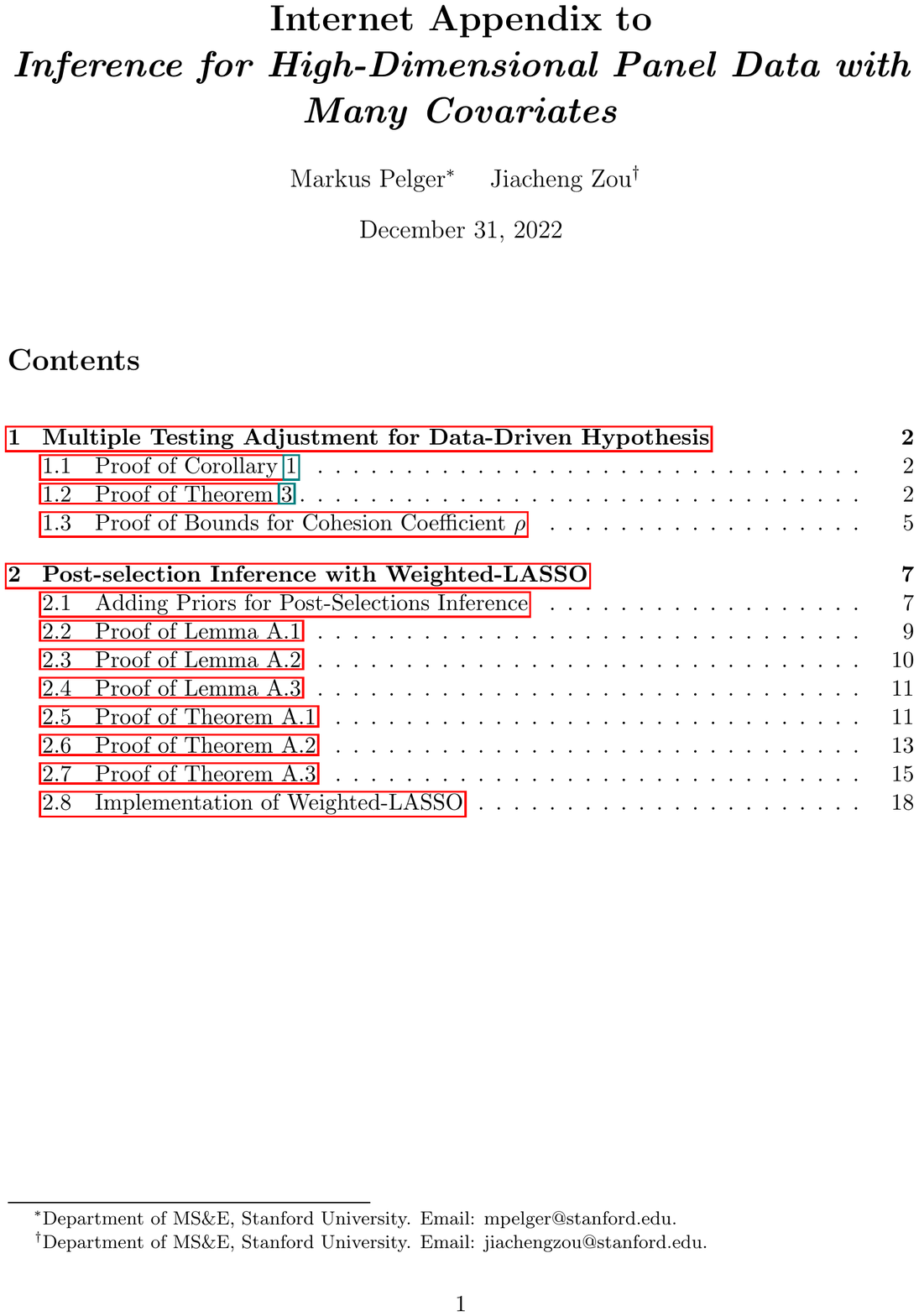}

\end{document}


\def\spacingset#1{\renewcommand{\baselinestretch}%
		{#1}\small\normalsize} \spacingset{1}

	
	\if0\blind
	{
		\title{\bf Internet Appendix to \\ \textit{Inference for High-Dimensional Panel Data with Many Covariates}}
		\author{Markus Pelger\thanks{Department of MS\&E, Stanford University. Email: mpelger@stanford.edu. }\hspace{.6cm}
			Jiacheng Zou\thanks{Department of MS\&E, Stanford University. Email: jiachengzou@stanford.edu. }\\
		}
		\date{December 31, 2022}
		\maketitle
	} \fi
	
	\if1\blind
	{
		\bigskip
		\bigskip
		\bigskip
		\begin{center}
			{\LARGE\bf Online Appendix of Conditional inference for\\high-dimensional panel data}
		\end{center}
		\author{Markus Pelger\hspace{.2cm}
			Department of MS\&E, Stanford University}
		and 
		{Jiacheng Zou\hspace{.2cm}
			Department of MS\&E, Stanford University}
		\maketitle
		\medskip
	} \fi
	
	\tableofcontents
	
	\newpage

	\section*{Overview of this Internet Appendix}

	The Internet Appendix collects the proofs and additional details for the statements of the main text. It is separated into two parts. 
	\begin{enumerate}
		\item The first part collects the proofs for the multiple testing adjustment for data-driven hypotheses. It contains the proof of Corollary \ref{col3}, Theorem \ref{thm4} and the bounds of the cohesion coefficient.
		\item The second part collects all the proofs for the post-selection inference with Weighted-LASSO. It starts with an illustrative discussion of how adding factor priors changes inference, when the covariates are orthogonal. We then prove the supportive lemmas and Theorems A.1 to A.3 of the main appendix. We conclude with suggestions for best practices in implementing the method.
	\end{enumerate}

	\section{Multiple Testing Adjustment for Data-Driven Hypothesis}
	%
	%
	%

	\subsection{Proof of Corollary \ref{col3}}
	
	\textbf{Proof:} The case of $\forall K:\gamma^*(K)>\gamma$ leads to a unique $K^*(\gamma)$ as we define it to be $J$.
	
	In the case where $\exists K:\gamma^*(K)\leq \gamma$, we show that the mapping $g:\gamma\mapsto K^*(\gamma)$ is a monotonically increasing function. The argmax of this mapping is not empty because $\exists K:\gamma^*(K)\leq \gamma$. Next, we show that it returns only one value, as $g$ is monotonically increasing in $\gamma$. Consider $\gamma\leq \gamma'$, then the $K^*(\gamma')$ satisfies
	\begin{equation}
		K^*(\gamma')=	\argmax_{0\leq K\leq J}\{\gamma^*(K)\leq\gamma'\}
		\geq 	\argmax_{0\leq K\leq J}\{\gamma^*(K)\leq\gamma\}=K^*(\gamma).
	\end{equation}
	This concludes our argument.

	\subsection{Proof of Theorem \ref{thm4}}
	
	\textbf{Proof:} The following proof draws parallels to Theorem 3 of \cite{rssb.12122}, with the key distinction that our approximated R\'enyi representations are conservative approximations rather than exact R\'enyi representations. 
	
	First, we define some notations that further break down the quantities introduced in the procedure of the step-down rejection of nested ordered family $H_{N}$ in the main text:
	\begin{itemize}
		\item ${N}^{\text{order}}_{k},{q}^{\text{order}}_k,{Z}^{\text{order}}_k$ are denoted for short as $\check{N}_k,\check{q}_k,\check{Z}_k$.
		\item The transformed p-values are denoted as $\check{Y}_k^{(n)}=-\ln(p^{(n)}_k)$.
		\item The active unit set for $k$th nested variable $\mathcal{K}_k$ is the collection of units that $k$th variable is active in. The test-related active unit set for $k$th nested variable is $\check{\mathcal{K}}_k=
		\bigcup_{k'\in\{k,k+1,...,J\}} \mathcal{K}_{k'}$, and the simultaneity count for $k$th nested variable is $ \check{N}_k=\sum_{j\in \check{\mathcal{K}}_k}|M_j| $.
		\item The approximated R\'enyi representation equals $\check{Z}_k=\sum_{j=k}^J\sum_{n\in\mathcal{K}_i}
		\frac{ \check{Y}_k^{(n)} }{\check{N}_1-\check{N}_{i+1}\oo\{i\neq J\}}$ and its transformed reversed order statistics is $\check{q}_k=\exp(-\check{Z}_k)$.
	\end{itemize}
	
	We show the FWER control for two cases:
	\subsubsection*{Case I: Global null is true $(s=0)$}\label{thm4p1}
	
	We start with the case when the global null is true, i.e. $\bm{H}_{N,0}$ is true and $s$=0. Then by Assumption 2 in the main text, the random variables $\check{Y}_k^{(n)}$ independently and identically follow the standard exponential distribution.
	
	The nested model implies that $k=1$ should be active for more units than $k=2$, and similar for higher order values. Leveraging this hierarchical structure, we construct a triangle of our $\check{Y}_k^{(n)}$. More specifically, we put the same $k$th value underneath the corresponding value of the other rows, but allow for arbitrary indexing schemes within each row:
	\begin{equation}
		\begin{split}
			\underbrace{\check{Y}_1^{\mathcal{K}_1(1)},\check{Y}_1^{\mathcal{K}_1(2)},...\q...\q...\q,\check{Y}_1^{\mathcal{K}_1(\check{N}_{1}-\check{N}_2)}}_{\mathcal{K}_1(1),...,\mathcal{K}_1(\check{N}_{1}-\check{N}_2)\in\mathcal{K}_1};\\
			\underbrace{\check{Y}_2^{\mathcal{K}_2(1)},\check{Y}_2^{\mathcal{K}_2(2)},...\q...\q,\check{Y}_2^{\mathcal{K}_J(\check{N}_{2}-\check{N}_3)}}_{\mathcal{K}_2(1),...,\mathcal{K}_2(\check{N}_{2}-\check{N}_3)\in\mathcal{K}_2};\\
			&\vdots\\
			\underbrace{\check{Y}_J^{\mathcal{K}_J(1)},\check{Y}_J^{\mathcal{K}_J(2)},...,\check{Y}_J^{(\mathcal{K}_J(\check{N}_J))}}_{\mathcal{K}_J(1),...,\mathcal{K}_J(\check{N}_J)\in\mathcal{K}_J}
		\end{split}
	\end{equation}
	
	Our procedure does not depend on the triangle shape of this construction. This means that we do not limit ourselves to the case where sequences within each row of this triangle are longer on the top than those on the bottom.
	
	If we count from the bottom row to top row and from right to left, then there are $\check{N}_J$ elements when we arrive at the left-most element of the last row $\check{Y}_J^{\mathcal{K}_J(1)}$, and there are $\check{N}_1$ elements when arrive at the left-most element of the top row $\check{Y}_1^{\mathcal{K}_1(1)}$. 
	
	By Assumption 2, under the null $\bm{H}_{N}$ we can invoke \cite{Rnyi1953OnTT} to show that
	\begin{equation}
		\begin{split}
			\bigg(
			\frac{\check{Y}_1^{\mathcal{K}_1(1)}}{1},...,
			\frac{\check{Y}_1^{\mathcal{K}_1(\check{N}_{1}-\check{N}_2)}}{\check{N}_{1}-\check{N}_2}
			,
			\frac{\check{Y}_2^{\mathcal{K}_2(1)}}{\check{N}_{1}-\check{N}_2+1}
			,...,
			\frac{\check{Y}_2^{\mathcal{K}_J(\check{N}_{2}-\check{N}_3)}}{\check{N}_{1}-\check{N}_3}
			,...,
			\frac{\check{Y}_J^{\mathcal{K}_J(1)}}{\check{N}_1-\check{N}_J+1}
			,,...,
			\frac{\check{Y}_J^{(\mathcal{K}_J(\check{N}_J))}}{\check{N}_1}
			\bigg)\\
			\stackrel{d}{=}
			\bigg(
			E_{\check{N}_1,\check{N}_1},...,E_{1,\check{N}_1}
			\bigg).
		\end{split}
	\end{equation}
	where $E_{j,m}$ is the $j$th-order statistic of $m$ standard exponential random variables. As a result, $\tilde{\check{Z}}_k$ defined below is an exact R\'enyi representation:
	\begin{equation}\label{38}
		\tilde{\check{Z}}_k=\sum_{i=k}^J(
		\frac{\check{Y}_i^{\mathcal{K}_i(1)}}{\check{N}_1-\check{N}_i+1}+
		...+
		\frac{\check{Y}_i^{(\mathcal{K}_i(\check{N}_i-\check{N}_{i+1}))}}{\check{N}_1-\check{N}_{i+1}\oo\{i\neq J\}}
		).
	\end{equation}
	Let $\tilde{\check{q}}_k=\exp(-\tilde{\check{Z}}_k)$. Our Assumption 2 in the main text satisfies the independent $p$-value assumption in \cite{2336545}, so the following rejection rule recovers the method of \cite{2336545} that has FWER control at level $\gamma$:
	\begin{equation}\label{renyi_oned}
		\tilde{\hat{k}}=\max\{k:    \tilde{\check{q}}_k\leq\frac{\gamma k}{J}\}.
	\end{equation}
	Our proof is based on the following steps:
	\begin{enumerate}
		\item [Step 1:] $\frac{ k}{J}\geq\frac{\check{N}_k}{JN}$ for each $k$.\NL
		This holds because some of the LASSO coefficients might not be active. By definition of the $\check{N}_k$'s, we have
		\begin{equation}
			\frac{k}{J}=\frac{kJ}{JN}\geq\frac{\sum_{i=k}^J|M_i|}{JN}=\frac{\check{N}_k}{JN} .
		\end{equation}
		It is worth pointing out that if $|M_j|$ is decreasing in $j$, we can further show that $\frac{k}{J}\geq \frac{\check{N}_k}{\check{N}_1} $. This can further increase the power, but we do not want to restrict ourselves to cases where this monotonicity of LASSO active count holds. Our setup enables us to accommodate the possibility that the data does not present a strict nested structure of covariates.
		\item [Step 2:] $\tilde{\check{Z}}_k\geq {\check{Z}}_k$ for each $k$.\NL
		By construction, our ${\check{Z}}_k$ only differs from $\tilde{\check{Z}}_k$ of (\ref{38}) in terms of the denominator in the sum, that is, each denominator in the sum is $\check{N}_1-\check{N}_{i+1}$, which is the largest possible entry of $\{\check{N}_1-\check{N}_i+1,...,\check{N}_1-\check{N}_{i+1}\}$.
		\item[Step 3:] $\tilde{\check{q}}_k$ and ${\check{q}}_k$ are both monotonically increasing in $k$.\NL
		Since $p^{(n)}_k$ is a $p$-value, it has support on $[0,1]$, so $\check{Y}^{(n)}_k=-\ln (p^{(n)}_k)\geq0$, and the monotonicity follows. 
		\item[Step 4:] For sequences $\{a_k\}$ and $\{b_k\}$, it holds that $\max\{k: a_k\leq c\}\geq \max\{k: b_k\leq c\}\geq\max\{k: b_k\leq d\} $, if $a_k\leq b_k$  for each $k$, both sequences are monotonically increasing, and $c\geq d$.\NL
		The first leg holds as $b_k\leq c$ implies $a_k\leq c$, but not vice versa. The second leg holds because $b_k\leq d$ implies $b_k\leq c$, but not vice versa.
	\end{enumerate}
	Step 2 and the fact that $x\mapsto \exp(-x)$ is monotonically decreasing yield that $\tilde{\check{q}}_k\leq {\check{q}}_k$ for each $k$. Then, we use the first leg of Step 4 to obtain the following:
	\begin{equation}\label{40}
		\tilde{\hat{k}}\geq\max\{k:    {\check{q}}_k\leq\frac{\gamma k}{J}\}.
	\end{equation}
	Combining Step 1, Step 3 and second leg of Step 4, we have:
	\begin{equation}\label{41}
		\max\{k:    {\check{q}}_k\leq\frac{\gamma k}{J}\}\geq \max\{k:    {\check{q}}_k\leq\frac{\gamma\check{N}_k}{JN}\}=\hat{k}.
	\end{equation}
	Combining (\ref{renyi_oned}), (\ref{40}), and (\ref{41}), we derive that the number of discoveries $\hat{k}$ in our rejection rule satisfies the following chain of inequalities:
	\begin{equation}
		\underbrace{\tilde{\hat{k}}=\max\{k:    \tilde{\check{q}}_k\leq\frac{\gamma k}{J}\}}_{\text{\cite{2336545}}}\geq\max\{k:    {\check{q}}_k\leq\frac{\gamma k}{J}\}\geq \underbrace{\max\{k:    {\check{q}}_k\leq\frac{\gamma\check{N}_k}{JN}\}=\hat{k}}_{\text{Our rejection}}.
	\end{equation}
	In the case of the global null $(s=0)$, any rejection is a false discovery. Since \cite{2336545}' method has false discovery FWER control of $\gamma$, our procedure has fewer rejections and has FWER control of $\gamma$ as well. This concludes the proof for Case I.

	\subsubsection*{Case II: Global null is false $(s\geq 1)$}\label{thm4p2}
	
	For the case when the global null is not true, there are two possibilities:
	\begin{enumerate}
		\item The first possibility is that our rejection is $\hat{k}\leq s$. Due to the nested nature of $\bm{H}_N$, this guarantees that $V=0$, that is, there is no false discovery, so the FWER control holds.
		\item The other possibility is that we have $\hat{k}=k>s$. Since our procedure is a step-down procedure, $\check{q}_k$ depends on $p^{(n)}_{k'}$ only for the $k'\geq k$ and on the $\check{N}_{k'}$'s, which are included in the LASSO selection event that we are conditioning on. So, there is no difference for the random event $\{\check{q}_k\leq\frac{\gamma\check{N}_k}{JN}\}$ between the $s\neq 0$ and $s=0$ ground truth. In other words, we can write:
		\begin{equation}\label{42}
			\PP_{H_D}(V\geq 1|s>0,\hat{k}>s)=\PP_{H_D}(V\geq 1|s=0,\hat{k}>s).
		\end{equation}
		The right-hand side of (\ref{42}) is smaller than $\gamma$ as shown in Case I.
	\end{enumerate}
	Combining Cases I and II, we conclude the proof.

	\subsection{Proof of Bounds for Cohesion Coefficient $\rho$}\label{app-sec:proof_bonf_global}
	
	Recall the definition of the cohesion coefficient $\rho =\left(\sum_{1\leq j\leq J:\mathcal{K}_j\neq\emptyset}\frac{|\mathcal{K}_j|}{N_j}   \right)^{-1}$. We want to show that $J^{-1}\leq \rho\leq 1$.
	
	\textbf{Lower bound: }The inverse of the summands in the definition of $\rho$ can be written as
	\begin{equation}\label{75.0}
		\frac{N_j}{|\mathcal{K}_j|}=\frac{|\mathcal{K}_j|+\sum_{j'\textrm{ co-active with }j}|\mathcal{K}_{j'}|}{|\mathcal{K}_j|}
		=1+\frac{\sum_{j'\textrm{ co-active with }j}|\mathcal{K}_{j'}|}{|\mathcal{K}_j|}.
	\end{equation}
	If there are no co-active covariates $j'$ with covariate $j$, then we have $\frac{\sum_{j'\textrm{ co-active with }j}|\mathcal{K}_{j'}|}{|\mathcal{K}_j|}=0$, which implies (\ref{75.0})$\geq 1$. Thus, the sum is bounded by
	\begin{equation}\label{eq:11}
		\begin{split}
			\sum_{1\leq j\leq J:\mathcal{K}_j\neq\emptyset}\frac{|\mathcal{K}_j|}{N_j}\leq 
			\sum_{1\leq j\leq J}1    \iff
			\rho^{-1}\leq 
			J \iff
			\rho\geq  J^{-1}.
		\end{split}
	\end{equation}
	
	\textbf{Upper bound: }For fixed $H_M$, we now show by induction that $\rho\leq 1$ or $\rho^{-1}\geq 1$. 
	First, we re-arrange the indices $j\in H_M$ based on $|\mathcal{K}_j|$ such that
	$$
	|\mathcal{K}_1|\geq |\mathcal{K}_2|\geq \cdots |\mathcal{K}_{|H_M|}| .
	$$
	Let $H_t$ denote the family of hypotheses that is constrained to only include $\{ j\in H_M: j\leq t\}$. We write $\rho^{(t)}$ as 
	\begin{equation}
		(\rho^{(t)})^{-1}:=\sum_{j\leq t}\frac{|\mathcal{K}_j|}{N_j^{(t)}}.
	\end{equation}
	
	We want to show that $(\rho^{|H_M|})^{-1}\geq 1$. It is sufficient to use induction for $t=1,2,...,|H_M|$. In other words, we show that $\rho^{-1}\geq 1$ by first looking at the family of hypotheses truncated up to the covariate with most active units, then the second most active units, etc.\\
	\begin{enumerate}
		\item[(i)] \textbf{Base case:} When $t=1$, clearly $|\mathcal{K}_1|=N_1$ so $\sum_{j\leq 1}\frac{|\mathcal{K}_j|}{N_j}= 1$.
		\item[(ii)] \textbf{Inductive case:} Suppose the claim is true for $t>1$. This implies
		\begin{equation}\label{78.1}
			(\rho^{(t)})^{-1}=\sum_{j\leq t}\frac{|\mathcal{K}_j|}{N_j^{(t)}}\geq 1.
		\end{equation}
		We study the next induction step:
		\begin{equation}\label{eqn:indiction}
			\begin{split}
				(\rho^{(t+1)})^{-1}
				&=\sum_{j\leq t+1}\frac{|\mathcal{K}_j|}{N_j^{(t+1)}}\\
				&=\sum_{j\leq t}\frac{|\mathcal{K}_j|}{N_j^{(t+1)}}+
				\frac{|\mathcal{K}_{t+1}|}{N_{t+1}^{(t+1)}}.
			\end{split}    
		\end{equation}
		Note that $1-\sum_{j\leq t}\frac{|\mathcal{K}_j|}{N_j^{(t)}}\leq 0$ by equation (\ref{78.1}), and thus it holds that
		\begin{equation}
			\begin{split}
				(\rho^{(t+1)})^{-1}
				&\geq 1-\sum_{j\leq t}\frac{|\mathcal{K}_j|}{N_j^{(t)}}+\sum_{j\leq t}\frac{|\mathcal{K}_j|}{N_j^{(t+1)}}+
				\frac{|\mathcal{K}_{t+1}|}{N_{t+1}^{(t+1)}}\\
				&=1+\sum_{j\leq t}|\mathcal{K}_j|(\frac{1}{N_j^{(t+1)}}-\frac{1}{N_j^{(t)}})+
				\frac{|\mathcal{K}_{t+1}|}{N_{t+1}^{(t+1)}}\\
				&=1+\sum_{j\leq t}|\mathcal{K}_j|\frac{-(N_j^{(t+1)}-N_j^{(t)})}{N_j^{(t+1)}N_j^{(t)}}+
				\frac{|\mathcal{K}_{t+1}|}{N_{t+1}^{(t+1)}}.
			\end{split}    
		\end{equation}
		We define $x_j:=N_j^{(t+1)}-N_j^{(t)}$ for $j$th covariate, which represents how many new co-active counts are added with $(t+1)$-th covariates. This is $|\mathcal{K}_{t+1}|+1$ if $j$ is co-active with $(t+1)$-th covariate, and 0 otherwise. Next, we define
		\begin{equation}
			f_j(x_j):=\frac{-(N_j^{(t+1)}-N_j^{(t)})}{N_j^{(t+1)}N_j^{(t)}}
			=\frac{-x_j}{(N_j^{(t)}+x_j)N_{j}^{t}}.
		\end{equation}
		It is straightforward to show that $\forall f_j$ we have $\frac{\partial f_j}{\partial x_j}\leq 0$, i.e. the sum $\sum_{j\leq t}|\mathcal{K}_j|f_j(x_j)$ is monotonically decreasing in $x_j$.
		
		The second part of equation \ref{eqn:indiction} can be written as
		
		\begin{equation}
			g(x)=\frac{|\mathcal{K}_{t+1}|}{N_{t+1}^{(t+1)}}
			=\frac{|\mathcal{K}_{t+1}|}{|\mathcal{K}_{t+1}|+\sum_{j\textrm{ co-active with }(t+1)}|\mathcal{K}_{j}|}.
		\end{equation}
		For non-zero values of $x$, the denominator of $g(x)$ is increasing, and hence $g(x)$ is decreasing in $x$.
		Combing these two arguments yields that $(\rho^{(t+1)})^{-1}$ is decreasing in $x$, and thus the inequality has a valid lower bound if we set $x=0$, which gives us $    (\rho^{(t+1)})^{-1}\geq 1$.
	\end{enumerate}
	
	Combining (i) and (ii), we conclude that $\rho^{-1}\geq 1$, and together with equation (\ref{eq:11}) we obtain the bounds
	\begin{equation}
		\rho\in[J^{-1},1].
	\end{equation}

	\section{Post-selection Inference with Weighted-LASSO}

	\subsection{Adding Priors for Post-Selections Inference}\label{sec2}
	
	We discuss how to make valid inference for $\hat{\beta}^{(n)}(\lambda,\omega) \in \mathbbm R^J$ given $\mathcal{M}^{(n)}$ when $n$ is fixed, which allows us to deal with high-dimensional covariate dimensions (large $J$). The multiple testing adjustment extends the analysis to a large number of cross-sectional units $N$. 
	As we only consider the conditional inference for a single cross-sectional unit, we drop the superscript $n$, that indexes units, and refer to the response variable as $Y\in\RR^T$, the active set as $M$, the LASSO selection event as $\mathcal{M}$, and the active sub-matrix of $X\in\RR^{T\times d}$ as $X_M$. The weights $\omega$ are fixed. \footnote{The PoSI literature, for examples \cite{markovic2018unifying}, has also studied the case where $\lambda$ is determined via cross-validation (CV). The resulting theory depends on the setup for CV and is asymptotic in the pre-selection data-generating process and a user-augmented randomizer. Its implementation requires an MCMC approach. The goal is different from our paper, as we provide easy-to-understand and generic post-selection inference $p$-values that focus on asymptotics for unknown variances.} We use $\hat{\beta}\in\RR^{|M|}$ to denote the LASSO fit in this section. Thus, the hypothesis can be written as
	\begin{equation}\label{10}
		H_{0,j}:\beta_j=0|\mathcal{M}.
	\end{equation}
	If we can calculate the valid $p$-value for a consistent statistic of $\beta_i$ conditioned on $\mathcal{M}$, we are able to reject $H_{0,i}$ of (\ref{10}) with the desired Type I error, which completes inference problem for a single unit.


	\subsubsection*{Illustrative case: orthogonal design}\label{sec:2.1}
	
	To provide some intuition, we consider the following simplified case, where an orthogonal collection of features and i.i.d. errors with known variances are available. As expected for orthogonal covariates, this leads to a simple closed-form solution.
	
	\begin{Assumption}[Low dimensional truth]\label{asu1}
		The data satisfies $Y=X_S\beta_S+\epsilon$ where $|S|=O(1)$ is much smaller than $J$ or $T$.
	\end{Assumption}
	\begin{Assumption}[Orthogonal design]\label{asu2}
		The features satisfy $\frac{X^\top X}{T}=diag(\kappa_1^2,...,\kappa_J^2)$.
	\end{Assumption}
	\begin{Assumption}[Gaussian residual with simple known variance]\label{asu3}
		$\epsilon_t\iid \mathcal{N}(0,\sigma^2)$, where $\epsilon_t$ is $t$-th element of $\epsilon$.
	\end{Assumption}
	
	We use the term \textbf{Orthogonal Design} (OD henceforth) to refer to the case where IA.Assumptions \ref{asu1}, \ref{asu2} and \ref{asu3} are met. We can write out the weighted LASSO optimization as
	\begin{equation}
		\begin{split}
			\hat{\beta}=\argmin_\beta		\ell(Y,X,\lambda,\beta,\omega)
			&=\frac{1}{2T}Y^\top Y+\frac{1}{2T}\beta^\top X^\top X\beta-\frac{1}{T}Y^\top X\beta+\lambda\sum_{i=1}^df_j(\beta_j,\omega_j)\\
			&=\frac{1}{2T}Y^\top Y+
			\underbrace{\sum_{i=1}^{d}(\frac{1}{2}\kappa_j^2\beta_j^2-\frac{1}{T}Y^\top X_j\beta_j+\lambda\sum_{i=1}^d f_j(\beta_j,\omega_j)}_{\bar{\ell}(Y,X,\lambda,\beta|\omega)} .
			\\
		\end{split}
	\end{equation}
	Note that only $\bar{\ell}(Y,X,\lambda,\beta|\omega)$ matters within $\ell(Y,X,\lambda,\beta|\omega)$, and $\bar{\ell}$ decouples across $i$'s,
	so that we can take sub-differentials of $\bar{\ell}$ with respect to $\beta$:
	\begin{equation}\label{4.0}
		\pa (\beta\mapsto \bar{\ell}(\beta))=\begin{cases}
			\kappa_j^2\beta_j-\frac{1}{T}Y^\top X_j& \omega_j=\infty\\
			\kappa_j^2\beta_j-\frac{1}{T}Y^\top X_j+\lambda\sgn(\beta_j)/\omega_j & \beta_j\neq 0,\omega_j<\infty\\
			\kappa_j^2\beta_j-\frac{1}{T}Y^\top X_j+\lambda v_j/\omega_j& \beta_j= 0,\omega_j<\infty
		\end{cases}.
	\end{equation}
	where the slacks satisfy $v_j\in[-1,1]$. The solution to this minimization problem equals
	\begin{equation}\label{5}
		\hat{\beta}_j=\begin{cases}
			\frac{1}{\kappa_j^2}(\frac{1}{T}Y^\top X_j-\frac{c}{\omega_j})&\text{if } \frac{1}{T}Y^\top X_j-\frac{\lambda}{\omega_j}>0\\
			\frac{1}{\kappa_j^2}(\frac{1}{T}Y^\top X_j+\frac{c}{\omega_j})&\text{if } \frac{1}{T}Y^\top X_j+\frac{\lambda}{\omega_j}<0\\
			0&o.w.,
		\end{cases}
	\end{equation}
	where the case of $\omega_j=\infty$ is included in the first or the second case of (\ref{5}). Since $\sgn(\frac{1}{T}Y^\top X_j-\frac{\lambda}{\omega_j})=\sgn(\hat{\beta}_j)$, the sign of $\hat{\beta}_j$ tells us in which of the three cases in (\ref{5}) we are, given the exogenous values of  $X$ and $\omega$. In other words, the $\sgn(\hat{\beta})$ values and set of indices of active variable $M$ identify the LASSO selection event, because we can recover $\hat{\beta}$ once we know them. Additional knowledge of $v$ is not needed for identifying $\hat{\beta}$. Hence, we denote by $\mathcal{M}=(M,s)$ the LASSO event that we condition on. Moreover, by observing (\ref{5}), it is intuitive to consider an ``adjusted'' estimator defined as
	\begin{equation}\label{14}
		\forall i\in[d]:\q\bar{\beta}_j=\hat{\beta}_j+\sgn(\hat{\beta_j})\cdot \frac{\lambda}{\kappa_j^2\omega_j}.
	\end{equation}
	We refer to $\bar{\beta}$ as ``one-step estimator'' and explain the motivation for this name below. Note that $\hat{\beta}_j$ is active\footnote{We refer to a LASSO estimated coefficient as active if it is not 0.} if and only if $\bar{\beta}_j$ is active, and for active $\bar{\beta}_j$, it contains an additive component of $\frac{1}{T}\cdot\frac{\epsilon_j^\top X_j}{\kappa_j^2}$ as shown in the proof of IA.Theorem \ref{thm0}. The distribution of $\bar{\beta}_j$ is truncated Gaussian. On the one hand it contains a linear combination of Gaussian random variables, which lead to Gaussian distribution. On the other hand, this Gaussian distribution has to be truncated based on the condition in (\ref{5}). Following this thread, we obtain the marginal distribution for $\bar{\beta}_j$:
	\begin{theorem}[Truncated Gaussian of OD]\label{thm0}
		With Assumptions  \ref{asu1}, \ref{asu2}, \ref{asu3} and conditional on $\mathcal{M}$, we have the following distribution associated with $\bar{\beta}_j$:
		\begin{equation}
			\begin{cases}
				\bar{\beta}_j\sim\mathcal{TN}(\beta_j,\frac{1}{T}\frac{\sigma^2}{\kappa_j^2};[V^-_j,V^+_j]) & j\in M\\
				\bar{\beta}_j=0 & otherwise
			\end{cases}.
		\end{equation}	
		Where the truncation intervals are 
		\begin{equation}
			[V^-_j,V^+_j]=\begin{cases}
				(-\infty,-\frac{\lambda}{\kappa_j^2\omega_j}]& \text{ if }\sgn(\hat{\beta_j})=-1\\
				[\frac{\lambda}{\kappa_j^2\omega_j},+\infty)& \text{ if }\sgn(\hat{\beta_j})=1\\
			\end{cases}.
		\end{equation}	
	\end{theorem}
	Note that the $\sgn(\hat{\beta_j})$ in the truncation are known quantities, because we condition on the LASSO output. This again reflects the perspective of our conditional inference framework: We do not presume that a hypothesis is written without seeing the data. Only after we have seen the LASSO output we can form our hypothesis, and the hypothesis will be about covariates that are $j\in M$. This makes the so the statement of IA.Theorem \ref{thm0} conditional on $\mathcal{M}$. IA.Theorem \ref{thm0} shows that by conditioning on the KKT sub-gradient equations we induce a set of $[V_j^-,V_j^+]$ truncations on the support of the parameters that would otherwise have a Gaussian distribution over entire Euclidean space.\NL
	
	\textit{Remark: }The results of IA.Theorem \ref{thm0} also highlights many known properties of the LASSO estimator. 
	\begin{itemize}
		\item First, we observe that $\bar{\beta}_j$ is a shifted version of the LASSO $\hat{\beta}_j$, which is commonly referred to as ``de-biasing'' of LASSO as suggested by \cite{10.1214/17-AOS1630}. With this shifted one-step estimator $\bar{\beta}_j$, Online Appendix Theorem \ref{thm0} also implies that $\bar{\beta}_j\stackrel{p}{\to}\beta_j$ as $T\to\infty$, and hence establishes consistency.
		
		\item Second, we note that the proper LASSO penalty scalar $\lambda$ needs to scale proportional to $\kappa_j^2$ to maintain the same fit. If we have a scaled version of $X$ as $\tilde{X}=bX$ and $\omega$ fixed, it is necessary to use $\tilde{\lambda}= b^2 \lambda $ to maintain the same LASSO estimate.

		\item Lastly, we see that the LASSO estimator can miss weak signals when $\lambda$ is mis-specified. For instance, when the true coefficient equals $\beta_j=0.5\lambda/\kappa_j^2$, the truncation would not admit this covariate and its LASSO estimate would be $\hat{\beta}_j=0$. Thus, we recommend to use a moderately sized $\lambda$ for the purpose of selecting a parsimonious model of potentially weak covariates that the explain time series, even if the larger value of $\lambda$ might directly yield a sparse model. In other words, the LASSO time series regression step should serve as a pre-screening tool that conducts a first dimensional reduction, and the inferential framework based on $p$-values that we develop in this paper provides the tool to fine-tune the model.
	\end{itemize}

	\subsection{Proof of Lemma A.1}
	
	This result follows from the definition and some simple algebraic manipulations. By the definition of $\bar{\beta}_M$ we can write: 
	\begin{equation}\label{14}
		\begin{split}
			\bar{\beta}_M
			&=\hat{\beta}_M+X_M^+\hat{\epsilon}_M\\
			&=\hat{\beta}_M+(X_M^\top X_M)^{-1}X_M^\top (Y-X_M\hat{\beta}_M)\\
			&=\hat{\beta}_M+(X_M^\top X_M)^{-1}X_M^\top Y-(X_M^\top X_M)^{-1}X_M^\top X_M\hat{\beta}_M\\
			&=(X_M^\top X_M)^{-1}X_M^\top Y\\
			&=X_M^+Y.
		\end{split}
	\end{equation}
	Recall that $X_M^+Y=\argmin_\beta\frac{1}{2T}\|Y-X_M\beta\|_2^2$ equals the OLS estimator. Hence, equation (\ref{14}) concludes the second half of Lemma A.1.\NL
	The first half of Lemma A.1 simply uses $e_j$ to map to $j$th coordinate. Given $\eta=(X_M^+)^\top e_j$ we have
	\begin{equation}
		\bar{\beta}_{j}=e_j^\top \bar{\beta}_M=e_j^\top X_M^+Y=((X_M^+)^\top e_j)^\top Y=\eta^\top Y.
	\end{equation}
	Thus, we have shown Lemma A.1.
	
	\subsection{Proof of Lemma A.2}
	We start by calculating the sub-differential of the regularization function evaluated at $\hat{\beta}$:
	\begin{equation}
		\partial(\beta\mapsto f(\beta,\omega))|_{\beta=\hat{\beta}}
		=\hat{r}\odot\omega^{-1}.
	\end{equation}
	where $\hat{r}$ is the sub-differential $\pa(\beta\mapsto\|\beta\|_1)$ evaluated at $\hat{\beta}$, $\odot$ is the element-wise multiplication of two vectors, $\omega^{-1}$ is the element-wise reciprocal of vector $\omega$, i.e. $\omega^{-1}=[\omega_1^{-1},...,\omega_J^{-1}]$ and $1/\infty=0$. Thus, we have the LASSO sub-gradient optimal condition:
	\begin{equation}
		X^\top(X\hat{\beta}-Y)+\lambda \hat{r}\odot\omega^{-1}=0.
	\end{equation}
	Without loss of generality we can assume that the active covariates are the first $|M|$ of the $d$ factors. Then, we can expand the KKT conditions based on $X=[X_M,X_{-M}]$, $\hat{\beta}=[\hat{\beta}_M;\bm{0}]$, $\hat{r}=[\hat{r}_M;\hat{r}_{-M}]$, and $\omega=[\omega_M;\omega_{-M}]$:
	\begin{equation}
		\begin{split}
			X_M^\top(X_M\hat{\beta}_M-Y)+\lambda\hat{r}_{M}\odot\omega_M^{-1}=0\\
			X_{-M}^\top(X_M\hat{\beta}_M-Y)+\lambda\hat{r}_{-M}\odot\omega_{-M}^{-1}=0\\
			\sgn(\hat{\beta}_M)=\hat{r}_M\\
			\|\hat{r}_{-M}\|_\infty<1
		\end{split}
	\end{equation}
	The last two lines are based on the sub-differentials of $\pa(\beta\mapsto\|\beta\|_1)$ evaluated at $\hat{\beta}_M$ and $\hat{\beta}_{-M}$, respectively. Recall that $s=\sgn(\hat{\beta}_M)\in\RR^{|M|}$ is the vector of signs. Since the KKT conditions are sufficient and necessary for a solution, we obtain that the KKT conditions are equivalent to the set of vectors $w\in\RR^{|M|}$ and $u\in\RR^{J-|M|}$ that satisfy 
	\begin{equation}\label{18}
		\begin{split}
			X_M^\top(X_Mw-Y)+\lambda s\odot\omega_M^{-1}=0\\
			X_{-M}^\top(X_Mw-Y)+\lambda u\odot\omega_{-M}^{-1}=0\\
			\sgn(w)=s\\
			\|u\|_\infty<1
		\end{split}
	\end{equation}
	Using only the first line of (\ref{18}), we solve for
	\begin{equation}
		w=(X_M^\top X_M)^{-1}(X_M^\top Y-\lambda s\odot\omega^{-1}_M).
	\end{equation}
	Recall $\mathcal{J}$ is the set of $j$'s corresponding to $\omega_j$'s that are infinite. For $\mathcal{J}\neq \emptyset$, the notation above is still valid and the segment corresponding to ${\mathcal{J}}$ simply becomes the usual OLS coefficients:
	\begin{equation}
		w_{\mathcal{J}}=(X_{\mathcal{J}}^\top X_{\mathcal{J}})^{-1}X_{\mathcal{J}}^\top Y.
	\end{equation}
	This implies that
	\begin{equation}
		\begin{split}
			X_Mw-Y
			=&X_M(X_M^\top X_M)^{-1}(X_M^\top Y-\lambda s\odot\omega^{-1}_M)-Y\\
			=&X_MX_M^+ Y-\lambda (X_M^+)^\top s\odot\omega^{-1}_M-Y\\
			=& -(I-P_M)Y-\lambda (X_M^+)^\top s\odot\omega^{-1}_M.
		\end{split}
	\end{equation}
	Plugging this back into the second line of (\ref{18}), we solve for
	\begin{equation}\label{20}
		u=\omega_{-M}\odot
		\bigg(
		X_{-M}^\top(X_M^+)^\top s\odot\omega^{-1}_M
		+\frac{1}{\lambda }X_{-M}^\top (I-P_M)Y\bigg).
	\end{equation}
	Note that (\ref{20}) does not lead to an ambiguity issue for $\infty\cdot (\infty)^{-1}$ in the case of infinity prior weights, because infinitely weighed covariates are guaranteed to be active, i.e. $\mathcal{J}\subseteq M$. To see this explicitly, we can without loss of generality assume that $\mathcal{J}$ is located at the top of $M$. Then the notation in (\ref{20}) is still valid and we can write the part with infinity prior weights as:
	\begin{equation}
		u=\omega_{-M}\odot
		\begin{bmatrix}
			\frac{1}{\lambda }X_{-M}^\top (I-P_{\mathcal{J}})Y\\
			X_{-M}^\top(X_{M-\mathcal{J}}^+)^\top s_{M-\mathcal{J}}\odot\omega^{-1}_{M-\mathcal{J}}
			+\frac{1}{\lambda }X_{-M}^\top (I-P_{M-\mathcal{J}})Y
		\end{bmatrix}.
	\end{equation}
	The remaining conditions in (\ref{18}) are the third and fourth lines, which are exactly $\sgn(w)=s$ and $\|u\|_\infty<1 $, respectively. This concludes the proof of Lemma A.2.

	\subsection{Proof of Lemma A.3}
	
	The quantity of interest is given by $\eta=(X_M^+)^\top e_j$, $\bar{\beta}_{j}=\eta^\top Y$ per our Lemma A.1. By Assumption \ref{asu_known}, $\Sigma$ is known, so indeed both $
	\xi=\Sigma\eta(\eta^\top \Sigma\eta)^{-1}$ and $z=(I-\xi\eta^\top )Y$ can be calculated once we observe $(X,Y)$.\NL
	We now show that $z$ is uncorrelated with $\eta^\top Y$. Suppose $I-\xi\eta^\top =\Gamma$, then we have:
	\begin{equation}\label{27}
		\begin{split}
			cov(z,\eta^\top Y)
			=&cov(\Gamma Y,\eta^\top Y)\\
			=&\Gamma cov(Y)\eta\\
			=&\Gamma\Sigma\eta\q\text{(by Assumption \ref{asu_known})}\\
			=&(I-\xi\eta^\top)\Sigma\eta\\
			=&(I-
			\Sigma\eta(\eta^\top \Sigma\eta)^{-1}\eta^\top)\Sigma\eta\q\text{(by defn. of $\xi$)}\\
			=&\Sigma\eta-\Sigma\eta\\
			=&0.
		\end{split}
	\end{equation}
	Since $z$ and $\eta^\top Y$ are both linear mappings of $Y$, they both have a Gaussian distribution by Assumptions \ref{asu_known}. Using (\ref{27}), we conclude that they are independent.

	\subsection{Proof of Theorem A.1}
	
	The proof below follows similar arguments as in \cite{tian2018selective}. Given our Lemma A.2, we can first rewrite the active constraints $\sgn(w(M,s,\omega))=s$ as the following linear system of inequalities:
	\begin{equation}\label{22}
		\begin{split}
			\{\sgn(w)=s\}
			&=\{diag(s)w>0 \}\\
			&=\{diag(s)(X_M^\top X_M)^{-1}(X_M^\top Y-\lambda s\odot\omega^{-1}_M)>0 \}\\
			&=\{diag(s)X_M^+Y>\lambda \cdot diag(s)(X_M^\top X_M)^{-1}s\odot\omega^{-1}_M\}\\
			&=\{A_1(M,s,\omega)Y<b_1(M,s,\omega)\},
		\end{split}
	\end{equation}
	where
	\begin{equation}
		A_1(M,s,\omega)=-diag(s)X_M^+,\q b_1(M,s,\omega)=-\lambda \cdot diag(s)(X_M^\top X_M)^{-1}s\odot\omega^{-1}_M.
	\end{equation}
	We use the results of Lemma A.2 and the $u$ defined therein. The inactive constraints $\|u\|_\infty<1 $ can also be reformulated into a linear system of inequalities. Since $u$ is of dimension $(J-|M|)$, we expand it with respect to the 1-vector $\bm{1}_{J-|M|}$ of the same dimension. This yields
	\begin{equation}\label{23}
		\begin{split}
			&\{\|u\|_\infty<1 \}\\
			=&\{-\bm{1}_{J-|M|}<\omega_{-M}\odot
			\bigg(
			X_{-M}^\top(X_M^+)^\top s\odot\omega^{-1}_M
			+\frac{1}{\lambda}X_{-M}^\top (I-P_M)Y\bigg)\}\\
			&\cap\{\omega_{-M}\odot
			\bigg(
			X_{-M}^\top(X_M^+)^\top s\odot\omega^{-1}_M
			+\frac{1}{\lambda}X_{-M}^\top (I-P_M)Y\bigg)<\bm{1}_{J-|M|}\}\q\text{(by definition of $\|\cdot\|_\infty$)}\\
			=&\{-\omega_{-M}^{-1}<
			X_{-M}^\top(X_M^+)^\top s\odot\omega^{-1}_M
			+\frac{1}{\lambda}X_{-M}^\top (I-P_M)Y<\omega_{-M}^{-1}\}\q\text{(by design $\omega>0$)}\\
			=&\{-\omega_{-M}^{-1}-X_{-M}^\top(X_M^+)^\top s\odot\omega^{-1}_M<
			\frac{1}{\lambda}X_{-M}^\top (I-P_M)Y<\omega_{-M}^{-1}-X_{-M}^\top(X_M^+)^\top s\odot\omega^{-1}_M\}\\
			=&\{A_2(M,s,\omega)Y<b_2(M,s,\omega)\},
		\end{split}
	\end{equation}
	where
	\begin{equation}\label{29}
		A_2(M,s,\omega)=\begin{bmatrix}
			\frac{1}{\lambda}X_{-M}^\top (I-P_M)\\
			-\frac{1}{\lambda}X_{-M}^\top (I-P_M)
		\end{bmatrix},\q
		b_2(M,s,\omega)=\begin{bmatrix}
			\omega_{-M}^{-1}-X_{-M}^\top(X_M^+)^\top s\odot\omega^{-1}_M\\
			\omega_{-M}^{-1}+X_{-M}^\top(X_M^+)^\top s\odot\omega^{-1}_M
		\end{bmatrix}.
	\end{equation}
	Note that (\ref{29}) is valid when $\mathcal{J}\neq\emptyset$, because $\mathcal{J}\subseteq M$. Now combining (\ref{22}) and (\ref{23}), we have written the KKT conditions $\{\sgn(w)=s,\|u\|_\infty<1\}$ into the form $\{AY\leq b\}$, where $A=[A_1;A_2]$ and $b=[b_1;b_2]$. 
	
	Given the Gaussian Assumption \ref{asu1}, we can directly invoke Lemma 5.1 of \cite{lee2016exact} to construct
	\begin{equation}
		\{AY\leq b\}=\{V^-(z)\leq \eta^\top Y\leq V^+(z),V^0(z)\geq 0 \},
	\end{equation}
	where 
	\begin{equation}
		\begin{split}
			V^{-}(z)=\max_{j:(A\xi)_j<0}\frac{b_j-(Az)_j}{(A\xi)_j}\\
			V^{+}(z)=\min_{j:(A\xi)_j>0}\frac{b_j-(Az)_j}{(A\xi)_j}\\
			V^0(z)=\min_{j:(A\xi)_j=0}b_j-(Az)_j.
		\end{split}
	\end{equation}
	Moreover, $(V^{-}(z),V^{+}(z),V^{0}(z))$ are functions of $z$, and we have argued that $z\indep \eta^\top Y$ in Lemma A.3. To make the final inferential statements about $\eta^\top Y$, we simply drop the $V^0$ conditions given that we have conditioned on $\tilde{\mathcal{M}}$ which contains $z$.\NL
	Our quantity of interest $\bar{\beta}_{j}=\eta^\top Y$ is distributed as
	\begin{equation}
		\begin{split}
			\bar{\beta}_{j}|\tilde{\mathcal{M}}&\stackrel{\mathcal{D}}{=}[\bar{\beta}_{M_{(l)}}|(\{AY\leq b\},z)]\\
			&\stackrel{\mathcal{D}}{=}[\eta^\top Y|(\{V^-(z)\leq \eta^\top Y\leq V^+(z),V^0(z)\geq 0 \},z)]\\
			&\stackrel{\mathcal{D}}{=}[\eta^\top Y|(\{V^-(z)\leq \eta^\top Y\leq V^+(z)\},z)]\\
			&\sim\mathcal{TN}(\beta_{j},\eta^\top \Sigma\eta;[V^{-}(z),V^{+}(z)] )
		\end{split}
	\end{equation}
	This concludes the theorem.

	\subsection{Proof of Theorem A.2}\label{pf:app-thm1}
	
The studentized quantity requires different arguments from Theorem A.1, because its denominator includes $\hat{\sigma}(Y)$, which depends on $Y$. Our strategy is to first show the conversion from the Weighted-LASSO with penalty $\lambda$ to the Square-Root LASSO with penalty $\tilde{\lambda}$. Then, we show that the truncation is of the form $C Y\leq\hat{\sigma}(Y)$. Lastly, we show the result of a truncated $t$-distribution by following similar arguments as for the Square-Root LASSO in Theorem 1 of \cite{tian2017selective} and by solving a set of non-linear inequalities. 
	
	First, we note equation (9) and Lemma 2 of \cite{tian2017selective} show that the conversion from LASSO to Square-Root LASSO depends on the slack variables from first-order conditions. We calculate the first-order conditions for the active variables from Square-Root LASSO, written in the same form of equation (9) of \cite{tian2017selective}:
	\begin{equation}\label{divi_form}
		\frac{X_M^\top (Y-X_M\hat{\beta}_M)}{\|Y-X_M\hat{\beta}_M\|_2}= \tilde{\lambda}\cdot s\odot\omega^{-1}_M,\q\wh\q s=\sgn(\hat{\beta}_M).
	\end{equation}

By Assumption A.1(b), the pseudo-inverse of $X_M$ : $X_M^+=(X_M^\top X_M)^{-1}X_M^\top$ is well-defined. Equation (\ref{divi_form}) allows us to invoke Lemma 2 of \cite{tian2017selective} to establish the mapping from Square-Root penalty $\tilde{\lambda}$ to LASSO penalty $\lambda$:
	\begin{equation}\label{frac_equation}
		\lambda = \tilde{\lambda}\hat{\sigma}(Y)\sqrt{\frac{T-|M|}{1-\tilde{\lambda}^2\left((X_M^+)^\top s\odot\omega^{-1}_M\right)^2}}.
	\end{equation}

	We solve for $\tilde{\lambda}$ from the fractional equation of \ref{frac_equation}, and complete the conversion to $\tilde{\lambda}$ from $\lambda$:
	\begin{equation}
		\tilde{\lambda}^2=\frac{\lambda^2}{\hat{\sigma}^2(Y)\cdot (T-|M|)+\|(X_M^+)^\top s\odot\omega^{-1}_M\|_2^2\lambda^2}.
	\end{equation}

	Next, we consider the distribution of the active coefficients $\hat{\beta}_M$. Lemma 2 of \cite{tian2017selective} states that the distribution is a truncated $t$-distribtion, but the truncation still needs to be calculated. The key is to characterize the condition $\sgn(\hat{\beta}_M)=s$ from (\ref{divi_form}), and we need to isolate $\hat{\beta}_M$. 
	
	First we use (\ref{divi_form}) to calculate the residuals from the projection with $\hat{\beta}_M$ as
	\begin{equation}
\|Y-X_M\hat{\beta}_M\|_2^2=\frac{\|(I-P_M)Y\|_2^2}{1-\tilde{\lambda}^2\left((X_M^+)^\top s\odot\omega^{-1}_M\right)^2}.
	\end{equation}

Note that $\|(I-P_M)Y\|_2^2=(T-|M|)\hat{\sigma}^2(Y)$, and substituting $\hat{\sigma}^2(Y)$ in (\ref{divi_form}) yields
	\begin{equation}
		\hat{\beta}_M=(X_M^\top X_M)^{-1}\left(X_M^\top Y-
		\hat{\sigma}(Y)\sqrt{\frac{T-|M|}{1-\tilde{\lambda}^2\left((X_M^+)^\top s\odot\omega^{-1}_M\right)^2}}		
		 \tilde{\lambda}\cdot s\odot\omega^{-1}_M\right).
	\end{equation}
	Using the same argument that we have applied before, we rewrite the equality condition $\sgn(\hat{\beta}_M)=s$ into an inequality equivalency that states $e_j^\top\hat{\beta}_Ms_j\geq 0$ for $j\in M$ and for one-hot vector $e_j$:
	\begin{equation}\label{36.sigY}
		\begin{split}
			&e_j^\top (X_M^\top X_M)^{-1}\left(X_M^\top Y-
			\hat{\sigma}(Y)\sqrt{\frac{T-|M|}{1-\tilde{\lambda}^2\left((X_M^+)^\top s\odot\omega^{-1}_M\right)^2}}		
			\tilde{\lambda}\cdot s\odot\omega^{-1}_M\right)\cdot  s_j\geq 0\\
			\iff& -e_j^\top X_M^+ Y\cdot s_j\leq- (\tilde{\lambda} s_j )\cdot  e_j^\top \left((X_M^\top X_M)^{-1}s\odot \omega^{-1}\right)\cdot\sqrt{\frac{T-|M|}{1-\tilde{\lambda}^2\left((X_M^+)^\top s\odot\omega^{-1}_M\right)^2}}\cdot	\hat{\sigma}(Y).
		\end{split}
	\end{equation}
	
	The second line in (\ref{36.sigY}) is an inequality with $Y$ on left-hand side and $\hat{\sigma}(Y)$ on right-hand side. We simplify it by referring to the intermediary quantities:
	\begin{equation}\label{eq_truncb_defn}
		\begin{split}
			b_j=- (\tilde{\lambda} s_j )\cdot  e_j^\top \left((X_M^\top X_M)^{-1}s\odot \omega^{-1}\right)\cdot\sqrt{\frac{T-|M|}{1-\tilde{\lambda}^2\left((X_M^+)^\top s\odot\omega^{-1}_M\right)^2}}.
		\end{split}
	\end{equation}

Alternatively, we can also write it in terms of $\lambda$:
	\begin{equation}\label{trunc_b_normal_lambda}
	\begin{split}
		b_j=- (\lambda s_j/\hat{\sigma}(Y) )\cdot  e_j^\top \left((X_M^\top X_M)^{-1}s\odot \omega^{-1}\right).
	\end{split}
\end{equation}

	Then, the inequality in (\ref{36.sigY}) is equivalent to the following quasi-linear inequality:
	\begin{equation}
		\begin{split}
			-s_j\cdot  e_j^\top X_M^+ Y\leq\hat{\sigma}(Y)\cdot b_j.
		\end{split}
	\end{equation}
	Note that $e_j^\top X_M^+\in\RR^{1\times T}$ is the $j$th row of $X_M^+$, so we stack $s_j\cdot  e_j^\top X_M^+$ into a $\RR^{|M|\times T}$ matrix across all $j\in M$, which is $diag(s)X_M^+$. Similarly, we stack $b_j$'s into a vector $b$. 
	
This enables us to rewrite (\ref{36.sigY}) in matrix form across $j\in M$:
	\begin{equation}\label{39}
		C Y\leq \hat{\sigma}(Y)\cdot b,\q\wh\q
			C=-diag(s) X_M^+.
	\end{equation}

Now, it remains for us to construct the specific form of the truncation by solving the quasi-linear inequalities (\ref{39}). By definition, $\eta=(X_M^+)^\top e_j$ so $\eta^\top \eta =((X_M^\top X_M )^{-1})_{jj}$. 

Let $d=\tr(I_T-P_M)$. We note that \cite{tian2017selective} uses another variance estimator $\hat{\sigma}_P^2(Y)=\frac{\|(I-P_M)Y\|_2^2}{d}$. We begin by showing
\begin{equation}
	\tr P_M
	=\tr( X_MX_M^+)
	=\tr \left(X_M(X_M^\top X_M)^{-1}X_M^\top\right)
	=\tr \left( (X_M^\top X_M)^{-1}X_M^\top X_M\right) = |M|.
\end{equation}

Thus, $d=T-|M|$, and we continue to use $\hat{\sigma}^2(Y)=\frac{\|(I-P_M)Y\|_2^2}{T-|M|}$, and our inequalities of (\ref{39})  match those of equation (15) of \cite{tian2017selective}.

Let $\eta'=\frac{\eta}{\|\eta\|_2}$. Clearly $\|\eta'\|=1$ and $\eta'(\eta')^\top=\eta\eta^\top/\|\eta\|_2^2$. In addition, we see that
\begin{equation}
P_M\eta=X_MX_M^+\eta=X_M\underbrace{(X_M^\top X_M)^{-1}X_M^\top}_{X_M^+}
\underbrace{((X_M^\top X_M)^{-1}X_M^\top)^\top e_j}_{\eta}=X_M(X_M^\top X_M)^{-1} e_j=\eta.
\end{equation}
\begin{equation}
CP_M=\underbrace{-  diag(s) X_M^+}_{C}
\underbrace{X_MX_M^+}_{P_M}=- diag(s) X_M^+=C.
\end{equation}
So we obtain $P_M\eta'=\eta'$ and $CP_M=C$, and we satisfy the requirements in equation (16) of \cite{tian2017selective} as well. 

Given Assumption A.1 and since our hypothesis is $\beta_M=0$, we obtain the post-selection law $\mathbb{M}_{(C,b,P)}$ that is the same as the premise of Theorem 1 in \cite{tian2017selective}. In particular, our $b,C,\eta', P_M$ would correspond to their quantities $b,C,\eta,P$ in \cite{tian2017selective}. Then, we construct $\nu=C\eta'$ and $\xi=C (P_M-\eta\eta^\top /\|\eta\|_2^2 )Y$, as well as the $W=\|(I-P_M)Y\|_2^2+\|\eta^\top Y\|_2^2/\|\eta\|_2^2$.

This allows us to conclude that
\begin{equation}
	\frac{\eta^TY-0}{\|\eta\|\hat{\sigma}(Y)}
	| (P_M-\eta\eta^T /\|\eta\|_2^2)Y,\|Y\|_2^2\stackrel{D}{=}\mathcal{TT}_{d;\Omega},\q\wh \Omega=
		\bigcap_{j\in M}\{t:t\sqrt{W}\nu_j+\xi_j\sqrt{d+t^2}\leq b_j\sqrt{rW}\}
	\end{equation}

	By our Lemma A.1, it holds that $\eta^\top Y=\bar{\beta}_j$. Since $(P_M-\eta\eta^T)Y$ and $\|Y\|_2^2$ are both measurable with respect to $\tilde{\mathcal{M}}$, we conclude that
	\begin{equation}\label{40.0}
		\frac{\bar{\beta}_j}{\|\eta\|\hat{\sigma}(Y)}| \tilde{\mathcal{M}}\stackrel{D}{=}\mathcal{TT}_{d;\Omega}.
	\end{equation}
	The $p$-value follows as consequence of (\ref{40.0}).

	\subsection{Proof of Theorem A.3}\label{proof_COL2}
	Denote the studentized coefficient $	\frac{\bar{\beta}_j}{\|\eta\|\hat{\sigma}(Y)}$ estimated with data up to time $T$ as $S_T$. Our proof strategy is as follows. First, we obtain the result of Theorem A.1 when $\Sigma=\sigma^2I$. Second, we show the distribution of $S_T$ converges in distribution to a truncated Gaussian as $T\to\infty$. Finally, we show that the truncation on $Y$ described by Theorem A.2 is contained in the truncation given in Theorem A.1, which means that the $p$-value calculated from the Gaussian with the truncation of Theorem A.1 satisfies Assumption 1 of the main text. 
	In more detail, the steps are as follows:
	\begin{enumerate}
		\item[(i)] First, $\bar{\beta}_j/\sqrt{\eta^T\Sigma\eta}\sim \mathcal{TN}(0,1;[V^{-}(z)/\sqrt{\eta^T\Sigma\eta},V^{+}(z)/\sqrt{\eta^T\Sigma\eta}])$ by Theorem A.1. In the case of $\Sigma=\sigma^2 I$, this becomes $\bar{\beta}_j/\|\eta\|_2\sigma \sim \mathcal{TN}(0,1;[V^{-}(z)/\|\eta\|_2\sigma,V^{+}(z)/\|\eta\|_2\sigma])$.
		
		\item[(ii)] Second, for $T\to\infty$, we obtain the following asymptotic results. It holds that $d=\tr(I-P_M)=T-|M|\to\infty$ as $T\to\infty$ and under Assumption A.3 we have $\hat{\sigma}^2(Y)\CP\sigma^2$ as $T\to\infty$. On the other hand, $d=\tr(I-P_M)=T-|M|\to\infty$ as $T\to\infty$. By Lemma 13 of \cite{10.1214/13-EJS815}, the distribution in (\ref{40.0}), that is not conditioned on the selection $\tilde{\mathcal{M}}$, satisfies 
		\begin{equation}\label{56}
			S_T\stackrel{D}{=} t_d\CD\mathcal{N}_{0,1},\q \text{ as }T\to\infty
		\end{equation}		
		\item[(iii)] Lastly, we establish the following claim about the truncation:\\		
		\textbf{Claim:} The truncation of $Y$ in Theorem A.2 is asymptotically same as the truncation of $Y$ in Theorem A.1, and the truncation of $Y$ in Theorem A.1 exists as $T\to\infty$.\\
		\textbf{Proof of the claim:} We consider the cases of active and non-active covariates separately.
		
		\textbf{Case 1: Truncation associated with active covariates}\\
		We start by looking at the truncation of $Y$ in equation (\ref{39}) due to active covariates, which according to Theorem A.2's (\ref{trunc_b_normal_lambda}) satisfies
		\begin{equation}\label{62}
			\begin{split}
				C Y&\leq \hat{\sigma}(Y)\cdot  \tilde{b}\\\wh&
				\begin{cases}
					C=-diag(s) X_M^+\\
										\tilde{b}_j=- (\lambda s_j/\hat{\sigma}(Y) )\cdot  e_j^\top \left((X_M^\top X_M)^{-1}s\odot \omega^{-1}\right)
				\end{cases}.
			\end{split}
		\end{equation}
		In Theorem A.1 we show the truncation corresponding to active covariates is
		\begin{equation}\label{63}
			\begin{split}
				A_1
				Y\leq b_1,\q\wh\q\begin{cases}
					A_1=-diag(s)X_M^+
					\\
					b_1=-\lambda \cdot diag(s)(X_M^\top X_M)^{-1}s\odot\omega^{-1}_M.
				\end{cases}
			\end{split}
		\end{equation}

		As $C=A_1$ and $b_1=\hat{\sigma}(Y)\tilde{b}$, the inequalites of (\ref{62}) and (\ref{63}) are equivalent for all $T$.
  
  It remains to check that the quantities in inequalities (\ref{63}) have limits as $T\to\infty$.
		 To begin with, we note that
		\begin{equation}
CY=A_1Y=-diag(s)X_M^+Y=-diag(s)\bar{\beta}_M.
		\end{equation}
The $j$th coordinate of $diag(s)\bar{\beta}_M$ is simply $s_j\bar{\beta}_j$ for $j\in M$, so left-hand side of the inequalities in (\ref{63}) exists as $T\to\infty$ as long as $\bar{\beta}_j$ exists. 
		
		On the other hand, the $j$th coordinate of $b_1$ is
\begin{equation}
b_1(j)
	= -\lambda s_je_j^\top(X_M^\top X_M)^{-1}s\odot\omega^{-1}_M
	= -\frac{1}{T}\lambda s_je_j^\top(\frac{1}{T}X_M^\top X_M)^{-1}s\odot\omega^{-1}_M
\end{equation}
		Since $\|s\odot\omega^{-1}_M\|_2\leq \sqrt{J}$, by Cauchy-Schwartz we have
		\begin{equation}\label{66}
			|b_1(j)|
			\leq  \frac{1}{T}\lambda \| (\frac{1}{T}X_M^\top X_M)^{-1}e_j\|_2\|s\odot\omega^{-1}_M\|_2
    \leq   \frac{\sqrt{J}}{T}\lambda \| (\frac{1}{T}X_M^\top X_M)^{-1}e_j\|_2
		\end{equation}
	By Assumption A.5(a), there exists a full-rank $G$ such that $\lim\limits_{T\to\infty}\frac{1}{T}X_M^\top X_M=G$, so
	\begin{equation}\label{exist_inv_G}
\lim\limits_{T\to\infty}(\frac{1}{T}X_M^\top X_M)^{-1}= G^{-1}.
	\end{equation}
Thus, $ (\frac{1}{T}X_M^\top X_M)^{-1}e_j$ converges to the $j$th row of $G^{-1}$, and $\| (\frac{1}{T}X_M^\top X_M)^{-1}e_j\|_2$ is bounded by $\sqrt{|M|}$ times the largest value of the $j$th row of $G^{-1}$, and hence finite.

 By Assumption A.5(b), $\lambda \sqrt{J}/T\to 0$.  Thus, (\ref{66}) gives us $b_1(j)\to 0$ as $T\to\infty$, and the right-hand side of the inequalities in (\ref{63}) exists as $T\to\infty$, In conclusion, the truncation associated with the activate covariates converges to the ones corresponding to the active covariates in Theorem A.1.
	
		\textbf{Case 2: Truncation associated with inactive covariates}\\
		We use equation (12) in Lemma 3 of \cite{tian2017selective} to write down the constraints from inactive-covariates. Let $\kappa= 
		(X_M^+)^\top s\odot\omega^{-1}_{M}$, then:
		\begin{equation}\label{eq-truncation-import}
			-\omega_{-M}^{-1}-X_{-M}^\top (X_M^+)^\top s\odot \omega_{M}^{-1}<\sqrt{\frac{1-\tilde{\lambda}^2\kappa^2_2}{\tilde{\lambda}^2}}\frac{X_{-M}(I-P_M)Y}{\|(I-P_M)Y\|_2}
			\leq 	\omega_{-M}^{-1}-X_{-M}^\top (X_M^+)^\top s\odot \omega_{M}^{-1}.
		\end{equation}
		Converting $\tilde{\lambda}$ to $\lambda$, we can rewrite the middle term as
		\begin{equation}
			\begin{split}
				&	\sqrt{\frac{1-\tilde{\lambda}^2\kappa^2}{\tilde{\lambda}^2}}\frac{X_{-M}(I-P_M)Y}{\|(I-P_M)Y\|_2}\\
				=	&\sqrt{\frac{1-	\frac{\lambda^2}{\hat{\sigma}^2(Y)\cdot (T-|M|)+\kappa^2\lambda^2}\kappa^2}{	\frac{\lambda^2}{\hat{\sigma}^2(Y)\cdot (T-|M|)+\kappa^2\lambda^2}}}\frac{X_{-M}(I-P_M)Y}{\|(I-P_M)Y\|_2}\q\text{(convert to $\lambda$)}\\
				=	&\sqrt{\frac{\|(I-P_M)Y\|_2^2+\kappa^2\lambda^2-\lambda^2}{\lambda^2}}\frac{X_{-M}(I-P_M)Y}{\|(I-P_M)Y\|_2}\\
				=	&\sqrt{\frac{\|(I-P_M)Y\|_2^2+\kappa^2\lambda^2-\lambda^2}{\|(I-P_M)Y\|_2^2\lambda^2}}\cdot X_{-M}(I-P_M)Y\q\text{(move into square-root)}\\
				=		&\sqrt{\lambda^{-2}+\frac{\kappa^2}{\|(I-P_M)Y\|_2^2}-\|(I-P_M)Y\|_2^{-2}}\cdot X_{-M}(I-P_M)Y.
			\end{split}
		\end{equation}
		By Assumption A.4, it holds that $\hat{\sigma}^2(Y)=\|(I-P_M)Y\|_2^2/(T-|M|)\stackrel{p}{\to}\sigma^2$, and hence $\|(I-P_M)Y\|_2^{-2}$ is $O_p(T^{-2})$. By Assumption 5(a) and (b), $\lambda^{-2}$ is $O(\frac{T}{\log T})$, dominating $\|(I-P_M)Y\|_2^{-2}$.
	
		Moreover, let $u=s\odot\omega^{-1}_{M}/\|s\odot\omega^{-1}_{M}\|_2\in\RR^{|M|}$, and note that $\|s\odot\omega^{-1}_{M}\|_2\leq \sqrt{J}$. Then, we have:
		\begin{equation}\label{70}
			\begin{split}
\frac{\kappa^2}{\|(I-P_M)Y\|_2^2}
&=
\frac{(s\odot\omega^{-1}_{M})^\top (X_M^+) (X_M^+)^\top s\odot\omega^{-1}_{M}
}{\hat{\sigma}^2(Y)(T-|M|)}
\\
&=
\frac{(s\odot\omega^{-1}_{M})^\top (X_M^\top X_M)^{-1} s\odot\omega^{-1}_{M}
}{\hat{\sigma}^2(Y)(T-|M|)}\\
&\leq
\frac{J}{T} \frac{u^\top (X_M^\top X_M)^{-1} u}{\hat{\sigma}^2(Y)(T-|M|)} .	\end{split}
		\end{equation}
			By Assumption A.5(a), we have $\lim\limits_{T\to\infty}(\frac{1}{T}X_M^\top X_M)^{-1}=G^{-1}$ and $u^\top G^{-1} u$ is bounded from above. By Assumption A.5(b), it holds that $J=O(T)$, and by Assumption A.4 $\hat{\sigma}^2(Y)\stackrel{p}{\to}\sigma^2$. Thus, the quantity in last line of (\ref{70}) satisfies:
\begin{equation}
	\frac{J}{T} \frac{u^\top (X_M^\top X_M)^{-1} u}{\hat{\sigma}^2(Y)(T-|M|)} =
\underbrace{\frac{J}{T(T-|M|)}}_{O(1/T)}
\underbrace{u^\top (\frac{1}{T}X_M^\top X_M)^{-1} u }_{\stackrel{T\to\infty}{\to} u^\top G^{-1} u, }
\underbrace{\frac{1}{\hat{\sigma}^2(Y)}}_{\stackrel{T\to\infty}{\to}\frac{1}{\sigma^2}}
  = o_p(T).\\
\end{equation}		
Therefore, $\lambda^{-2}$ dominates $\frac{\kappa^2}{\|(I-P_M)Y\|_2^2}
$ as well, when as $T\to\infty$, and the first term of the product converges to
			 \begin{equation}
\sqrt{\lambda^{-2}+\frac{\kappa^2}{\|(I-P_M)Y\|_2^2}-\|(I-P_M)Y\|_2^{-2}}\to\lambda^{-1}.
			 \end{equation}
	
		On the other hand, in Theorem A.1 the truncation corresponding to active covariates equals
		\begin{equation}
			-\omega_{-M}^{-1}-X_{-M}^\top (X_M^+)^\top s\odot \omega_{M}^{-1}<\lambda^{-1}{X_{-M}(I-P_M)Y}
			\leq \omega_{-M}^{-1}-X_{-M}^\top (X_M^+)^\top s\odot \omega_{M}^{-1}.
		\end{equation}
		
		This shows that the truncation for inactive covariates (\ref{eq-truncation-import}) converges to the one corresponding to inactive covariates in Theorem A.1
	\end{enumerate}
	
	We now combine the two steps. For $T\to\infty$, (iii) concludes that the truncation converges to the one in Theorem A.1 when $T\to\infty$. The truncation is a finite intersection of convex sets with an interior, given we observe a fitted LASSO from solving the KKT conditions. Combining it with (i) and (ii), we conclude the distribution result
	\begin{equation}\label{eq:converge_t}
		\frac{\bar{\beta}_j}{\|\eta\|\hat{\sigma}(Y)}| \tilde{\mathcal{M}}\CD\mathcal{TN}_{0,1;V^{-}(z)/\|\eta\|_2\sigma,V^{+}(z)/\|\eta\|_2\sigma}.
	\end{equation}
	For any random variable $X$ and its corresponding cumulative distribution function (cdf) $F(\cdot)$, it holds that $F(X)\sim \text{Unif}[0,1]$. Thus, by (\ref{eq:converge_t}), under the null and as $T\to\infty$, it holds that $\Phi^{\Omega}( \frac{\bar{\beta}_j}{\|\eta\|\hat{\sigma}(Y)})\CD \textrm{Unif}[0,1]$, where $\Omega=[V^{-}(z)/\|\eta\|_2\sigma,V^{+}(z)/\|\eta\|_2\sigma]$ and $\Phi^{\Omega}$ is the cdf of a standard normal distribution truncated with $\Omega$.


\subsection{Implementation of Weighted-LASSO}

\subsubsection*{Calculation of $p$-values using Theorem A.3}

Numerical stability is a practical concern when applying estimation methods. As we calculate potentially the far tail of a truncated Gaussian CDF, we propose a set of best practices for fast and stable calculations with our methods:
\begin{enumerate}
	\item Use the logarithmic transformation of $p$-values instead of $p$-values.\\
	This is motivated by three reasons:
	\begin{enumerate}
		\item Given finite machine accuracy (such as 16-bit float numeric storage) we are susceptible to arithmetic underflow when our $\bar{\beta}$'s are in the far tail and the $p$-values are very close to 0. The logarithmic transformation addresses this issue.
		\item We need to distinguish between small $p$-values because eventually our procedures need to compare $N_jp_j$ for different covariates $j$.
		\item We can save time and preserve numerical accuracy by avoiding an additional log transform step when calculating R\'enyi statistics.
	\end{enumerate}
	\item Use an approximation when the two tails are vastly different.\\
	Without loss of generality, we consider the case of $\bar{\beta}_{M(i)}>0$ and we label the two building blocks of our proposed $p$-value as
	\begin{equation}
		\begin{split}
			LeftTail=\ln\left(F_{TN}-(
			\frac{\bar{\beta}_{M(i)}}{\|\eta\|_2\hat{\sigma}})\right)\\
			RightTail=\ln\left(
			F^C_{TN}(\frac{\bar{\beta}_{M(i)}}{\|\eta\|_2\hat{\sigma}})\right).
		\end{split}
	\end{equation}
	The correct log $p$-value are
	\begin{equation}
		\ln( p)=\ln\left(\exp(LeftTail)+\exp(RightTail)\right).
	\end{equation}
	We propose to consider the following accurate approximation, which speeds up the calculations:
	\begin{equation}
		\ln( p)=
		\begin{cases}
			\ln\left(\exp(LeftTail)+\exp(RightTail)\right) &|LeftTail-RightTail|<\zeta\\
			\max\{LeftTail,RightTail\} &o.w.\\
		\end{cases}
	\end{equation}
	In our empirical study, we use $\zeta=10$ as the largest $N_j\ll10000$ in our data, and $\log \left(\max_j(N_j)p_j\right)\leq 10+\log p_j$.
\end{enumerate}

\subsubsection*{Additional Implementation Details}

There are additional implementational details for LASSO. We summarize what we consider good practice below:
\begin{enumerate}
	\item Search range for the $\ell_1$ penalty scalar $\lambda$: \\We propose candidates by a log-linear sequence of numbers multiplied by $
	\frac{\log J}{\sqrt{|T_{selection}|}}$, inspired by \cite{10.1214/17-AOS1630}.
	
	\item Cross-validation (CV henceforth): \\We use 5-fold random splits of the training data to cross-validate across candidate set of $\lambda$'s.
	
	\item Criteria for CV: \\Our cross-validation follows the one-standard-deviation rule for selecting parsimonious models, that is, the largest choice of $\lambda$ within one standard error of minimizing the squared errors. This is the default setting of popular implementations like \texttt{glmnet} and argued for in \S3.4 of \cite{hastie2009elements}.
	
	\item Refit after CV: \\We refit the sparse model using the entire training data after selecting the best $\lambda$.
\end{enumerate}

\bibliographystyle{econometrica}
{\small
	\bibliography{main}
}